%% file: ms.tex
\documentclass[journal,twoside]{IEEEtran}
		\usepackage[utf8]{inputenc}
		\usepackage{cite}
		\usepackage{graphicx}
		\usepackage{tikz}
			\usetikzlibrary{shapes.geometric}
			\usetikzlibrary{calc}
		\usepackage{tikz-3dplot}
		\usepackage{datatool}
		\usepackage{pgfplots,pgffor}
			\pgfplotsset{compat=newest,every axis/.append style={font=\small},}
		\usepackage{amsmath,amssymb,amsthm}
		\usepackage{url}
		\usepackage{etoolbox}
		\usepackage{xpatch}
		\usepackage{algorithmic}
			\makeatletter
			\xpatchcmd{\algorithmic}{\itemsep\z@}{\itemsep=8pt plus4pt}{}{}
			\makeatother
		\usepackage{array}
		\usepackage[caption=false,font=normalsize,labelfont=sf,textfont=sf]{subfig}
	
		\usepackage{subfiles}
		\usepackage{hyperref}

		\IfFileExists{macros.tex}{\input{macros.tex}}{\input{../macros.tex}}
		
\begin{document}

			\title{Cell Detection by Functional Inverse Diffusion and Non-negative Group Sparsity---Part II: Proximal Optimization and Performance Evaluation}
			\author{
				\IEEEauthorblockN{Pol del Aguila Pla, \emph{Student Member, IEEE}, and Joakim Jaldén, \emph{Senior Member, IEEE}\thanks{
				Manuscript received September 21, 2017; revised March 31, 2018 and July
				9, 2018; accepted August 24, 2018. Date of publication September 3, 2018;
				date of current version September 14, 2018. The associate editor coordinating
				the review of this manuscript and approving it for publication was Prof. Mark
				A. Davenport. This work was supported in part by Mabtech AB and in part by
				the Swedish Research Council (VR) under Grant 2015-04026. \emph{(Corresponding
				author: Pol del Aguila Pla.)} }\thanks{
				The authors are with the Department of Information Science and Engineering,
				School of Electrical Engineering and Computer Science, KTH Royal
				Institute of Technology, Stockholm 11428, Sweden (e-mail:, poldap@kth.se;
				jalden@kth.se).} \thanks{
				This paper has supplementary downloadable material available at http://
				ieeexplore.ieee.org, provided by the authors. The material includes detailed
				derivations of some key steps and further experimental results. This material is
				$533$ kB in size.} \thanks{
				Color versions of one or more of the figures in this paper are available online
				at http://ieeexplore.ieee.org.}\thanks{
				\textbf{Author's own archival version.} Digital Object Identifier of the original manuscript: 10.1109/TSP.2018.2868258.}}
				}
		\makeatletter
		\ifCLASSOPTIONpeerreview
			\markboth{IEEE TRANSACTIONS ON SIGNAL PROCESSING, VOL. 66, NO. 20, OCTOBER 15, 2018 
			}%
			{\MakeUppercase{\@title}}
		\else
			\markboth{IEEE TRANSACTIONS ON SIGNAL PROCESSING, VOL. 66, NO. 20, OCTOBER 15, 2018 
			}%
			{DEL AGUILA PLA AND JALD\'EN: CELL DETECTION BY FUNCTIONAL INVERSE DIFFUSION AND NON-NEGATIVE GROUP SPARSITY---PART II}
		\fi
		\makeatother
		\IEEEpubid{\begin{minipage}{\textwidth}
			    \vspace{20pt}
			    \begin{center}
			      1053-587X~\copyright~2018 IEEE. Translations and content mining are permitted for academic research only. 
			      Personal use is also permitted, but republication/redistribution requires IEEE permission. 
			      See http://www.ieee.org/publications standards/publications/rights/index.html for more information.
			    \end{center}
			   \end{minipage}
		}
	
	\maketitle

	\begin{abstract}
		\input{\secs/abstract_2}
	\end{abstract}

		\ifCLASSOPTIONpeerreview
		\begin{center} 
			\bfseries EDICS Category: 3-BBND 
		\end{center}
		\fi
		\IEEEpeerreviewmaketitle
	
			\providebool{tot}
			\booltrue{tot}

	
		\section{Introduction} \label{sec:Intro}

\subfile{\secs/Intro_2}

		\section{Accelerated Proximal Gradient for Weighted Group-Sparse Regularized Inverse Diffusion} \label{sec:algs}

\subfile{\secs/Algs.tex}

		\section{Numerical Results} \label{sec:NumRes}

\subfile{\secs/NumRes}

		\section{Discussion} \label{sec:Discussion}

\subfile{\secs/Discussion}

		\appendix[Constraints and Regularization, Proximal Operators]

\subfile{\secs/AppProxOps}

		\section*{Acknowledgments}
		
		Professor Radu Bot, through his lecture series \emph{Recent Advances in Numerical Algorithms for Convex Optimization} given at KTH Royal Institute of Technology in May 2016,
		was of great help in the development of this research. Doctor Christian Smedman provided expert labeling of synthetic data.
		The authors would like to thank Professor Mário A. T. Figueiredo for first pointing them to \cite{Pustelnik2017} during the workshop \emph{Generative models, parameter learning and sparsity} at the Isaac Newton Institute for Mathematical Sciences.
		The excellent team of anonymous reviewers provided feedback 
		that improved the presentation of our results considerably. 
		Of particular relevance were their pointing
		to optimal transport and the EMD, and their suggestion of simpler proof techniques for Lemma~\ref{prop:ConjFunctional}.

		\IEEEtriggeratref{42} 

		\bibliographystyle{IEEEtranMod}
		\bibliography{IEEEabrv,\bib/multi_deconv}

		
		\vspace{-108pt}
		
		\begin{IEEEbiography}[{\includegraphics[width=1in,height=1.25in,clip,keepaspectratio]{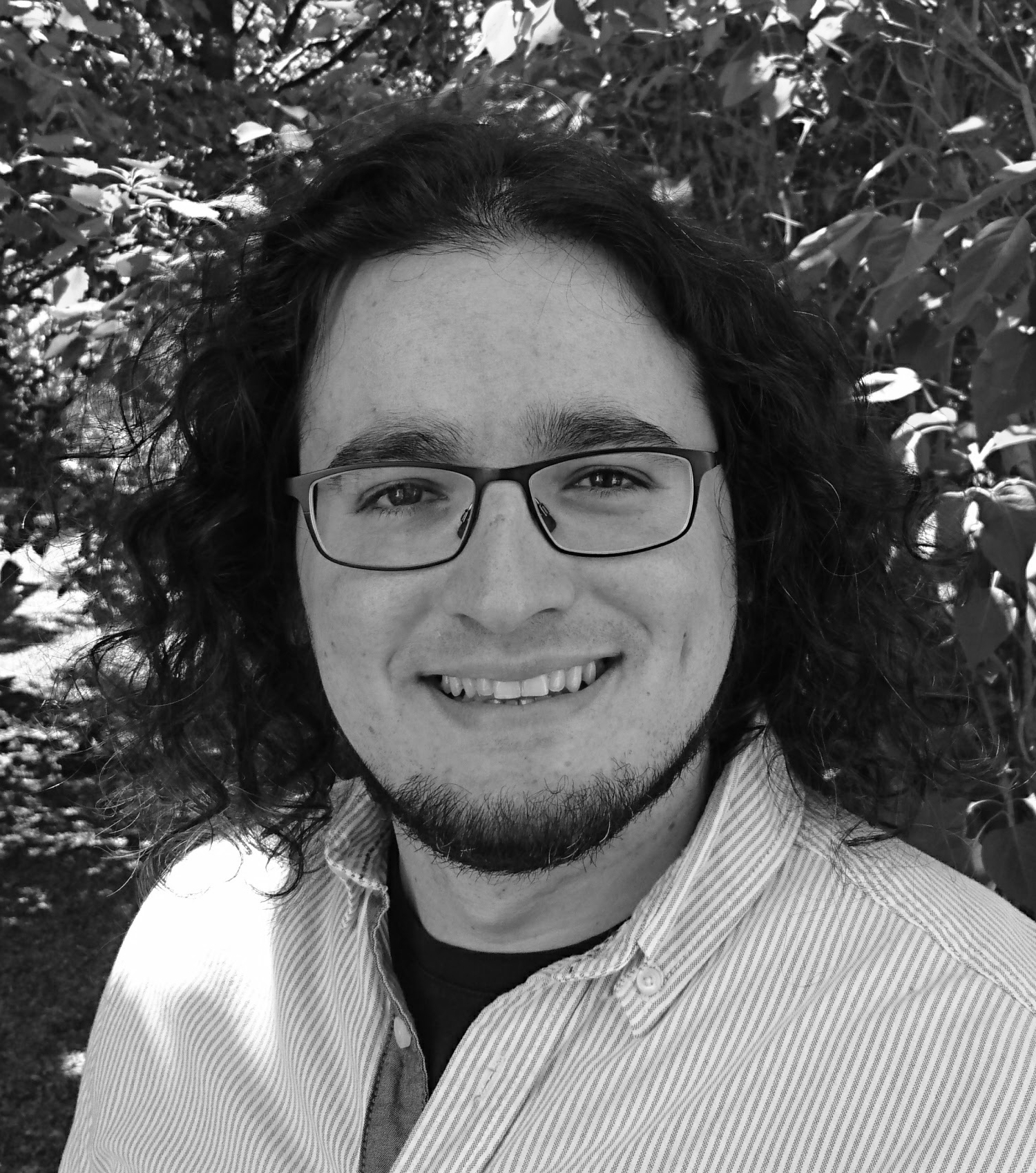}}]{Pol del Aguila Pla}
			a (S’15) received a double degree in
telecommunications and electrical engineering from the Universitat Polit\`ecnica de Catalunya, Barcelona,
Spain, and the Royal Institute of Technology (KTH),
Stockholm, Sweden, in 2014. Since August 2014,
he is currently working toward the Ph.D. degree
in electrical engineering and signal processing with
KTH under the supervision of Joakim Jalden. His
Ph.D. work includes the research collaboration with
Mabtech AB that led to the results published here
and the development of the ELISPOT and Flourospot
reader Mabtech IRIS\textsuperscript{TM} . Since August 2015, he is a Reviewer for the \textsc{IEEE
Transactions on Signal Processing}. During 2017, he received a number of
grants to support the international promotion of his research in inverse problems
for scientific imaging, including a 2017 KTH Opportunities Fund scholarship,
a Knut and Alice Wallenberg Jubilee appropriation, an Aforsk Foundation’s
scholarship for travel and a 2017 Engineering Sciences grant from the Swedish
Academy of Sciences (KVA, ES2017-0011).
		\end{IEEEbiography}	
		
		\vspace{-95pt}
		
		\begin{IEEEbiography}[{\includegraphics[width=1in,height=1.25in,clip,keepaspectratio]{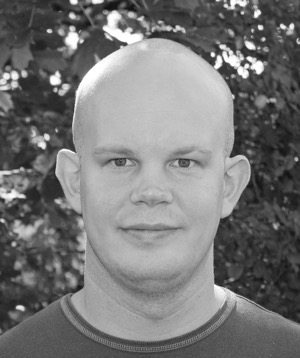}}]{Joakim Jald\'{e}n} 
			(S’03–M’08–SM’13) received the
M.Sc. and Ph.D. degrees in electrical engineering
from the Royal Institute of Technology (KTH),
Stockholm, Sweden, in 2002 and 2007, respectively.
From July 2007 to June 2009, he held a postdoctoral
research position with the Vienna University
of Technology, Vienna, Austria. He also studied at
Stanford University, Stanford, CA, USA, from
September 2000 to May 2002, and worked at ETH,
Zurich, Switzerland, as a Visiting Researcher, from
August to September, 2008. In July 2009, he returned
to KTH, where he is currently a Professor of signal processing. His recent work
includes work on signal processing for biomedical data analysis, and the automated
tracking of (biological) cell migration and morphology in time-lapse
microscopy in particular. Early work in this field was awarded a conference
best paper Award at IEEE ISBI 2012, and subsequent work by the group has
been awarded several Bitplane Awards in connection to the ISBI cell tracking
challenges between 2013 and 2015. He was an Associate Editor for the \textsc{IEEE
Communications Letters} between 2009 and 2011, and an Associate Editor
for the \textsc{IEEE
Transactions on Signal Processing} between 2012 and 2016.
Since 2013, he has been a member of the IEEE Signal Processing for Communications
and Networking Technical Committee, where he is currently a
Vice-Chair. Since 2016, he has also been responsible for the five year B.Sc and
M.Sc. Degree Program in electrical engineering with KTH.

For his work on MIMO communications, he has been awarded the IEEE
Signal Processing Societies 2006 Young Author Best Paper Award, the Distinguished
Achievement Award of NEWCOM++ Network of Excellence in
Telecommunications 2007–2011, and the best student conference paper Award
at IEEE ICASSP 2007. He is also the recipient of the Ingvar Carlsson Career
Award issued in 2009 by the Swedish Foundation for Strategic Research.
		\end{IEEEbiography}

\end{document}

%% file: macros.tex
	\def\natu{\mathbb{N}}
	\def\reals{\mathbb{R}}
	\def\integ{\mathbb{Z}}
	
	
	
	
	\newcommand{\Proj}[2]{\operatorname{P}_{#1}\left[ #2 \right]}
	\newcommand{\ProjOp}[1]{\operatorname{P}_{#1}}
	\newcommand{\supp}[1]{\operatorname{supp}\left( #1 \right)}
	\renewcommand{\complement}[1]{#1^{\mathsf{c}}}
	\def\prox{\operatorname{prox}}
	\def\sigmax{\sigma_{\max}}
	\def\dsigma{\tilde{\sigma}}
	\def\dsigmax{\tilde{\sigma}_{\max}}
	\def\ka{\kappa_\mathrm{a}}
	\def\kd{\kappa_\mathrm{d}}
	\def\pos{\mathbf{r}}
	\def\ae{\mathrm{a.e.}}
	\def\obs{d_{\mathrm{obs}}}
	\def\dobs{\tilde{d}_{\mathrm{obs}}}
	
	\def\xp{x^*_{\mathrm{p}}}
	\def\xn{x^*_{\mathrm{n}}}
	\def\wninf{\left\| w \right\|_{\mathrm{L}^\infty\left(\reals^2\right)}}
	\def\dint{\mathrm{d}}
	

	\newcommand{\listint}[1]{\lbrace 1,2,\dots,#1\rbrace}
	\newcommand{\matrices}[1]{\operatorname{\mathbb{T}}\left( #1 \right)}
	\newcommand{\matricesP}[1]{\operatorname{\mathbb{T}_+}\left( #1 \right)}
		\def\DenSpace{\mathcal{D}}
		\newcommand{\prodDen}[2]{ \left( #1|#2 \right)_{\DenSpace} }
		\newcommand{\normDen}[1]{ \left\|#1\right\|_{\DenSpace} }
		\def\ASpace{\mathcal{A}}
		
		\newcommand{\normA}[1]{ \left\|#1\right\|_{\ASpace} }
		\newcommand{\Leb}[2]{\mathrm{L}^#1\left( #2 \right)}
		\newcommand{\LebTwo}[1]{\Leb{2}{#1}}
		\newcommand{\LebOne}[1]{\Leb{1}{#1}}
		\newcommand{\normLebTwo}[2]{ \left\| #2 \right\|_{\LebTwo{#1}} }
		\newcommand{\normLebOne}[2]{ \left\| #2 \right\|_{\LebOne{#1}} }
		\newcommand{\prodLebTwo}[3]{ \left( #2|#3 \right)_{\LebTwo{#1}} }
		\newcommand{\LebTwoP}[1]{\mathrm{L}^2_+\left( #1 \right)}
		\newcommand{\opers}[2]{ \mathcal{L}\left(#1,#2\right) }
		
		\def\X{\mathcal{X}}
		\newcommand{\prodX}[2]{ \left( #1|#2 \right)_{\X} }
		\newcommand{\linfunX}[2]{ \left\langle #1,#2 \right\rangle_{\X} }
		\newcommand{\normX}[1]{ \left\|#1\right\|_{\X} }
		\newcommand{\normXs}[1]{ \left\|#1\right\|_{\X^*} }
		\newcommand{\clellips}[3]{ \bar{\mathcal{B}}^{#1}_{#2}\left({#3}\right)}
		
	\newcommand{\changetoAlg}{ 
		\renewcommand{\figurename}{Alg.}
	}
	\newcommand{\changetoFig}{
		\renewcommand{\figurename}{Fig.}
	}
	
	\makeatletter%
	\@ifclassloaded{beamer}%
		{}%
		{
		\newtheorem{definition}{Definition}
		\newtheorem{lemma}{Lemma}
		\newtheorem{property}{Property}
		\newtheorem{theorem}{Theorem}
		\newenvironment{algorithm}
			{
				\changetoAlg
				\begin{figure}
			}
			{
				\end{figure}
				\changetoFig
			}
		}
		\makeatother%

		\def\figs{figs}
		\def\secs{secs}
		\def\bib{.}
	

	

%% file: secs/abstract_2.tex
In this two-part paper, we present a novel framework and methodology to analyze data
from certain image-based biochemical assays, e.g., ELISPOT and Fluorospot assays.
In this second part, we focus on our algorithmic contributions. We provide an algorithm
for functional inverse diffusion that solves the variational problem we posed in Part I.
As part of the derivation of this algorithm, we present the proximal operator for the non-negative
group-sparsity regularizer, which is a novel result that is of interest in itself, also in 
comparison to previous results on the proximal operator of a sum of functions.
We then present a discretized approximated implementation of our algorithm and evaluate it
both in terms of operational cell-detection metrics and in terms of distributional optimal-transport metrics.

%% file: secs/Intro_2.tex
	\IEEEPARstart{S}{ource} localization (SL) arises in application fields in which a number of point-sources emit some measurable signal,
	e.g., chemical compounds 
	\cite{Olivo-Marin2002,Matthes2005,Hamdi2007,Rebhahn2008,Pan2010a,Smal2010,Kimori2010,Ram2012,Hamdi2012,Zhao2014,Basset2015,Kervrann2016},
	sound \cite{Ehrenfried2006,Markovic2015}, 
	light \cite{Starck2002,Giovannelli2005} 
	or heat \cite{Ternat2012,Zhang2016}, and one wants to recover their location. Typically, the measured
	signal is a map of these locations observed through a linear operator, such as 
	convolution \cite{Ehrenfried2006,Zhao2014} or diffusion \cite{Matthes2005,Hamdi2007,Hamdi2012}. 
	This is the second part of a paper that considers SL in the context of cell detection in image-based biochemical assays. In this case, 
	the source locations are explicit in the source density rate, the reactive term in a reaction-diffusion-adsorption-desorption
	system, from which a single image of the adsorped density at the end of the experiment is measured. For more details on the exact setup,
	its biological application or the physics involved, see Part~I~\cite[Section II]{AguilaPla2017}.

	In Part~I~\cite[Section III]{AguilaPla2017}, we proposed the following optimization problem to detect particle-generating 
	(active) cells in this setting,
	\begin{IEEEeqnarray}{c} \label{eq:InvDif:Regularised}
		\!\!\!\!\!\min_{a\in\ASpace} \left[\! \normDen{Aa - \obs}^2+ \underbrace{\delta_{\ASpace_+}(a) + \!
		\lambda \overbrace{\normLebOne{\reals^2}{\normLebTwo{\reals_+}{\xi a_\pos}}}^{f_1(a)}}_{f(a)} \!\right]\!\!.
	\end{IEEEeqnarray}	
	Here, the non-negative quantity $a$ we aim to recover is the post adsorption-desorption source density rate (PSDR). The PSDR is an equivalent to the 
	source density rate (SDR) where the information on adsorption and desorption have been summarized. In fact, it characterizes the generation of particles across a plane, that 
	we represent by the locations 
	$\pos\in\reals^2$, and a third non-negative dimension $\sigma\geq 0$ that expresses the distance each generated particle has diffused from its origin. 
	Here, non-negative group-sparsity, induced by the regularizer $f$ in \eqref{eq:InvDif:Regularised}, plays a fundamental role. This is because a certain form of grouping \cite{Yuan2006} is key for the end application, 
	i.e., cell detection, but $a$ has to remain non-negative at all times to preserve its physical meaning. 
	In particular, it is important that the different values of $a_\pos(\sigma)=a(\pos,\sigma)$ across the different distances $\sigma$ for a certain position $\pos$ are grouped, because
	they represent the same potential active cell generating particles which are captured either closer to the cell (low $\sigma$s) or further away (large $\sigma$s). To our knowledge, previous techniques for promoting group-sparsity,
	e.g., \cite{Teschke2007,Fornasier2008}, can not handle non-negativity constraints.
	
	\IEEEpubidadjcol	
	
	The Hilbert space where this PSDR lies is defined as
		$\ASpace = \left\lbrace 
			a \in  \LebTwo{\Omega} : \supp{a} \subseteq \supp{\mu} \times [0,\sigmax] 
		\right\rbrace$, 
	where $\Omega=\reals^2 \times \reals_+$, $\mu$ is a $(0,1)$-indicator function of a bounded set $\supp{\mu}$ in which cells can physically lie, and $\sigmax=\sqrt{2DT}$ is given by the physical
	parameters of the assay. 
	The image observation $\obs$ lies in a weighted $\LebTwo{\reals^2}$ space defined as
		$\DenSpace=\left\lbrace d:\reals^2\rightarrow \reals : \prodDen{d}{d} < +\infty \right\rbrace$, 
	where $\prodDen{d_1}{d_2} = \prodLebTwo{\reals^2}{w^2 \cdot}{\cdot}$ and $w\in \mathrm{L}_+^\infty\left(\reals^2\right)$ is a non-negative bounded weighting function that
	penalizes errors at different locations according to sensor properties.
	The bounded linear operator $A\in\opers{\ASpace}{\DenSpace}$ represents the forward diffusion process, that maps an $a$ onto $\obs$, and
	was derived in \cite[Section II-B]{AguilaPla2017} as the mapping $a \mapsto \int_{0}^{\sigmax} G_\sigma a_\sigma \dint\sigma$, where $G_\sigma$
	is the convolutional operator with 2D rotationally invariant Gaussian kernel of standard deviation $\sigma$.
	The diffusion operator $A$ was extensively characterized in Part~I~\cite[Section III-B]{AguilaPla2017}.
	Finally, the group-sparsity regularizer includes a non-negative bounded weighting function $\xi\in\mathrm{L}_+^{\infty}\left[ 0, \sigmax\right]$ that allows 
	incorporating further prior knowledge in terms of the relative importance of each value of $\sigma$.
	
	In this paper, we derive an accelerated proximal gradient (APG) algorithm to solve \eqref{eq:InvDif:Regularised}. Namely, we combine the characterization of the diffusion operator $A$
	we presented in Part~I~\cite[Section III-B]{AguilaPla2017} with the derivation of the proximal operator of the non-negative group-sparsity regularizer, i.e., of $f$
	in \eqref{eq:InvDif:Regularised}. Furthermore, we present an efficient
	implementation of a discretization of the resulting algorithm and provide thorough performance evaluation on synthetic data, complementing the real data example in Part~I~\cite[Section V-A]{AguilaPla2017}.

	\subsection{Proximal operator of a sum of functions} \label{sec:proxofsums}
	
		Proximal methods for convex optimization \cite{Beck2009,Bauschke2011,Parikh2014} are now prevalent in the signal processing, inverse problems and machine 
		learning communities \cite{Yu2013,Bonettini2015,Pustelnik2017}. This is mainly due to their first-order nature, i.e., that the intermediate
		variables they entail have at most the same dimensionality as the variable one seeks, and to their ability to handle complex, non-smooth shapes
		of the functional to optimize. Consequently, applications are characterized by high-dimensional parameters with rich structure and non-smooth 
		penalizations. 
		
		In the most generic setting, the problems solved by these methods are of the form
		\begin{IEEEeqnarray}{c} \label{eq:GenericOptProblem-AccProxGrad}
			\underset{ x \in\X}{\min} \left[ g(Bx) + f(x) \right]  \,,
		\end{IEEEeqnarray}
		where $f:\X\rightarrow\bar{\reals}$ and $g:\mathcal{G}\rightarrow\bar{\reals}$ are proper, convex, and lower semi-continuous and the domains $\X$ and $\mathcal{G}$ 
		are two real Hilbert spaces. On one hand, a smoothness assumption is made on $g$, namely, that it
		is Fréchet differentiable in $\mathcal{G}$ and has a ${\beta}^{-1}$-Lipschitz continuous Fréchet derivative $\nabla g:\mathcal{G}\rightarrow \mathcal{G}^*$
		for some $\beta>0$. On the other hand, no further structure is imposed on $f$, which can be non-smooth and discontinuous.
		Finally, the bounded linear operator $B\in\opers{\X}{\mathcal{G}}$ has an adjoint $B^*\in\opers{\mathcal{G}}{\X}$ and operator norm 
		$\|B\|_{\opers{\X}{\mathcal{G}}}$.

		The term \emph{proximal} that encompasses these methods
		relates to the proximal operator of the function $f$, which is necessary for a fundamental step in the iterations defined by these algorithms. The proximal operator 
		is a mapping $\prox_{\gamma f}:\X\rightarrow \X$ such that 
		\begin{IEEEeqnarray}{c} \label{eq:prox_def}
			\prox_{\gamma f}(x) = \arg \min_{y\in \X}\left[ \normX{y-x}^2 + 2\gamma f(y) \right]\,.
		\end{IEEEeqnarray}	
		
		A case that generates special interest is that in which $f$ is constructed as a sum of two non-smooth components 
		\cite{Combettes2007,Chaux2009,Yu2013,Pustelnik2017,Gu2017}. In particular, \cite[Proposition 12]{Combettes2007} proved that if $\X=\reals$ and
		$f=f_1 + \delta_\mathcal{Z}$, with $\delta_\mathcal{Z}$ the $(\infty,0)$-indicator function of a closed convex subset $\mathcal{Z}\subset\X$, then
		$\prox_f = \ProjOp{\mathcal{Z}} \circ \prox_{f_1}$, where $\circ$ represents composition and $\ProjOp{\mathcal{Z}}$ the projection onto $\mathcal{Z}$. 
		In the context of the derivation of the proximal operator of the non-negative group-sparsity regularizer $f$ in \eqref{eq:InvDif:Regularised}, we provide a contrasting result.
		In particular, in the appendix to this paper, we prove that if $\X=\mathrm{L}^2$, $\mathcal{Z}=\X_+$, and $f_1=\normX{\cdot}$, the inverse order applies, i.e., 
		$\prox_f = \prox_{f_1} \circ \ProjOp{\mathcal{Z}}$\footnote{After the acceptance of this paper, we were informed that this claim, made in our Lemma~\ref{prop:ProxScaPosContrNorm}, was covered by 
		the broader result \cite[Proposition 2.2]{Briceno-Arias2009}. During the revision of this paper, \cite{Combettes2018,Yukawa2018} also presented statements equivalent or encompassing Lemma~\ref{prop:ProxScaPosContrNorm}. }. 
		Combining this result with the separable sum property  allows us to prove that this same order is
		applicable when $f=\lambda f_1+\delta_\mathcal{Z}$ for some $\lambda \geq 0$, $f_1$ is the group-sparsity regularizer with non-overlapping groups, and $\X$ and $\mathcal{Z}$ are as above.
		Besides allowing us to solve \eqref{eq:InvDif:Regularised}, 
		the proximal operator for the non-negative group sparsity 
		regularizer facilitates the use of group-sparsity in other fields
		that inherently require non-negativity constraints, e.g., 
		classification, text mining, environmetrics, speech recognition and
		computer vision \cite{Chen2009,Combettes2016}. 
		
	\subsection{Notation} \label{ssec:Notation}
	
		When sets and spaces of numbers are involved, we use either standard notation such as
		$\reals_+=\left[ 0, +\infty\right)$, $\bar{\reals}=\reals \cup \lbrace -\infty, +\infty \rbrace$ 
		and $\bar{\reals}_+=[0,+\infty]$
		or capital non-Latin letters. When discussing locations in $\reals^2$,
		we note them as bold face letters, e.g., $\pos \in \reals^2$.
		
		When discussing functional sets and spaces, we use capital calligraphic notation, such as $\X$ for a generic 
		normed space, $\|\cdot\|_{\X}$ for its norm, and $\left( \cdot | \cdot \right)_{\X}$ for its inner product if $\X$ is also a Hilbert space.
		For any functional space $\X$, $\X_+\subset \X$ is the cone of non-negative functionals, and for any functional $f$, $f_+$ is its positive part, i.e.,
		$	f_+(y) = \max\lbrace f(y), 0 \rbrace$, for any $y$ in its domain $\mathcal{Y}$. The support of the functional $f$ is written as
			$\supp{f} = \lbrace y\in\mathcal{Y}: f(y) \neq 0 \rbrace\subset \mathcal{Y}$.
		For any set $\mathcal{Z} \subseteq \X$, its $(\infty,0)$-indicator function is the function $
		\delta_\mathcal{Z}:\X \rightarrow \lbrace 0,+\infty \rbrace$
		such that $\delta_\mathcal{Z}(x) = 0$ if $x \in \mathcal{Z}$ and $\delta_\mathcal{Z}(x) = +\infty$ if
		$x \in \complement{\mathcal{Z}}=\X\setminus\mathcal{Z}$.
		
		When discussing operators, if $\mathcal{Z}$ is some normed space, we write $\opers{\X}{\mathcal{Z}}$ for 
		the space of linear continuous operators from $\X$ to $\mathcal{Z}$, with norm $\|\cdot\|_{\opers{\X}{\mathcal{Z}}}$.
		We will note operators as $A$ or $B$, e.g., $ B \in \opers{\X}{\mathcal{Z}}$.
		Adjoints are noted
		as $B^*\in \opers{\mathcal{Z}}{\X}$
		
		When discussing matrices and tensors, the space of real $M$-by-$N$ matrices for some $M,N\in\natu$ is $\matrices{M,N}$, while its element-wise 
		positive cone is $\matricesP{M,N}$. For a specific matrix $\tilde{f}\in\matrices{M,N}$, we specify it as a group of its elements 
		$\left\lgroup \tilde{f}_{m,n} \right\rgroup$ for $m\in\listint{M}$ and $n\in\listint{N}$. 
		For tensors, we work analogously by adding appropriate indexes, e.g., $\tilde{f}\in\matrices{M,N,K}$ and $\left\lgroup \tilde{f}_{m,n,k} \right\rgroup$
		for $k\in\listint{K}$.
		
		When presenting our statements, we refer to them as properties if they are not novel, but are necessary 
		for clear exposition,
		lemmas if they contain minor novel contributions and theorems if they constitute major novel contributions.
		
	\ifbool{tot}{}{
		\bibliographystyle{IEEEtran}
		\bibliography{IEEEabrv,\bib/multi_deconv}
	}

%% file: secs/Algs.tex
			In this section, we propose to use the accelerated proximal gradient algorithm (APG algorithm or FISTA \cite{Beck2009}) to solve 
			\eqref{eq:InvDif:Regularised}. Because the APG algorithm is posed in generic Hilbert spaces, we do not need to discretize the problem in order
			to derive and pose our algorithm. In other words, the proposed algorithm will work directly on the abstract parameter space 
			$\ASpace$. For an introduction on optimization in function spaces, see \cite{Luenberger1969}. 
			Any implementation, however, will require some form of discretization. In our case, we choose the simple discretization presented in 
			Part~I~\cite[Section IV]{AguilaPla2017}.
			
	\subsection{Accelerated proximal gradient} 
		\label{subsec:APG}
		\input{\secs/Algs_Generic}
		
	\subsection{Accelerated proximal gradient for weighted group-sparse-regularized inverse diffusion} 
		\label{subsec:APG4us}
		\input{\secs/Algs_InvDif}

	\subsection{Discretization of the APG for inverse diffusion} 
		\label{subsec:APG4us_discrete}
		\input{\secs/Algs_Discr}

	\ifbool{tot}{}{
		\bibliographystyle{IEEEtran}
		\bibliography{IEEEabrv,\bib/multi_deconv}
	}

%% file: secs/Algs_Generic.tex
	The APG algorithm, i.e., Alg.~\ref{algs:GenericAccProxGrad}, was proposed in \cite{Beck2009} to solve \eqref{eq:GenericOptProblem-AccProxGrad} with 
	$t:\natu\rightarrow \reals_+$ such that $t(i)=1/2+[\displaystyle 1/4+t^2(i-1)]^{1/2}$, $\forall i\geq1$ and $t(0) = 1$. 
	\cite{Beck2009} proved that this algorithm yields a sequence of objective values 
	$f(x_i) + g(B x_i)$ with a convergence rate towards the minimum value of $\mathcal{O}\left(i^{-2}\right)$. 
	\cite{Chambolle2015} proposed modifying the update rule of the Nesterov acceleration term to $t(i)=(i+a-1)/a$, $\forall i \geq 1$, for some $a>2$. This modification 
	preserves the convergence rate of the objective values, and additionally grants weak convergence of the iterates, i.e., 
	$x_i \rightarrow  x_{\mathrm{opt}}$ weakly (see \cite{Luenberger1969} for more on weak convergence). As discussed 
	in \cite{Chambolle2015}, convergence is observed empirically with the sequence proposed by \cite{Beck2009} too, and thus, 
	a choice between the two methods will be mainly based on observed empirical results.
	
	\begin{algorithm}
		\begin{algorithmic}[1]
		\vspace{.5pt}\hrule height 1pt \vspace{.5pt}
		\REQUIRE An initial $x^{(0)}\in\X$
		\ENSURE A solution $ x_{\mathrm{opt}} \in \X$ that solves   \eqref{eq:GenericOptProblem-AccProxGrad}
		\vspace{.5pt}\hrule height .5pt \vspace{.5pt}
			\item[] 
			\STATE $y^{(0)} \leftarrow x^{(0)}$, $i\leftarrow 0$
			\REPEAT
				\STATE $i\leftarrow i+1$, $\alpha \leftarrow \frac{t(i-1) - 1}{t(i)}$
				\STATE $x^{(i)} \leftarrow \prox_{\frac{\beta}{\|B\|^2} f}\left[ y^{(i-1)} - \frac{\beta}{\|B\|^2} B^* \nabla g\left( B y^{(i-1)} \right) \right] $ 
				\label{algs:GenericAccProxGrad:ImportantStep}
				\STATE $y^{(i)} \leftarrow x^{(i)} + \alpha\left( x^{(i)} - x^{(i-1)}\right)$
			\UNTIL{ convergence }
			\STATE $ x_{\mathrm{opt}}\leftarrow x^{(i)}$
			\item[] \item[]		
		\vspace{.5pt}\hrule height 1pt \vspace{.5pt}
		\end{algorithmic}
		\caption{Accelerated Proximal Gradient to find $ x_{\mathrm{opt}}$ that solves   \eqref{eq:GenericOptProblem-AccProxGrad}
			with function-value convergence rate $\mathcal{O}\left(i^{-2}\right)$. 
			To simplify exposition, we identify $\mathcal{G}$ with its dual $\mathcal{G}^*$ and write $\nabla g(By)$ to refer to its representation in
			$\mathcal{G}$. We also note $\|B\|^2=\|B\|^2_{\opers{\X}{\mathcal{G}}}$. Moreover, the ratio $\|B\|^2/\beta$ here is representing the best Lipschitz
			continuity constant for $\nabla\left( g \circ B \right)$, but can be replaced by any constant upper bound of this quantity and the convergence rate $\mathcal{O}\left(i^{-2}\right)$ will still be preserved.} 
			\label{algs:GenericAccProxGrad}
	\end{algorithm}

	In summary, in order to solve a problem of the form \eqref{eq:GenericOptProblem-AccProxGrad} using the APG algorithm, one needs to identify or 
	upper-bound $\|B\|_{\opers{\X}{\mathcal{G}}}$, find an expression for $B^*$ and $\nabla g$, identify $\beta$, and be able to obtain
	$\prox_{\gamma f}(x)$ for any $x\in \X$.

%% file: secs/Algs_InvDif.tex
	In this section, we introduce the results that allow us to solve   \eqref{eq:InvDif:Regularised} using the APG algorithm.
	First, note that \eqref{eq:InvDif:Regularised} is of the form \eqref{eq:GenericOptProblem-AccProxGrad} by
	identifying, with respect to the notation in Section~\ref{sec:proxofsums}, the Hilbert spaces
	$\mathcal{G}=\DenSpace$, $\X=\ASpace$, the operator $B={A:\ASpace \rightarrow \DenSpace}$,
	the lower semi-continuous non-smooth convex function ${f:\ASpace\rightarrow\bar{\reals}}$ such that
	\begin{IEEEeqnarray}{c}\label{eq:PositiveGroupSparsity} \label{eq:ffunction}
		f(a) = \delta_{\ASpace_+}(a) + \lambda \normLebOne{\reals^2}{\normLebTwo{\reals_+}{\xi a_\pos}},
	\end{IEEEeqnarray}
	$\forall a \in \ASpace$, and the lower semi-continuous, Fréchet-differentiable convex function ${g:\DenSpace\rightarrow \reals}$ such that
	\begin{IEEEeqnarray}{c} \label{eq:gfunction}
		g(d) =  \normDen{d - \obs}^2,\forall d \in \DenSpace\,.
	\end{IEEEeqnarray}
	Consequently, to derive the APG algorithm to solve \eqref{eq:InvDif:Regularised}, we use some of the results we obtained in 
	Part~I~\cite[Section III-B]{AguilaPla2017}
	on the diffusion operator $A$, namely, a bound on its norm and the expression for its adjoint. Furthermore, we need to
	find a $\beta>0$ such that $\nabla g$ is ${\beta}^{-1}$-Lipschitz continuous, and provide a way to compute 
	$\prox_{\gamma f}(a)$ for any $\gamma>0$ and $a\in\ASpace$.

	We start by characterizing the behavior of the smooth function $g$ in \eqref{eq:gfunction}. The result in Property~\ref{prop:FrechDer}
	follows finite-dimensional intuition and specifies this behavior completely. 
	\begin{property}[Fréchet derivative of the squared-norm loss] \label{prop:FrechDer}
		Consider the functional $g:\DenSpace \rightarrow \reals$ in \eqref{eq:gfunction}. Then, $g$ has
		a Fréchet derivative $\nabla g:\DenSpace \rightarrow \DenSpace^*$ such that $\nabla g(d)$ is represented in $\DenSpace$ by $2(d-\obs)$, $\forall d \in \DenSpace$.
		Additionally, $\forall d_1,d_2 \in \DenSpace$
		\begin{IEEEeqnarray*}{c}
			\|\nabla g(d_1) - \nabla g(d_2)\|_{\DenSpace^*} = \normDen{2 d_1 - 2 d_2} = 2 \normDen{d_1-d_2}
		\end{IEEEeqnarray*}
		and, thus, $\nabla g$ is $2$-Lipschitz continuous, i.e., $\beta = 1/2$. Here, $\DenSpace^*$ is the dual space of 
		$\DenSpace$, where $\nabla g(d)$ resides. See \cite{Luenberger1969}
		for more on dual spaces and Fréchet derivatives.
	\end{property}
	
	Now, we turn our attention towards the non-smooth function $f$ in \eqref{eq:ffunction}. 
	Deriving a closed form expression for $\prox_{\gamma f}(a)$ is the most complex result
	required to use the APG algorithm. This is mainly because the proximal operator does not generally decompose well
	for sums of functions, and no previous result indicates that $\prox_{\gamma f}(a)$ can be computed in closed form.
	The appendix of this paper is dedicated to proving our contribution in 
	Theorem~\ref{theorem:genvers}, which provides an expression for $\prox_{\gamma f}$ in the most generic setting possible.
	To simplify the exposition of this result, let $\aleph = \supp{\xi}$ and 
	$\complement{\aleph} = [0,\sigmax]\setminus \aleph$. These two sets distinguish values of $\sigma$
	at which the recovered PSDR $a$ in \eqref{eq:InvDif:Regularised} is influenced by the weighted group-sparsity 
	regularization, i.e., $\sigma \in \aleph$, from values of $\sigma$ at which it is not, i.e., $\sigma \in \complement{\aleph}$. 
	Consider also, for any $a\in\ASpace$, two functions $a_{\aleph},a_{\complement{\aleph}}:\Omega\rightarrow\reals_+$ 
	such that 
	\begin{IEEEeqnarray*}{c}
		\supp{a_{\aleph}}\subset \reals^2\times\aleph,\,\, \supp{a_{\complement{\aleph}}}\subset \reals^2 \times\complement{\aleph},
	\end{IEEEeqnarray*}
	and $a = a_{\aleph} + a_{\complement{\aleph}}$,
	which provides a way for us to refer to these two distinct regions of the PSDR $a$.
	\begin{theorem}[Proximal operator of the non-negative weighted group-sparsity regularizer] \label{theorem:genvers}
		Consider the functional $f:\ASpace \rightarrow \bar{\reals}$ in \eqref{eq:PositiveGroupSparsity}. 
		For any $\gamma,\lambda >0$, if $p=\prox_{\gamma f}(a)$, then,
		\begin{IEEEeqnarray*}{c}
			p_\pos = \left[a_{\pos}\right]_+ - \Proj{\clellips{}{\xi}{\lambda \gamma}}{\left[a_{\aleph,\pos}\right]_+}\,.
		\end{IEEEeqnarray*}
		Here, $\ProjOp{\clellips{}{\xi}{\lambda \gamma}}$ is the projection onto $\clellips{}{\xi}{\lambda \gamma}$,
		the closed ellipsoid of $\xi^{-1}$-weighted norm under $\lambda \gamma$. This convex set and the projection onto it are further
		discussed in the appendix. Following the convention used in \eqref{eq:InvDif:Regularised}, for each $\pos \in \reals^2$,
		we have $p_\pos,a_{\aleph,\pos} : [0,\sigmax] \rightarrow \reals_+$ such that $a_{\aleph,\pos}(\sigma) = a_\aleph(\pos,\sigma)$ 
		and $p_\pos(\sigma) = p(\pos,\sigma)$ for any $\sigma \in [0,\sigmax]$.
	\end{theorem}
	The interpretation of this result is a direct parallel with the interpretation of the FISTA iterations in the known framework
	of $\ell^1$-regularized inverse problems, i.e., an iterative shrinkage-thresholding effect. 
	To see this, consider a specific $\pos \in \reals^2$ at which $w(\pos)> 0$, and analyze the iteration of Step~\ref{algs:GenericAccProxGrad:ImportantStep} 
	in Alg.~\ref{algs:GenericAccProxGrad}.
	The iteration of the proximal operator in Theorem~\ref{theorem:genvers} will keep shrinking $\left[a_{\aleph,\pos}\right]_+$ by subtracting its projection
	onto the ellipsoid $\clellips{}{\xi}{\lambda \gamma}$. If the gradient step inside does not raise this contribution again due to its
	importance for the minimization of $g$, at some point we will have $\left[a_{\aleph,\pos}\right]_+\in\clellips{}{\xi}{\lambda \gamma}$,
	which will result in a thresholding effect, because then applying the proximal operator in Theorem~\ref{theorem:genvers} again will yield  $\left[a_{\aleph,\pos}\right]_+=0$.
	In this context, we can read Theorem~\ref{theorem:genvers} as a statement that the non-negativity constraint and the weighted norm 
	in \eqref{eq:ffunction} decouple, neither affecting the optimality of iterative shrinkage-thresholding for inducing sparsity.
	
	Expressing the projection in Theorem~\ref{theorem:genvers} in closed form for a generic $\xi$ and for each $\pos\in\reals^2$,
	however, is not trivial. In Property~\ref{prop:ProjEli} in the appendix, we generalize a well-known finite dimensional result that 
	states that this projection can not generally be fully determined in closed form and, thus, iterative procedures should be used for each $\pos \in \reals^2$.
	Although this establishes an interesting research direction to obtain algorithms that solve \eqref{eq:InvDif:Regularised} in its more generic form, we opt
	here for limiting our choice of $\xi$. In particular, we select only its support $\aleph$ and we let $\xi = 1~\ae$ in $\aleph$. This simplifies the 
	projection onto the closed ellipsoid $\clellips{}{\xi}{\lambda \gamma}$, which becomes the simple closed ball of norm smaller than $\lambda \gamma$ in 
	$\LebTwo{\aleph}$ (see Property~\ref{prop:ball} in the appendix). For this particular case, Theorem~\ref{theorem:ballvers} states a 
	closed-form expression for $\prox_{\gamma f}(a),\forall a \in \ASpace$, completing the list of required results to use the APG algorithm.
	\begin{theorem}[Proximal operator of the non-negative group-sparsity regularizer on $\aleph$] \label{theorem:ballvers}
		Consider the functional $f:\ASpace \rightarrow \bar{\reals}$ in \eqref{eq:PositiveGroupSparsity}. 
		Let $\xi = 1~\ae$ in $\aleph$. Then,
		$\forall \gamma,\lambda > 0$, if $p=\prox_{\gamma f}(a)$,
		\begin{IEEEeqnarray*}{c}
			p_\pos = \left[a_{\complement{\aleph},\pos}\right]_+ +
						\left[a_{\aleph,\pos}\right]_+ 
						\left( 1 - \frac{\gamma \lambda}{\normLebTwo{\aleph}{\left[a_{\aleph,\pos}\right]_+}} \right)_+\,.
		\end{IEEEeqnarray*}
		Here, $a_{\aleph,\pos}$ and $p_\pos$ are defined as in Theorem~\ref{theorem:genvers} and 
		$a_{\complement{\aleph},\pos}$ is defined mutatis mutandis.
	\end{theorem}
	
	This result, jointly with the bound on the diffusion operator's norm derived in Part~I~\cite[Section III-B]{AguilaPla2017}, is summarized in the proposed 
	algorithm for inverse diffusion, i.e., Alg.~\ref{algs:AccProxGradforRegInvDif}. This algorithm establishes a reference from which different 
	discretization schemes can lead to different implementable algorithms for inverse diffusion and cell detection. 
	A relevant observation here is that, precisely because $\prox_{\gamma f} = \prox_{\gamma \lambda f_1} \circ \ProjOp{\ASpace_+}$, where $f_1$ is the 
	group-sparsity regularizer as in \eqref{eq:InvDif:Regularised}, the implementation of $\prox_{\gamma f}$ is decomposed in the non-negative projection in
	Step~\ref{algs:AccProxGradforRegInvDif:nneg} and the subsequent group-sparsity shrinkage-thresholding in Step~\ref{algs:AccProxGradforRegInvDif:Shrink}.
	
	\begin{algorithm}
		\begin{algorithmic}[1]
		\vspace{.5pt}\hrule height 1pt \vspace{.5pt}
		\REQUIRE An initial $a^{(0)}\in\ASpace$, an image observation $\obs\in\DenSpace$
		\ENSURE A solution $ a_{\mathrm{opt}} \in \ASpace$ that solves   \eqref{eq:InvDif:Regularised}
		\vspace{.5pt}\hrule height .5pt \vspace{.5pt}
			\item[] 
			\STATE $b^{(0)} \leftarrow a^{(0)}$, $i\leftarrow 0$
			\REPEAT
				\STATE $i\leftarrow i+1$, $\alpha \leftarrow \frac{t(i-1) - 1}{t(i)}$
				\STATE $\displaystyle a^{(i)} \leftarrow \left[ b^{(i-1)} -  \eta  A^* \left(A b^{(i-1)} - \obs\right)\right]_+$ \label{algs:AccProxGradforRegInvDif:nneg}
				\STATE $\displaystyle a^{(i)}_{\aleph} \leftarrow  a^{(i)}_{\aleph}
					\left(1 - \frac{\eta}{2} \lambda \normLebTwo{\aleph}{a^{(i)}_{\pos,\aleph}}^{-1}\right)_+$ \label{algs:AccProxGradforRegInvDif:Shrink}
				\STATE $b^{(i)} \leftarrow a^{(i)} + \alpha\left( a^{(i)} - a^{(i-1)}\right)$
			\UNTIL{ convergence }
			\STATE $ a_{\mathrm{opt}}\leftarrow a^{(i)}$
			\item[] \item[]				
		\vspace{.5pt}\hrule height 1pt \vspace{.5pt}
		\end{algorithmic}
		\caption{Accelerated Proximal Gradient to find $ a_{\mathrm{opt}}$ that solves \eqref{eq:InvDif:Regularised}
			with function-value convergence rate $\mathcal{O}\left(i^{-2}\right)$ when 
			$\xi=1~\ae$ in $\aleph$. Here, $\eta = \sigmax^{-1} \wninf^{-2}$ is used for 
			clarity of exposition.} 
		\label{algs:AccProxGradforRegInvDif}
	\end{algorithm}

%% file: secs/Algs_Discr.tex
		
	In Part~I~\cite[Section IV]{AguilaPla2017} we presented a discretization scheme that establishes approximation rules for 
	any element in $\ASpace$ by an element of $\matrices{M,N,K}$, and for any element in $\DenSpace$ by an element of $\matrices{M,N}$. 
	Here, $M$ and $N$ are the number of pixels in each dimension, and $K$ is the number of discretization points for the 
	$\sigma$-dimension. This discretization scheme also enables us to obtain discrete versions of the diffusion operator $A$ and its adjoint
	$A^*$, and yields Alg.~\ref{algs:disc_AccProxGradforRegInvDif} as a practical implementation of 
	Alg.~\ref{algs:AccProxGradforRegInvDif}. In Alg.~\ref{algs:disc_AccProxGradforRegInvDif}, $\dobs$, $\tilde{w}$, $\tilde{\mu}$ and
	$\tilde{a}$ are discretizations of $\obs$, $w$, $\mu$ and $a$, $\tilde{\aleph}$ is the set of indexes $k$ that represent portions of the $\sigma$-dimension that lie
	inside $\aleph$, and $\tilde{g}_k$ are the doubly spatially integrated Gaussian kernels, as specified in 
	Part~I~\cite[Section IV]{AguilaPla2017}.
	In Alg.~\ref{algs:disc_AccProxGradforRegInvDif}, Steps~\ref{line:iniacc}, \ref{line:acc1}, and \ref{line:acc2} take care of the Nesterov acceleration of the proximal 
	gradient algorithm by using the momentum in its convergence path, Step~\ref{line:forwardanddiff} computes the diffusion operator and 
	evaluates the prediction error, Step~\ref{line:adjointandpos} computes the adjoint operator, completes the gradient step, and enforces 
	the positivity constraint, and Steps~\ref{line:shrinkth1} and \ref{line:shrinkth2} implement the group-sparsity shrinkage-thresholding.
				
	\begin{algorithm}
		\begin{algorithmic}[1]
		\vspace{.5pt}\hrule height 1pt \vspace{.5pt}
		\REQUIRE An initial $\tilde{a}^{(0)}\in\matrices{M,N,K}$, a discrete image observation $\dobs\in\matrices{M,N}$
		\ENSURE A discrete approximation $\tilde{a}_{\mathrm{opt}} \in \matrices{M,N,K}$ to the solution of \eqref{eq:InvDif:Regularised}, i.e., the solution to \eqref{eq:optidisc}
		\vspace{.5pt}\hrule height .5pt \vspace{.5pt}
			\item[] 
			\STATE $\tilde{b}^{(0)} \leftarrow \tilde{a}^{(0)}$, $i\leftarrow 0$ \label{line:iniacc}
			\REPEAT \vspace{5pt}
				\STATE $i\leftarrow i+1$, $\alpha \leftarrow \frac{t(i-1) - 1}{t(i)}$ \label{line:acc1}
				\STATE $\displaystyle \tilde{d}^{(i)} \leftarrow \sum_{k=1}^{K} \tilde{g}_k \circledast \tilde{b}^{(i-1)}_k - \dobs$ \label{line:forwardanddiff}
				\FOR{ $ k=1 $ \TO $K$}  \vspace{6pt}
					\STATE $\displaystyle \tilde{a}_k^{(i)} \leftarrow \left[ \tilde{b}_k^{(i-1)} - \eta \tilde{\mu} \odot \left(\tilde{g}_k \circledast \left[\tilde{w}^2 \odot \tilde{d}^{(i)}\right] \right) \right]_+$ \label{line:adjointandpos}
				\ENDFOR
				\STATE $\displaystyle \tilde{p} \leftarrow \left( 1 - \frac{\eta}{2} \lambda \left[ \sqrt{\sum_{k\in\tilde{\aleph}} \left(\tilde{a}^{(i)}_{k}\right)^2}\right]^{-1} \right)_+ $ \label{line:shrinkth1}
				\FOR{ $ k\in\tilde{\aleph}$ } \vspace{5pt}
					\STATE  $\displaystyle \tilde{a}_k^{(i)} \leftarrow \tilde{p} \odot \tilde{a}_k^{(i)} $ \label{line:shrinkth2}
				\ENDFOR
				\STATE $\tilde{b}^{(i)} \leftarrow \tilde{a}^{(i)} + \alpha\left( \tilde{a}^{(i)} - \tilde{a}^{(i-1)}\right)$ \label{line:acc2}
			\UNTIL{ convergence }
			\STATE $ \tilde{a}_{\mathrm{opt}} \leftarrow \tilde{a}^{(i)}$
			\item[] \item[]				
		\vspace{.5pt}\hrule height 1pt \vspace{.5pt}
		\end{algorithmic}
		\caption{ Algorithm to find a discrete approximation $\tilde{a}_{\mathrm{opt}} \in \matrices{M,N,K}$ to the solution of \eqref{eq:InvDif:Regularised}, i.e., the solution to \eqref{eq:optidisc}.
				Here, $\eta$ is as in Alg.~\ref{algs:AccProxGradforRegInvDif}, $\circledast$ refers to discrete zero-padded same-size convolution,
				and all matrix powers and products are element-wise.} 
		\label{algs:disc_AccProxGradforRegInvDif}
	\end{algorithm}
	
	Many of the choices implicit in the discretization scheme presented in Part~I~\cite[Section IV]{AguilaPla2017} were derived from an intuitive goal,
	i.e., that the properties present in the function spaces are preserved after discretization. In this 
	manner, the discretized adjoint is the adjoint of the discretized operator, and the proximal operator is preserved,
	because the discretized and continuous norms are equivalent in an inner-approximation sense. As a result, Alg.~\ref{algs:disc_AccProxGradforRegInvDif} is an APG 
	algorithm too, and it can be proven to solve the discretized equivalent to \eqref{eq:InvDif:Regularised}, i.e., \eqref{eq:optidisc},
	for $\tilde{a}\in\matricesP{M,N,K}$ (see Part~I~\cite[Equation (24)]{AguilaPla2017}).
	\begin{IEEEeqnarray}{c} \label{eq:optidisc}
			\!\!\!\!\min_{\tilde{a}}
			\left\lbrace 
				\left\| \dobs\! -\! \sum_{k=1}^{K} \tilde{g}_k \circledast \tilde{a}_k \right\|_{\tilde{w}}^2
				\! + \! \lambda \sum_{m,n} \sqrt{\sum_{k\in\tilde{\aleph}} \tilde{a}_{m,n,k}^2 }
				\right\rbrace\!
	\end{IEEEeqnarray}

%% file: secs/NumRes.tex
	In this section, we provide empirical validation of the optimization framework we presented in Part~I~\cite[Section III]{AguilaPla2017}, 
	i.e., 
	\eqref{eq:InvDif:Regularised}, and of the theoretical results in Section~\ref{sec:algs}. We do this through the evaluation of an efficient 
	approximated implementation of Alg.~\ref{algs:disc_AccProxGradforRegInvDif} we present in Section~\ref{sec:Impl}. In particular,
	in Section~\ref{sec:NumRes:Sim} we specify how we use the observation model we presented in
	Part~I~\cite[Section II-B]{AguilaPla2017} to generate realistic synthetic data, in which the location and total secretion of each of the 
	active cells is known. On that data, we evaluate our approach in two different ways. First, in Section~\ref{sec:DetNumRes}, we provide detection performance metrics after 
	simple post-processing, and compare that to the detection performance of a human expert on similarly generated data. This,
	jointly with our results on real data in Part~I~\cite[Section V-A]{AguilaPla2017}, validates our 
	proposal for use in practical scenarios. Second, in Section~\ref{sec:DistrNumRes}, we evaluate the output 
	$\tilde{a}_{\mathrm{opt}}$ of Alg.~\ref{algs:disc_AccProxGradforRegInvDif} by interpreting its accumulated sum over $k$ as a 2D discrete
	particle distribution, and comparing it to the one given by the true simulated PSDR $\tilde{a}$ using optimal-transport theory.
	
	\subsection{Implementation, Kernel approximations} \label{sec:Impl}  \label{ssec:KerApp}

\subfile{\secs/Impl.tex}

	\subsection{Data simulation} \label{sec:NumRes:Sim}
	
			We simulated image data from a physical system that follows the reaction-diffusion-adsorption-desorption process 
			we presented in Part~I~\cite[Section II-A]{AguilaPla2017} with the parameters specified in Tab.~\ref{tab:SimPars}.
			Here, we have that 
			1) $\ka$, $\kd$, $D$ and $T$ are physical parameters characterizing the biochemical assays, 
			2) $M$, $N$ and $\Delta_\mathrm{pix}$ determine the spatial discretization of a supposed camera, as detailed in 
				Part~I~\cite[Section IV]{AguilaPla2017},
			3) $N_t$ determines the number of discretization points in time used to generate the SDR $s(\pos,t)$, 
			4) $K_\mathrm{g}$ determines the number of uniform discretization intervals of the PSDR $a(\pos,\sigma)$ in the 
			$\sigma$-dimension during data generation, 
			5) $J$ determines the number of terms to which we truncate the infinite sum that expresses $\varphi(\tau,t)$, the
			function that relates the SDR $s$ to the PSDR $a$, as we exposed in Part~I~\cite[Section II-C, Lemma 2]{AguilaPla2017}, and 
			6) $\dsigma_{\mathrm{b}}$ determines the standard deviation of the discretized Gaussian kernel used to simulate an imperfect 
			optical system, as presented in Part~I~\cite[Section II-D]{AguilaPla2017}.
			
			For each considered active cell, say, in a location $(m,n)$, we generated a random discrete SDR $\tilde{s}_{m,n}$ in the form of a 
			square pulse in time. In particular, 
			we drew uniform activation (particle generation initiation) and deactivation (particle generation finalization) times in the 
			interval $(1,6)~\mathrm{h}$, and we chose the amplitude of the square pulse by uniformly drawing the total amount of generated 
			particles between a certain maximum and its half. This was done for $50$ different sets of uniformly-drawn, pixel-centered, 
			active-cell locations for each considered number $N_\mathrm{c}$ of active cells in an image.
			
			We then used our contribution in Part~I~\cite[Theorem 2]{AguilaPla2017} to obtain the PSDR 
			$\tilde{a}\in\matrices{M,N,K_\mathrm{g}}$
			from the resulting SDR $\tilde{s}\in\matrices{M,N,N_t}$. Details on the exact procedure to do so can be found in the supplementary
			material to this paper. Then, we computed the ideal discretized measurement by applying the discretized diffusion 
			operator $\tilde{A}$ to it. Note here that in synthesis, the kernels $\tilde{g}_k$ were not approximated.
			We then simulated the effect of an imperfect optical system by convolution with a discretized, i.e. spatially integrated,
			Gaussian kernel with a standard deviation of $\dsigma_\mathrm{b}$, and rescaled the image to keep the intensity in the range 
			$[0,1]$. We then incorporated additive white Gaussian noise of the variance that corresponded to that of the statistical model for 
			quantization in	the range $[0,1]$ with a number of bits $b$, i.e. $2^{-2b}/12$. 
			Finally, we clipped the resulting image to the range $[0,1]$ and re-scaled it to the range $[0,255]$.
			It is worthwhile to mention here that extensive analysis carried out on real data has
			suggested that the Gaussian assumption is sensible. Moreover, no magnification is 
			usually employed in the image capture for the described biochemical assays. This implies high photon counts, which theoretically supports the
			Gaussian assumption over the Poisson assumption, more common in low-photon-count applications such as microscopy.
						
			\begin{figure}
				\renewcommand{\figurename}{Tab.}
				\centering
				\begin{tabular}{c|c|c|c|c}
					$\ka~\left[\mathrm{m}\mathrm{s}^{-1}\right]$ & $\kd~\left[\mathrm{s}^{-1}\right]$ & $D~\left[\mathrm{m}^{2}\mathrm{s}^{-1}\right]$ & $T~\left[\mathrm{h}\right]$  \\ \hline
					$10^{-7}$                                    & $10^{-4}$                          & $3\cdot10^{-12}$                                     & $8$                         
				\end{tabular} 
				
				\vspace{5pt}
				
				\begin{tabular}{c|c|c|c|c|c|c}
					$\Delta_{\mathrm{pix}}~\left[\mu\mathrm{m}\right]$ & $M$   & $N$    & $K_\mathrm{g}$  & $J$  & $N_t$  & $\dsigma_\mathrm{b}~\left[\mathrm{pix}\right]$ \\ \hline
					$6.45$                                             & $512$ & $512$  & $30$            & $10$ & $10^3$ & $ 2.28 $  
				\end{tabular}
				\caption{\label{tab:SimPars}
					Parameters used for data-generation in our simulations. Note that the selection of these parameters has been done within 
					realistic ranges (see, among others, \cite{Karulin2012} for Elispot examples). Nonetheless, $\ka$ and $\sigma_\mathrm{b}$
					have been specifically adjusted to match the aspect of the real Fluorospot data that was available. 
					Finally, note that with these physical parameters, $\dsigmax=\sqrt{2 D T}/\Delta_{\mathrm{pix}}\approx 64.5$.
				}
				\renewcommand{\figurename}{Fig.}
			\end{figure}
			
			Throughout this section, we will present results obtained by this data-generation procedure in twelve different scenarios,
			in which three different cell densities (few, average, many) and four different noise levels ($\mathrm{NL}$) are considered. For details on their
			characterization in terms of $N_\mathrm{c}$ and $b$, see Tab.~\ref{tab:testedConfs}.

			\begin{figure}
				\renewcommand{\figurename}{Tab.}
				\centering
				\begin{tabular}{c|c|c|c}
									& Few   & Average   & Many \\ \hline
					$N_\mathrm{c}$	& $250$ & $750 $    & $1250$
				\end{tabular}
				
				\vspace{5pt}
				
				\begin{tabular}{c|c|c|c|c}
						& $\mathrm{NL}~1$ & $\mathrm{NL}~2$ &$\mathrm{NL}~3$ &$\mathrm{NL}~4$ \\ \hline
					$b$ & $	10$             & $8$               & $6$              & $4$
				\end{tabular}
				\caption{\label{tab:testedConfs}
					Characteristic parameters of the $12$ different scenarios considered in the simulations, 
					formed by four noise levels ($\mathrm{NL}$) and three cell densities.
				}
				\renewcommand{\figurename}{Fig.}
			\end{figure}
		
		\subsection{Performance evaluation and Numerical results} \label{sec:res:res}
		
\begin{figure*}
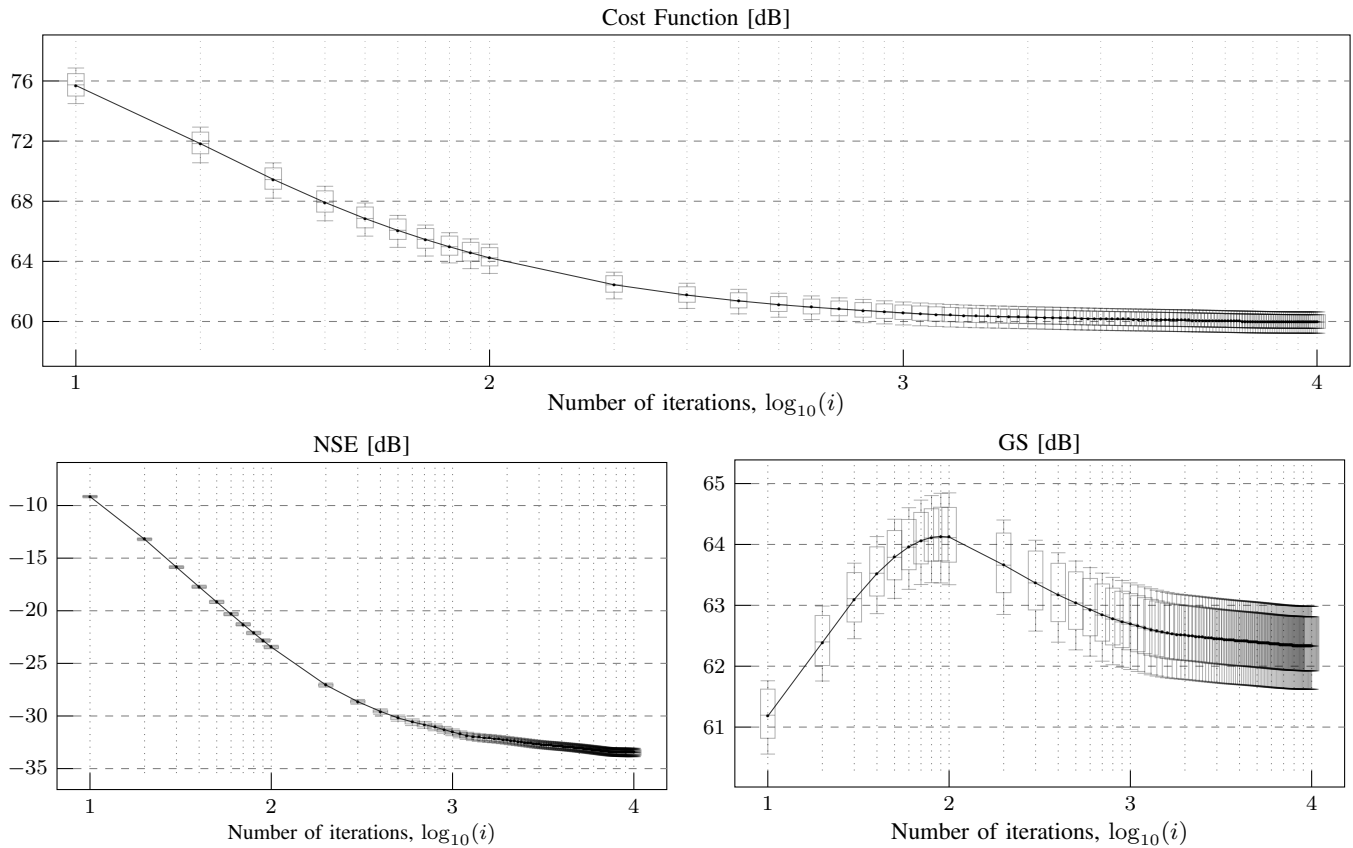

				\centering 
				\def\w{5.5} \def\h{0.2}
				\input{\figs/conv_CF} 
				
				\def\w{2.41}
				\def\h{0.14}
				\input{\figs/conv_NSE} 
				\def\h{0.81}
				\input{\figs/conv_GS}
				\vspace{-10pt}
				\caption{\label{fig:conv} Statistics of the optimization metrics' convergence with 
					the number of iterations $i$. 
					Showing the normalized prediction's square error 
					$\mathrm{NSE} = \|A a - \obs \|^2_{\DenSpace}/\|\obs\|^2_{\DenSpace}$, 
					the value of the group sparsity regularizer (GS), and the value of the cost function in \eqref{eq:InvDif:Regularised}.
					Comprising results from $50$ images with $N_\mathrm{c}=750$ cells and noise level $3$ (see Tab.~\ref{tab:testedConfs})
					when analyzed with Alg.~\ref{algs:disc_AccProxGradforRegInvDif} with the parameters in Tab.~\ref{tab:parstorun}
					and $\lambda=0.5$. For each quantity, a dot and the line illustrate mean behavior, whiskers indicate the evolution
					of the $10$th and $90$th percentiles, and the box indicates the evolution of the $25$th, $50$th and $75$th percentiles. }
			\end{figure*}

			The empirical evaluation of Alg.~\ref{algs:disc_AccProxGradforRegInvDif} can be addressed in terms of diverse metrics. 
			On one hand, one could focus on metrics characteristic of the optimization framework itself, i.e., the prediction's 
			square error, the group-sparsity level in the solution, or the value of the cost function from \eqref{eq:InvDif:Regularised}
			and the rate at which it decreases. Fig.~\ref{fig:conv} exemplifies the statistics of these quantities during convergence.
			These metrics, however, have already been studied theoretically and do not hold operational
			meaning in terms of performance on the task at hand, i.e., SL on data from reaction-diffusion-adsorption-desorption systems.
			On the other hand, detection metrics such as precision and recall, or their compromise, the F1-score, directly characterize 
			SL performance, and are therefore naturally operational. Therefore, when presenting results to validate the 
			operational value of our algoriths, we will use the F1-score after $I=10^4$ iterations, relying on
			convergence. For example, in Fig.~\ref{fig:vsNC} we compare our algorithm's F1-Score to that of an expert human labeler on
			synthetic data for some specific experimental conditions. Nonetheless, pure detection metrics like the F1-score can not be 
			obtained simply from the value 
			$\tilde{a}_{\mathrm{opt}}$ our algorithm provides, and some post-processing is necessary. 
			Therefore, any attempt at evaluating our approach in this manner will be influenced by the specificities of the 
			chosen post-processing. In this context, optimal transport theory and the earth mover's distance (EMD) \cite{Rubner2000}
			offer an interesting alternative. In particular, the $\mathrm{EMD}$ is an interpretable objective metric between any two discrete 
			distributions of the same total weight. In other words, it not only evaluates the location at which each spatial peak in 
			the recovered $\tilde{a}_{\mathrm{opt}}$ is, but also their relative contribution to the total amount of particles. In
			conclusion, then, when evaluating our results in terms of the accuracy of the information they provide about the spatial 
			distribution of particle generation, we will use the $\mathrm{EMD}$ as our preferred metric.

			\subsubsection{Operational evaluation and detection results} \label{sec:DetNumRes}
			
			\def\TP{\mathrm{TP}} \def\FP{\mathrm{FP}} \def\FN{\mathrm{FN}}
			Consider an SL detector that, given an observation $\dobs$, provides a list of positions
			$\left\lbrace \pos_l \right\rbrace_{l=1}^{L}\subset\reals^2$ and a co-indexed list of positive numbers 
			(pseudo-likelihoods) $\left\lbrace p_l \right\rbrace_{l=1}^{L}\subset\reals_+$.
			Then, for a given tolerance $\varrho>0$, we will evaluate each position $\pos_{l}$ in decreasing order of pseudo-likelihood $p_l$,
			and consider it a correct detection if a previously unmatched true cell location $\pos_{\mathrm{c}}$ can be found inside the ball with diameter $\varrho$
			centered at $\pos_l$. If that is the case, the closest such true cell location will 
			not be paired with any further $\pos_l$s.
			Then, if we refer to $\TP$, $\FP$ and $\FN$ as the number of correct detections, incorrect detections, and cells that were not 
			detected, respectively, the precision $\mathrm{pre}$, recall $\mathrm{rec}$ and F1-score $\mathrm{F1}$ are defined as 
			\begin{IEEEeqnarray*}{c}
				\mathrm{pre} = \frac{\TP}{\TP + \FP},\, \mathrm{rec} = \frac{\TP}{\TP + \FN},\mbox{ and }
				\mathrm{F1} = \frac{2 \, \mathrm{pre} \cdot \mathrm{rec}}{\mathrm{pre}+ \mathrm{rec}}.
			\end{IEEEeqnarray*}
			Note, then, that the F1-score is a number in the range $[0,1]$ that establishes a compromise between the 
			probability of a detection being correct (precision) and the probability of a true cell being found (recall).
			Throughout the rest of the paper, we will use $\varrho=3~\mathrm{pix}$ as our tolerance for the localization of active cells.
			Note here that, as mentioned in \cite[Section II-A]{AguilaPla2017}, the cells under consideration are tens of $\mu\mathrm{m}$s in diameter, and so a tolerance 
			of $\varrho \Delta_\mathrm{pix}=19.5~\mu\mathrm{m}$ should be considered extremely accurate.
						
			Obtaining a set of detections $\left\lbrace (\pos_l, p_l) \right\rbrace_{l=1}^{L}$ from the output of 
			Alg.~\ref{algs:disc_AccProxGradforRegInvDif} can be done in multiple ways.
			In an ideal case, i.e., with the perfect reconstruction of $\tilde{a}$, we would simply use
			$\left\lbrace \tilde{\pos}_l \right\rbrace_{l=1}^{L} = \bigcup_{k=1}^K \supp{\tilde{a}_k}$, where the
			support of a matrix is the set of indexes $\tilde{\pos}\in\integ^2$ where its elements are not zero. In this case,
			the value of $p_l$ would not have any impact, and the obtained F1-score would be $1$.
			In real cases, in which an imperfect reconstruction $\tilde{a}_{\mathrm{opt}}$ includes approximation and numerical errors, we propose to first compute a 
			pseudo-likelihood for each pixel, corresponding to the contribution of each pixel to the 
			overall group-sparsity regularizer, i.e. the matrix $\tilde{p} = \left(\sum_{k\in\tilde{\aleph}} \tilde{a}^{2}_{k}\right)^{1/2}$.
			We then propose to build a list of candidate detections $\left\lbrace \tilde{\pos}_q \right\rbrace_{q=1}^{Q}$
			formed by the local maxima (with $8$-connectivity) in $\tilde{p}$, and discard those with pseudo-likelihood
			$p_q = \tilde{p}_{\tilde{\pos}_q}$ under a certain threshold, i.e., for some $\delta>0$, 
			$\left\lbrace (\tilde{\pos}_l, p_l) \right\rbrace_{l=1}^{L} = \left\lbrace (\tilde{\pos}_q, p_q) : p_q > \delta \right\rbrace$.
			In practice, we pick the $\delta$ that yields the best F1-score given the known true data.
			This mimics real application, in which experts select the threshold that best fits their criterion by visual inspection of the
			results overlaid on the image data.
			This same evaluation
			by connected maxima detection and optimal thresholding can be applied to other methods in which the pseudo-likelihood 
			image $\tilde{p}$ is generated differently. 
			Finally, note that although different heuristics could generate a better set 
			of detections and pseudo-likelihoods $\left\lbrace \pos_l, p_l \right\rbrace_{l=1}^{L}$, our focus here is in showing that the PSDR $\tilde{a}$ recovered
			from Alg.~\ref{algs:disc_AccProxGradforRegInvDif} provides the means for robust and reliable SL.
		
\begin{figure}
						\centering
						\input{\figs/vs_lam_NC750NL3_new} 
						\vspace{-20pt}
						\caption{\label{fig:choicelam}
							Statistics of the obtained F1-scores for different methods to obtain $\tilde{p}$. 
							Dependence on the regularization parameter 
							$\lambda$. Those methods that do not use a regularization parameter appear centered in the figure.
							The statistics are reported in the whiskers-box plot as in Fig.~\ref{fig:conv}.
						}
					\end{figure}
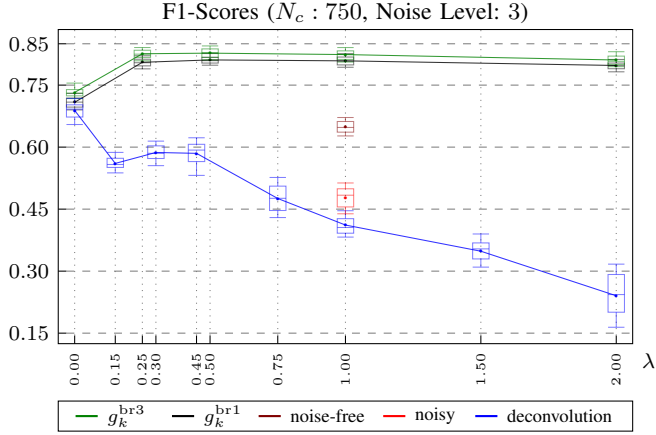

\begin{figure}
						\centering
						\input{\figs/vs_NC_NL3+4_new_withChristian}
						\vspace{-20pt}
						\caption{\label{fig:vsNC}
							Statistics of the obtained F1-scores for different methods.
							Dependence on the total number of cells	in the simulated image, at the two highest noise levels.
							For noise level 3, performance obtained by an expert human labeler on one image of each density.
							Regularization parameters $\lambda=0.5$ and 
							$\lambda_\mathrm{d}=0$ chosen for their respective methods due to the results in Fig.~\ref{fig:choicelam}. 
							The statistics are reported in the whiskers-box plot as in Fig.~\ref{fig:conv}.
						}
					\end{figure}
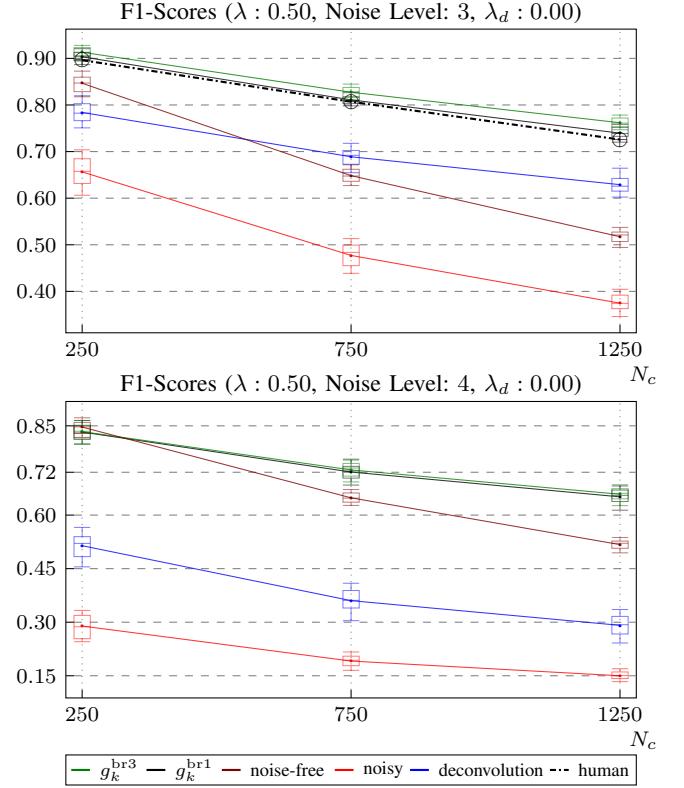

\begin{figure}
					\centering
					\input{\figs/vs_NL_NC250+1250_new}
					\vspace{-20pt}
					\caption{\label{fig:vsNL}
						Statistics of the obtained F1-scores for different methods. Dependence on the noise level, at the lowest
						and highest cell densities. The method that does not depend on noise appears centered in the figure.
						Regularization parameters' choice and reporting of statistics consistent with Fig.~\ref{fig:choicelam}.
					}
				\end{figure}
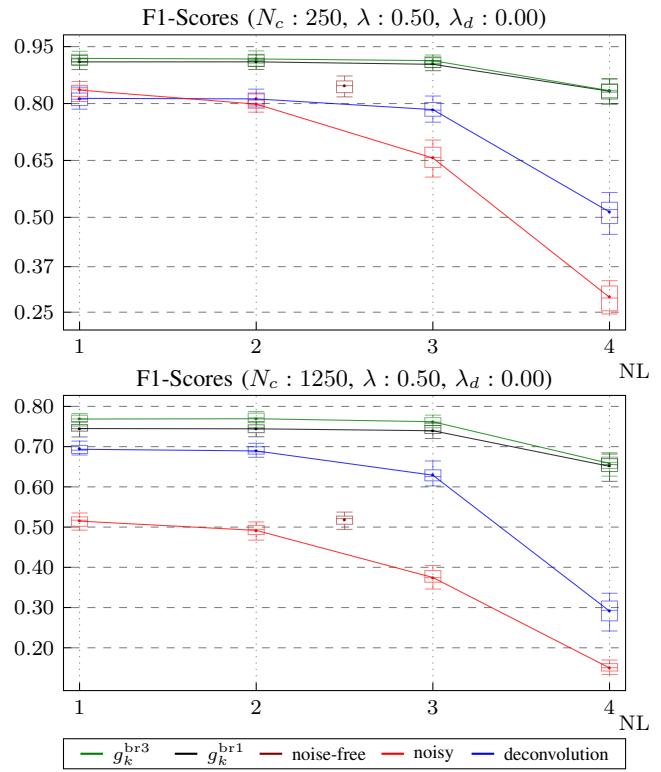

			
			In Fig.~\ref{fig:choicelam}, \ref{fig:vsNC}, and \ref{fig:vsNL}, we report statistics on the 
			results of using $\tilde{p}$ computed as above when 
			Alg.~\ref{algs:disc_AccProxGradforRegInvDif} is used with the parameters in 
			Tab.~\ref{tab:parstorun} and the sequence $t:\natu\rightarrow\reals_+$ suggested in \cite{Beck2009} (see Section~\ref{subsec:APG} for details).
			Note here that $K=8$ implies that the discretization in the $\dsigma$-dimension used
			in the analysis is much coarser than that used in the generation of the data, i.e., $K_\mathrm{g}=30$.
			Note also that due to the large amount of decisions involved in choosing the
			discretization of the $\dsigma$-dimension, this was done manually by trial-and-error and 
			always maintaining SL performance in mind. In this sense, the lowest $\dsigma$s were 
			discretized with more detail, since they allow for a more accurate localization of the active
			cells' position. Finally, note that $\tilde{w}(\tilde{\pos})=1$ and 
			$\tilde{\mu}(\tilde{\pos})=1$ were used in the context of the simulated data.
			
			To provide a fair evaluation, we compare the obtained results with different proposals for 
			$\tilde{p}$. As a baseline for comparison, we obtain the results of using a noise-free version
			of the observed image $\dobs$ as $\tilde{p}$. Because under the observation model $Aa=\obs$, 
			isolated active cells lead to monomode profiles in $\obs$ around the true cell location, 
			this will provide a reference on how detection is affected by interactions between different
			active cells. At the same time, this will also provide an upper bound on the performance of 
			any denoising-centered approach. Similarly, we will also obtain the results of using 
			$\tilde{p}=\dobs$ directly, which will provide a reference on how detection is affected by 
			additive noise. Finally, we also provide the results of obtaining $\tilde{p}$ from a 
			sparsity-based deconvolution scheme on $\dobs$ that aims to invert the blur introduced by the 
			optics. This latter approach is implemented by using $I=10^4$ iterations of 
			Alg.~\ref{algs:disc_AccProxGradforRegInvDif} with $K=1$ and $g_1$ the same kernel used
			to simulate the optical imperfections, and with the step-length $\eta$ optimized to
			obtain the empirically best results, in terms of both performance and robustness,
			i.e., $\eta\approx 0.44$.
				For the analysis of Fig.~\ref{fig:choicelam}, \ref{fig:vsNC}, and \ref{fig:vsNL}, we will consider that the difference between
				two quantities is statistically significant if the $10$th empirical percentile of one of the two quantities is above the $90$th 
				empirical percentile of the other. 
				
			\begin{figure}
				\renewcommand{\figurename}{Tab.}
				\centering
				\begin{tabular}{c|c|c|c}
					$K$ & $ \dsigma_0, \dsigma_1, \dots, \dsigma_8$ & $I$ & $ \tilde{\aleph} $  \\ \hline
					$8$ & $ 2.3, 5, 9, 13, 23, 33, 43, 53, 67 $  & $10^4$ & $\listint{8}$
				\end{tabular}
				\caption{
					Parameters used for Alg.~\ref{algs:disc_AccProxGradforRegInvDif} in all of the simulations presented in this 
					paper. Note here that the choice of the grid in the $\dsigma$-dimension is coherent with the observation model under
					an imperfect optical system derived in \cite[Section II-D]{AguilaPla2017}, i.e., with respect to Tab.~\ref{tab:SimPars},
					$\dsigma_0\approx\dsigma_{\mathrm{b}}$ and $\dsigma_K \approx \dsigmax + \dsigma_{\mathrm{b}}$. 
				\label{tab:parstorun}}
				\renewcommand{\figurename}{Fig.}
			\end{figure}			
			
			
					Both the sparsity-based deconvolution scheme and our own approach rely on an hyper-parameter $\lambda$ that needs 
					to be selected. As is common in sparsity-based optimization frameworks, this choice is made here experimentally. 
					Fig.~\ref{fig:choicelam} shows the statistics of the F1-score for the considered methods as a function of $\lambda$
					for $N_\mathrm{c}=750$ and the third noise level considered in Tab.~\ref{tab:testedConfs}. Methods
					that do not depend on $\lambda$ are additionally reported for comparison.
					Fig.~\ref{fig:choicelam} suggests that the choice of 
					regularizer in the optimization framework \eqref{eq:InvDif:Regularised} proposed in Part I \cite[Section III]{AguilaPla2017}
					is beneficial for SL. Indeed, regardless of the approximation used,
					any of the tested values for the regularization parameter $\lambda$
					yield significantly better F1-scores than $\lambda=0$. 
					Furthermore,
					the results in Fig.~\ref{fig:choicelam} indicate that \eqref{eq:InvDif:Regularised}
					is robust to the choice of regularization parameter $\lambda$, showing practically 
					no change in performance across a whole order of magnitude, i.e., from $\lambda=0.15$ to $\lambda=2$.
					In contrast, the results for sparsity-based deconvolution indicate that $\ell^1$-regularization is not appropriate in
					this setting, and that, if used, the choice of regularization parameter will be critical to the obtained performance.
					Additionally, Fig.~\ref{fig:choicelam} also validates our rank-one approximation strategy, as using $g_k^{\mathrm{br3}}$ yields only
					non-significant improvements on the performance obtained by using $g_k^{\mathrm{br1}}$ while triplicating the computational cost.					
					
					These conclusions, i.e., 1) that the regularizer chosen in \eqref{eq:InvDif:Regularised} is adequate for inverse diffusion for SL,  
					2) that the proposed optimization framework is robust to the choice of regularization parameter, and 3) that the differences in performance
					when using a rank-one and a rank-three approximation of the kernels are non-significant, are preserved
					throughout the remaining eleven scenarios characterized by combinations of the parameters in Tab.~\ref{tab:testedConfs}.
					Replicates of Fig.~\ref{fig:choicelam} for all possible combinations are reported in the supplementary material.
					Finally, using Fig.~\ref{fig:choicelam}, we decide that for the remainder of our analysis we will use $\lambda = 0.5$
					for our approach and $\lambda_\mathrm{d}=0$ for deconvolution.
				
					Of the two factors under consideration that affect SL performance, interference between several sources seems to be 
					the hardest to address. Indeed, in Fig.~\ref{fig:vsNC} we see that all considered methods, including an expert human labeler,
					decay steeply in detection performance when dealing with higher densities of active cells. This is to be expected, because the closer 
					any two active cells are, the more indistinguishable they will be on the spots that result in the observed image. In particular, we observe that 
					the performance of the expert human labeler decays with $N_\mathrm{c}$ at a similar or at a steeper rate as that obtained by our methodology. 
					This seems to indicate that there is a common limiting factor to these performances, to which our methodology is at least as robust as a domain
					expert. Further, Fig.~\ref{fig:vsNC} indicates that, for the tested cell densities, the human labeler consistently performs within the 10th and 
					the 90th percentile of the results obtained by our approximated implementation of Alg.~\ref{algs:disc_AccProxGradforRegInvDif}, exhibiting 
					no significant differences. Nonetheless, the gap between the performance obtained by using $g_k^{\mathrm{br3}}$ in 
					Alg.~\ref{algs:disc_AccProxGradforRegInvDif} and that obtained by the expert human labeler does clearly increase with $N_\mathrm{c}$, which 
					suggests that substantial differences could have been observed at higher cell densities, i.e., for $N_\mathrm{c}>1250$. Note here that due to 
					the considerable amount of time required to manually label each image, only three synthetic images were manually labeled by the human expert, one 
					of each cell density and at noise level 3. Further, note that the expert human labeler was unaware of the total expected number of cells in each 
					image.
					
					Fig.~\ref{fig:vsNC} also shows that Alg.~\ref{algs:disc_AccProxGradforRegInvDif} provides significantly better SL performance than 
					even picking local maxima in a noise-free version of the image.
					Indeed, only when our approach is exposed to a noise level $4$ and we consider the lowest cell density does a noise-free image yield similar 
					performances. This suggests that our approach is capable, through a noisy observation of $\dobs$, of breaking apart clusters that would not 
					exhibit local maxima at the active cells' locations even in a noise-free observation. 
					\begin{figure} 
						\centering
						\includegraphics[width=\linewidth,keepaspectratio=true]{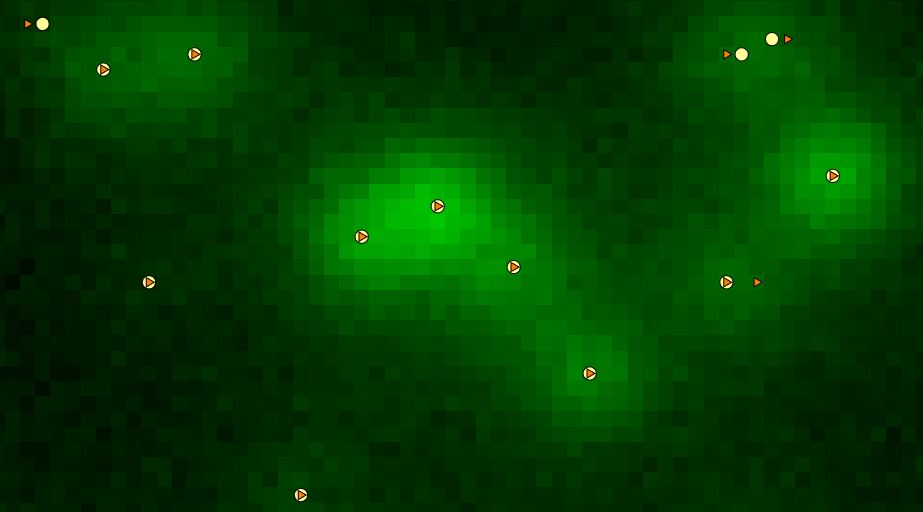}
						\footnotesize (a) Detection results (yellow circles) and true active cells' positions (orange triangles) \\ \vspace{2pt}
						\includegraphics[width=\linewidth,keepaspectratio=true]{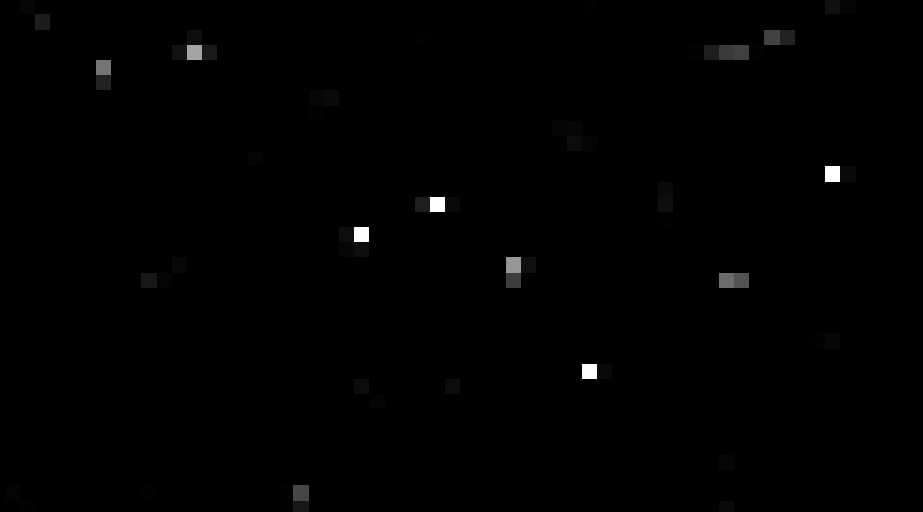}
						\\ \vspace{2pt}
						\footnotesize (b) $\tilde{p} = \sqrt{\sum_{k\in\tilde{\aleph}} \tilde{a}^{2}_{k}}$ obtained from
						Alg.~\ref{algs:disc_AccProxGradforRegInvDif} with the parameters in Tab.~\ref{tab:parstorun},
						$\lambda = 0.5$, and using $g_k^{\mathrm{br1}}$, at increased 
						luminosity.
						\\ \vspace{-2pt}
						\caption{ \label{fig:clusterexample}
							Example of SL performance on a section of a simulated image with $N_\mathrm{c}=1250$ and noise level $4$ (see Tab.~\ref{tab:testedConfs}).
						}
					\end{figure}
					An example of this capacity of breaking clusters that do not exhibit maxima at the sought locations is illustrated in 
					Fig.~\ref{fig:clusterexample}, where a section of a simulated image in the worst considered scenario 
					(noise level $4$, $N_\mathrm{c}=1250$) is shown, together with its true active cell locations and the obtained detections.
					
					In conclusion then, although Alg.~\ref{algs:disc_AccProxGradforRegInvDif} is still incapable of telling apart cells that
					are arbitrarily close, it is well equipped to accurately detect active cells from spots generated by their combined 
					secretion. In fact, Fig.~\ref{fig:vsNC} suggests that better approximations of the kernels $g_k$ yield increased robustness
					in this sense, vouching for the proposed optimization framework \cite[Section III]{AguilaPla2017}, i.e., \eqref{eq:InvDif:Regularised}.
					In Fig.~\ref{fig:clusterexample}, note that most of the correctly detected cells are detected in the exact same pixel they were located, and all 
					others are at a distance of one single pixel. This accuracy of the obtained locations has been observed consistently throughout our 
					experimentation.
					
				Finally, Fig.~\ref{fig:vsNC} also reveals a great advantage of Alg.~\ref{algs:disc_AccProxGradforRegInvDif}, i.e., robustness to
				additive noise. Indeed, Fig.~\ref{fig:vsNL} confirms that Alg.~\ref{algs:disc_AccProxGradforRegInvDif}, 
				and our optimization framework exhibit an unparalleled robustness to additive noise, 
				regardless the considered cell density.

		\subsubsection{Distributional evaluation and results} \label{sec:DistrNumRes}
		
		  Consider now $\tilde{p} = \sum_{k=1}^{K_\mathrm{g}} \sqrt{\sigmax/K_{\mathrm{g}}}\tilde{a}_{k}$, the true spatial 
		  distribution of released particles within the discretization scheme of \cite[Section IV]{AguilaPla2017}, approximating
		  the term $\int_0^{\sigmax} a\,\dint\sigma$ .
		  The ultimate objective of any source localization and characterization technique is to recover $\tilde{p}$ perfectly.
		  Indeed, if one obtains $\tilde{p}$, one knows exactly how many particles were released from each location, and thus, 
		  the exact location of each source and their relative importance. For the evaluation of source localization and characterization
		  algorithms, then, it is natural to consider whether an interpretable metric between $\tilde{p}$ and a recovered or estimated 
		  spatial density $\hat{p}$ is available.
		  
		  \def\i{\mathbf{i}} \def\j{\mathbf{j}}
		  The $\mathrm{EMD}$ \cite{Rubner2000} plays this role when two discrete distributions have the same total weight, i.e. if 
		  $\sum_{m,n=1}^{M,N}\tilde{p}_{m,n}= \sum_{m,n=1}^{M,N}\hat{p}_{m,n}$. In the following, we will consider this condition to be verified,
		  and practically normalize both $\tilde{p}$ and $\hat{p}$ to the same overall weight, i.e., the true number of cells $N_\mathrm{c}$. 
		  The $\mathrm{EMD}$ can be interpreted as the minimal average displacement required to transform the estimated distribution $\hat{p}$ into the 
		  real distribution $\tilde{p}$. To formalize how the $\mathrm{EMD}$ is computed, consider $\mathcal{I}=\supp{\hat{p}}\subset\integ^2$ and 
		  $\mathcal{J}= \supp{\tilde{p}}\subset\integ^2$, and consider the Euclidean distance (in $\mathrm{pix}$) between any two locations 
		  $\i\in\mathcal{I}$ and $\j\in\mathcal{J}$, i.e., $\left\|\i - \j\right\|_2$. 
		  Then, the following linear program 
		  \begin{IEEEeqnarray*}{rl}
		    \min_{\tilde{f}\in\matrices{|\mathcal{I}|,|\mathcal{J}|}} &
		    \,\,\sum_{\i\in\mathcal{I}} \sum_{\j\in\mathcal{J}} \tilde{f}_{\i,\j} \left\|\i - \j\right\|_2 \\
		    \mbox{subject to } & \tilde{f}_{\i,\j} \geq 0, \forall (\i,\j) \in \mathcal{I}\times\mathcal{J},\, 
								 \sum_{\j\in\mathcal{J}} \tilde{f}_{\i,\j} \leq \hat{p}_{\i}, \forall \i\in\mathcal{I} \\ &
								 \sum_{\i\in\mathcal{I}} \tilde{f}_{\i,\j} \leq \tilde{p}_{\j}, \forall \j\in\mathcal{J} \mbox{, and }
								 \sum_{\i\in\mathcal{I}} \sum_{\j\in\mathcal{J}} \tilde{f}_{\i,\j} = N_\mathrm{c}\,,
		  \end{IEEEeqnarray*}
		  is known as Monge-Kantorovich transportation problem.
		  In our context, it determines the density of particles $f_{\i,\j}$ that has to be moved from each location $\i$ to each location 
		  $\j$ so that $\hat{p}$ becomes $\tilde{p}$ with the minimal amount of overall work, understood as the product between the densities
		  of particles and the distances they have to be moved. As a result, one can derive the average distance the density of particles 
		  has been transported (in $\mathrm{pix}$), i.e., the EMD, as
		  \begin{IEEEeqnarray*}{c}
			\mathrm{EMD} = \sum_{\i\in\mathcal{I}} \sum_{\j\in\mathcal{J}} \frac{\tilde{f}_{\i,\j}}{N_\mathrm{c}} \left\|\i - \j\right\|_2\,,
		  \end{IEEEeqnarray*}
		  for the $\tilde{f}\in\matrices{|\mathcal{I}|,|\mathcal{J}|}$ that solved the linear program above.
		  
		  \begin{figure}
		    \centering
		    \input{\figs/EMDsvsNoise_integral}
		    \vspace{-20pt}		    
		    \caption{Statistics of the EMD between $\tilde{p}$ and $\hat{p}$ resulting from Alg.~\ref{algs:disc_AccProxGradforRegInvDif}
				using the best rank-one approximation to the discretized kernels and the parameters in Tab.~\ref{tab:parstorun}.
				Estimated from $50$ images with $N_\mathrm{c}=250$ cells. Dependence on noise level.\label{fig:EMD}}
		  \end{figure}
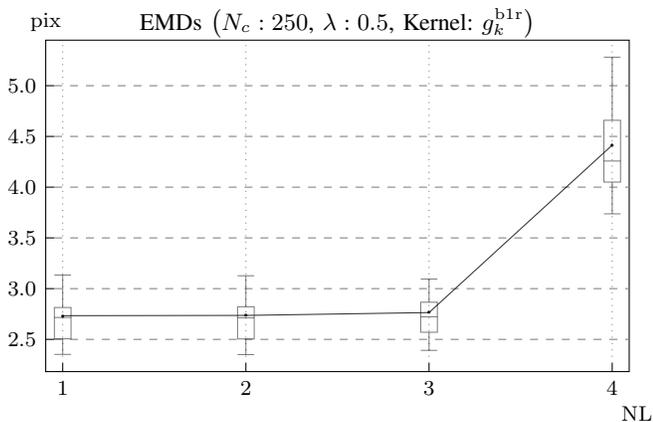
		  In our setting, we used $\hat{p}=\sum_{k=1}^{K}\sqrt{\Delta_k}\tilde{a}_{\mathrm{opt},k}$, where $\tilde{a}_\mathrm{opt}$ is the 
		  PSDR recovered from Alg.~\ref{algs:disc_AccProxGradforRegInvDif} with the same configuration as in the previous section and 
		  $\lambda=0.5$. We solved the transportation problem above using CVX \cite{Grant2014} with the MOSEK \cite{ApS2017} solver and
		  reported the statistics of the obtained $\mathrm{EMD}$s for $50$ images with $N_\mathrm{c}=250$ for each of the four different noise levels of 
		  Tab.~\ref{tab:testedConfs} in Fig.~\ref{fig:EMD}. There, we can observe that for the three first noise levels, the $\mathrm{EMD}$ consistently 
		  stays below $3~\mathrm{pix}$. This result is remarkable, because it does not only include the displacement of the highest peaks 
		  of particle secretion density, but also any errors in the relative scalings between different locations, and even cells that have 
		  been omitted or falsely detected. Further, the behavior of the $\mathrm{EMD}$ with respect to the noise level confirms what we observed in 
		  Fig.~\ref{fig:vsNL}, in which the progressive increase of the noise level seems to have no effect up to a breaking point.
		  Fig.~\ref{fig:vsNL} and Fig.~\ref{fig:EMD}, then, appear to suggest that this breaking point is more related to the inverse problem
		  at hand than to any metric in particular.
		
		\subsubsection{Recovery of the third dimension}
		  To finalize this section, we transcend the purpose of localization and present in Fig.~\ref{fig:aRecovery} two examples of the 
		  recovery of the PSDR's behavior over $\dsigma$, i.e., $a_\pos(\dsigma)$ for some location $\pos \in \reals^2$.
		  There, we can see that the quality of this recovery depends highly on the simulation conditions. On one hand, in complete absence
		  of interference, i.e., in an image with $N_\mathrm{c}=1$, and with noise level $1$, this recovery partly captures some of the traits
		  of the real curve. In particular, although it exhibits important errors for lower $\dsigma$s, it correctly captures the decay of the
		  amount of secretion from $k=3$ onwards. On the other hand, with many interfering sources ($N_\mathrm{c}=1250$) and noise level $4$,
		  the information on the $\dsigma$-dependence of the PSDR is completely lost. Although we will not explore this any further in this 
		  paper, note that the positive weighting function $\xi$ in \eqref{eq:InvDif:Regularised} introduced in Part~I~\cite{AguilaPla2017} 
		  could be used to correct systematic errors in the estimation of the PSDR and its profile over $\dsigma$, like the apparent 
		  systematic overestimation of the first bin in Fig.~\ref{fig:aRecovery}. 

				\begin{figure}
					\centering 
					\def\w{0.108}
					\def\h{0.44175}
					\input{\figs/recovery_asigma_1}
					\def\h{0.33225}
					\input{\figs/recovery_asigma_2}
					\vspace{-10pt}
					\caption{\label{fig:aRecovery}
						Two extreme examples of the recovery of $a_\pos(\dsigma)$ in simulated spots in different simulation conditions. 
						In blue, $a_\pos(\dsigma)$ that is used to simulate the particular spot, with generation parameters as in Tab.~\ref{tab:testedConfs}. 
						In red, $\hat{a}_{\mathrm{opt},\pos}(\dsigma)$ that is recovered by Alg.~\ref{algs:disc_AccProxGradforRegInvDif} with the parameters in Tab.~\ref{tab:parstorun},
						$\lambda = 0.5$, and using the kernel approximations  $g_k^{\mathrm{br1}}$.
						Above, recovery for a cell in an image with $N_\mathrm{c}=1$ and noise level $1$. Below, recovery for a well-detected cell in an image 
						with $N_\mathrm{c}=1250$ and noise level $4$. The two profiles were normalized to integrate to the same total secretion.
					}
				\end{figure}

%% file: secs/Impl.tex
	The main driving factor of the computational cost of Alg.~\ref{algs:disc_AccProxGradforRegInvDif} is the 
	$2K$ convolutions with 2D kernels $\tilde{g}_k$ at each iteration in Steps~\ref{line:forwardanddiff} and \ref{line:adjointandpos}. 
	Although efficiently parallelizable in GPUs, 2D convolution is still an expensive operation.  
	Recall from Part~I~\cite[Section IV]{AguilaPla2017} that the discretized filters $\tilde{g}_k$ are given by
	\begin{IEEEeqnarray}{c} \label{eq:gk}
		\tilde{g}_k\!\left[ (m, n) \right] = \frac{1}{\sqrt{\Delta_k}}\int_{\dsigma_{k-1}}^{\dsigma_k} \omega_{\dsigma}(m) \omega_{\dsigma}(n) \dint\dsigma\,,
	\end{IEEEeqnarray}
	for some $\omega_{\dsigma}:\integ\rightarrow\reals_+$, where $\Delta_k=\dsigma_k-\dsigma_{k-1}$ is the width of the $\dsigma$-dimension bin represented by $k$.
	This expression suggests that $\tilde{g}_k$ is close to being separable, at least for small values of $\Delta_k$. 
	In the particular choice of parameters for our analysis on synthetic data, detailed in Section~\ref{sec:NumRes:Sim}, Tab.~\ref{tab:parstorun}, 
	the smallest value of $\lambda_1/\sum \lambda_l$, where $\lambda_l$ are the decreasingly sorted singular values of a given kernel $\tilde{g}_k$, 
	was $97.72\,\%$, while the smallest value of $(\lambda_1+\lambda_2+\lambda_3)/\sum \lambda_l$ was $99.99\,\%$. 
	In this context, we propose to approximate the 2D kernels $\tilde{g}_k$ by separable 2D kernels, i.e., rank-one kernels. Thus, 
	we will approximate each convolution with a 2D kernel by $2$ successive convolutions with 1D kernels, substantially reducing the computational effort. 
	Note, however, that regardless the approximation, the $2K$ convolutions per iteration will still remain the bottleneck of Alg.~\ref{algs:disc_AccProxGradforRegInvDif}, and thus,
	further efforts on the reduction of the computational burden should involve efficient techniques to implement these convolutions.
	
	In Section~\ref{sec:res:res}, we report the results of approximating $\tilde{g}_k$ as $g_k^{\mathrm{br1}}$, the best rank-one approximation in terms of the
	Frobenius norm. 
	$g_k^{\mathrm{br1}}$ can be obtained numerically by using singular value decomposition on the original kernel $\tilde{g}_k$. In the supplementary material to 
	this paper,
	we discuss two simpler rank-one approximations and report their performance, which was significantly worse than that of $g_k^{\mathrm{br1}}$ in almost every scenario. 
	In order to quantify the loss in performance due to the rank-one approximation, we will also include in Section~\ref{sec:res:res} the results using $g_k^{\mathrm{br3}}$, the best 
	rank-three approximation in terms of the Frobenius norm, which approximates every convolution with a 2D kernel by combining $6$ convolutions with 1D 
	kernels.

%% file: figs/vs_lam_NC750NL3_new.tex
\begin{tikzpicture}[xscale=3.6,yscale=5.5]
	\def\ww{0.030303}\def\bw{0.030303}\def\xm{0.030303}\def\ym{\xm*3.6/5.5}

	\draw (0.000000-2*\xm,0.164238-2*\ym) -- (2.000000+2*\xm,0.164238-2*\ym) -- (2.000000+2*\xm,0.844806+2*\ym) -- (0.000000-2*\xm,0.844806+2*\ym) -- cycle;
\node at (2.000000+2*\xm,0.164238-2*\ym) [anchor=north west]{\footnotesize $\lambda$};

	\foreach \x in {0.00,0.15,0.25,0.30,0.45,0.50,0.75,1.00,1.50,2.00}{ \draw (\x,0.164238-\ym) -- (\x,0.164238-2*\ym) node[anchor=east,rotate=90]{\tiny $\x$}; \draw[thin,dotted,gray] (\x,0.164238-\ym) -- (\x,0.844806+2*\ym); }
	\foreach \y in  {0.15,0.30,0.45,0.60,0.75,0.85}{ \draw (0.000000-\xm,\y) -- (0.000000-2*\xm,\y) node[anchor=east]{\footnotesize $\y$};  \draw[thin,dashed,gray] (0.000000-\xm,\y) -- (2.000000+2*\xm,\y); }
\node at (1.000000,0.844806+4*\ym) {\small F1-Scores ($N_c: 750$, Noise Level: $ 3$)};

		\draw[black,opacity = 0.5] (0.000000-\ww,0.690056) -- (0.000000+\ww,0.690056);\draw[black,opacity = 0.5] (0.000000-\ww,0.729949) -- (0.000000+\ww,0.729949);
\draw[black,opacity = 0.5] (0.000000,0.690056) -- (0.000000,0.696804);\draw[black,opacity = 0.5] (0.000000,0.718676) -- (0.000000,0.729949);		\draw[black,opacity = 0.5] (0.000000-\bw,0.696804) -- (0.000000+\bw,0.696804) -- (0.000000+\bw,0.718676) -- (0.000000-\bw,0.718676) -- cycle;
		\draw[black,opacity = 0.5] (0.000000-\bw,0.708881) -- (0.000000+\bw,0.708881);
		\node[color=black] at (0.000000,0.708447) {.};
		\draw[black,opacity = 0.5] (0.250000-\ww,0.789343) -- (0.250000+\ww,0.789343);\draw[black,opacity = 0.5] (0.250000-\ww,0.821045) -- (0.250000+\ww,0.821045);
\draw[black,opacity = 0.5] (0.250000,0.789343) -- (0.250000,0.796586);\draw[black,opacity = 0.5] (0.250000,0.812635) -- (0.250000,0.821045);		\draw[black,opacity = 0.5] (0.250000-\bw,0.796586) -- (0.250000+\bw,0.796586) -- (0.250000+\bw,0.812635) -- (0.250000-\bw,0.812635) -- cycle;
		\draw[black,opacity = 0.5] (0.250000-\bw,0.804175) -- (0.250000+\bw,0.804175);
		\node[color=black] at (0.250000,0.805195) {.};
		\draw[black,opacity = 0.5] (0.500000-\ww,0.798175) -- (0.500000+\ww,0.798175);\draw[black,opacity = 0.5] (0.500000-\ww,0.826384) -- (0.500000+\ww,0.826384);
\draw[black,opacity = 0.5] (0.500000,0.798175) -- (0.500000,0.803371);\draw[black,opacity = 0.5] (0.500000,0.817159) -- (0.500000,0.826384);		\draw[black,opacity = 0.5] (0.500000-\bw,0.803371) -- (0.500000+\bw,0.803371) -- (0.500000+\bw,0.817159) -- (0.500000-\bw,0.817159) -- cycle;
		\draw[black,opacity = 0.5] (0.500000-\bw,0.809255) -- (0.500000+\bw,0.809255);
		\node[color=black] at (0.500000,0.810948) {.};
		\draw[black,opacity = 0.5] (1.000000-\ww,0.793199) -- (1.000000+\ww,0.793199);\draw[black,opacity = 0.5] (1.000000-\ww,0.826082) -- (1.000000+\ww,0.826082);
\draw[black,opacity = 0.5] (1.000000,0.793199) -- (1.000000,0.798867);\draw[black,opacity = 0.5] (1.000000,0.817923) -- (1.000000,0.826082);		\draw[black,opacity = 0.5] (1.000000-\bw,0.798867) -- (1.000000+\bw,0.798867) -- (1.000000+\bw,0.817923) -- (1.000000-\bw,0.817923) -- cycle;
		\draw[black,opacity = 0.5] (1.000000-\bw,0.807997) -- (1.000000+\bw,0.807997);
		\node[color=black] at (1.000000,0.808698) {.};
		\draw[black,opacity = 0.5] (2.000000-\ww,0.782186) -- (2.000000+\ww,0.782186);\draw[black,opacity = 0.5] (2.000000-\ww,0.812060) -- (2.000000+\ww,0.812060);
\draw[black,opacity = 0.5] (2.000000,0.782186) -- (2.000000,0.789658);\draw[black,opacity = 0.5] (2.000000,0.803407) -- (2.000000,0.812060);		\draw[black,opacity = 0.5] (2.000000-\bw,0.789658) -- (2.000000+\bw,0.789658) -- (2.000000+\bw,0.803407) -- (2.000000-\bw,0.803407) -- cycle;
		\draw[black,opacity = 0.5] (2.000000-\bw,0.797205) -- (2.000000+\bw,0.797205);
		\node[color=black] at (2.000000,0.797480) {.};
		\draw[black,opacity=0.8] (0.000000,0.708447) -- (0.250000,0.805195) -- (0.500000,0.810948) -- (1.000000,0.808698) -- (2.000000,0.797480);

		\draw[black!50!green,opacity = 0.5] (0.000000-\ww,0.717456) -- (0.000000+\ww,0.717456);\draw[black!50!green,opacity = 0.5] (0.000000-\ww,0.754871) -- (0.000000+\ww,0.754871);
\draw[black!50!green,opacity = 0.5] (0.000000,0.717456) -- (0.000000,0.724031);\draw[black!50!green,opacity = 0.5] (0.000000,0.739198) -- (0.000000,0.754871);		\draw[black!50!green,opacity = 0.5] (0.000000-\bw,0.724031) -- (0.000000+\bw,0.724031) -- (0.000000+\bw,0.739198) -- (0.000000-\bw,0.739198) -- cycle;
		\draw[black!50!green,opacity = 0.5] (0.000000-\bw,0.730580) -- (0.000000+\bw,0.730580);
		\node[color=black!50!green] at (0.000000,0.732362) {.};
		\draw[black!50!green,opacity = 0.5] (0.250000-\ww,0.808824) -- (0.250000+\ww,0.808824);\draw[black!50!green,opacity = 0.5] (0.250000-\ww,0.840942) -- (0.250000+\ww,0.840942);
\draw[black!50!green,opacity = 0.5] (0.250000,0.808824) -- (0.250000,0.817126);\draw[black!50!green,opacity = 0.5] (0.250000,0.834382) -- (0.250000,0.840942);		\draw[black!50!green,opacity = 0.5] (0.250000-\bw,0.817126) -- (0.250000+\bw,0.817126) -- (0.250000+\bw,0.834382) -- (0.250000-\bw,0.834382) -- cycle;
		\draw[black!50!green,opacity = 0.5] (0.250000-\bw,0.824911) -- (0.250000+\bw,0.824911);
		\node[color=black!50!green] at (0.250000,0.825872) {.};
		\draw[black!50!green,opacity = 0.5] (0.500000-\ww,0.811042) -- (0.500000+\ww,0.811042);\draw[black!50!green,opacity = 0.5] (0.500000-\ww,0.844806) -- (0.500000+\ww,0.844806);
\draw[black!50!green,opacity = 0.5] (0.500000,0.811042) -- (0.500000,0.817280);\draw[black!50!green,opacity = 0.5] (0.500000,0.837631) -- (0.500000,0.844806);		\draw[black!50!green,opacity = 0.5] (0.500000-\bw,0.817280) -- (0.500000+\bw,0.817280) -- (0.500000+\bw,0.837631) -- (0.500000-\bw,0.837631) -- cycle;
		\draw[black!50!green,opacity = 0.5] (0.500000-\bw,0.828628) -- (0.500000+\bw,0.828628);
		\node[color=black!50!green] at (0.500000,0.827332) {.};
		\draw[black!50!green,opacity = 0.5] (1.000000-\ww,0.805782) -- (1.000000+\ww,0.805782);\draw[black!50!green,opacity = 0.5] (1.000000-\ww,0.840876) -- (1.000000+\ww,0.840876);
\draw[black!50!green,opacity = 0.5] (1.000000,0.805782) -- (1.000000,0.812147);\draw[black!50!green,opacity = 0.5] (1.000000,0.832865) -- (1.000000,0.840876);		\draw[black!50!green,opacity = 0.5] (1.000000-\bw,0.812147) -- (1.000000+\bw,0.812147) -- (1.000000+\bw,0.832865) -- (1.000000-\bw,0.832865) -- cycle;
		\draw[black!50!green,opacity = 0.5] (1.000000-\bw,0.824307) -- (1.000000+\bw,0.824307);
		\node[color=black!50!green] at (1.000000,0.823889) {.};
		\draw[black!50!green,opacity = 0.5] (2.000000-\ww,0.794668) -- (2.000000+\ww,0.794668);\draw[black!50!green,opacity = 0.5] (2.000000-\ww,0.830818) -- (2.000000+\ww,0.830818);
\draw[black!50!green,opacity = 0.5] (2.000000,0.794668) -- (2.000000,0.800000);\draw[black!50!green,opacity = 0.5] (2.000000,0.820154) -- (2.000000,0.830818);		\draw[black!50!green,opacity = 0.5] (2.000000-\bw,0.800000) -- (2.000000+\bw,0.800000) -- (2.000000+\bw,0.820154) -- (2.000000-\bw,0.820154) -- cycle;
		\draw[black!50!green,opacity = 0.5] (2.000000-\bw,0.807684) -- (2.000000+\bw,0.807684);
		\node[color=black!50!green] at (2.000000,0.810740) {.};
		\draw[black!50!green,opacity=0.8] (0.000000,0.732362) -- (0.250000,0.825872) -- (0.500000,0.827332) -- (1.000000,0.823889) -- (2.000000,0.810740);

		\draw[blue,opacity = 0.5] (0.000000-\ww,0.654692) -- (0.000000+\ww,0.654692);\draw[blue,opacity = 0.5] (0.000000-\ww,0.717711) -- (0.000000+\ww,0.717711);
\draw[blue,opacity = 0.5] (0.000000,0.654692) -- (0.000000,0.672811);\draw[blue,opacity = 0.5] (0.000000,0.702362) -- (0.000000,0.717711);		\draw[blue,opacity = 0.5] (0.000000-\bw,0.672811) -- (0.000000+\bw,0.672811) -- (0.000000+\bw,0.702362) -- (0.000000-\bw,0.702362) -- cycle;
		\draw[blue,opacity = 0.5] (0.000000-\bw,0.691941) -- (0.000000+\bw,0.691941);
		\node[color=blue] at (0.000000,0.688855) {.};
		\draw[blue,opacity = 0.5] (0.150000-\ww,0.537692) -- (0.150000+\ww,0.537692);\draw[blue,opacity = 0.5] (0.150000-\ww,0.587252) -- (0.150000+\ww,0.587252);
\draw[blue,opacity = 0.5] (0.150000,0.537692) -- (0.150000,0.550854);\draw[blue,opacity = 0.5] (0.150000,0.572981) -- (0.150000,0.587252);		\draw[blue,opacity = 0.5] (0.150000-\bw,0.550854) -- (0.150000+\bw,0.550854) -- (0.150000+\bw,0.572981) -- (0.150000-\bw,0.572981) -- cycle;
		\draw[blue,opacity = 0.5] (0.150000-\bw,0.557984) -- (0.150000+\bw,0.557984);
		\node[color=blue] at (0.150000,0.560582) {.};
		\draw[blue,opacity = 0.5] (0.300000-\ww,0.555199) -- (0.300000+\ww,0.555199);\draw[blue,opacity = 0.5] (0.300000-\ww,0.614510) -- (0.300000+\ww,0.614510);
\draw[blue,opacity = 0.5] (0.300000,0.555199) -- (0.300000,0.572474);\draw[blue,opacity = 0.5] (0.300000,0.603053) -- (0.300000,0.614510);		\draw[blue,opacity = 0.5] (0.300000-\bw,0.572474) -- (0.300000+\bw,0.572474) -- (0.300000+\bw,0.603053) -- (0.300000-\bw,0.603053) -- cycle;
		\draw[blue,opacity = 0.5] (0.300000-\bw,0.584993) -- (0.300000+\bw,0.584993);
		\node[color=blue] at (0.300000,0.587084) {.};
		\draw[blue,opacity = 0.5] (0.450000-\ww,0.531620) -- (0.450000+\ww,0.531620);\draw[blue,opacity = 0.5] (0.450000-\ww,0.622621) -- (0.450000+\ww,0.622621);
\draw[blue,opacity = 0.5] (0.450000,0.531620) -- (0.450000,0.565056);\draw[blue,opacity = 0.5] (0.450000,0.606792) -- (0.450000,0.622621);		\draw[blue,opacity = 0.5] (0.450000-\bw,0.565056) -- (0.450000+\bw,0.565056) -- (0.450000+\bw,0.606792) -- (0.450000-\bw,0.606792) -- cycle;
		\draw[blue,opacity = 0.5] (0.450000-\bw,0.592872) -- (0.450000+\bw,0.592872);
		\node[color=blue] at (0.450000,0.584931) {.};
		\draw[blue,opacity = 0.5] (0.750000-\ww,0.429634) -- (0.750000+\ww,0.429634);\draw[blue,opacity = 0.5] (0.750000-\ww,0.526591) -- (0.750000+\ww,0.526591);
\draw[blue,opacity = 0.5] (0.750000,0.429634) -- (0.750000,0.446693);\draw[blue,opacity = 0.5] (0.750000,0.505626) -- (0.750000,0.526591);		\draw[blue,opacity = 0.5] (0.750000-\bw,0.446693) -- (0.750000+\bw,0.446693) -- (0.750000+\bw,0.505626) -- (0.750000-\bw,0.505626) -- cycle;
		\draw[blue,opacity = 0.5] (0.750000-\bw,0.476186) -- (0.750000+\bw,0.476186);
		\node[color=blue] at (0.750000,0.475774) {.};
		\draw[blue,opacity = 0.5] (1.000000-\ww,0.382363) -- (1.000000+\ww,0.382363);\draw[blue,opacity = 0.5] (1.000000-\ww,0.446185) -- (1.000000+\ww,0.446185);
\draw[blue,opacity = 0.5] (1.000000,0.382363) -- (1.000000,0.391770);\draw[blue,opacity = 0.5] (1.000000,0.426948) -- (1.000000,0.446185);		\draw[blue,opacity = 0.5] (1.000000-\bw,0.391770) -- (1.000000+\bw,0.391770) -- (1.000000+\bw,0.426948) -- (1.000000-\bw,0.426948) -- cycle;
		\draw[blue,opacity = 0.5] (1.000000-\bw,0.405596) -- (1.000000+\bw,0.405596);
		\node[color=blue] at (1.000000,0.411518) {.};
		\draw[blue,opacity = 0.5] (1.500000-\ww,0.309720) -- (1.500000+\ww,0.309720);\draw[blue,opacity = 0.5] (1.500000-\ww,0.389889) -- (1.500000+\ww,0.389889);
\draw[blue,opacity = 0.5] (1.500000,0.309720) -- (1.500000,0.329545);\draw[blue,opacity = 0.5] (1.500000,0.368282) -- (1.500000,0.389889);		\draw[blue,opacity = 0.5] (1.500000-\bw,0.329545) -- (1.500000+\bw,0.329545) -- (1.500000+\bw,0.368282) -- (1.500000-\bw,0.368282) -- cycle;
		\draw[blue,opacity = 0.5] (1.500000-\bw,0.353834) -- (1.500000+\bw,0.353834);
		\node[color=blue] at (1.500000,0.348193) {.};
		\draw[blue,opacity = 0.5] (2.000000-\ww,0.164238) -- (2.000000+\ww,0.164238);\draw[blue,opacity = 0.5] (2.000000-\ww,0.316939) -- (2.000000+\ww,0.316939);
\draw[blue,opacity = 0.5] (2.000000,0.164238) -- (2.000000,0.200640);\draw[blue,opacity = 0.5] (2.000000,0.291871) -- (2.000000,0.316939);		\draw[blue,opacity = 0.5] (2.000000-\bw,0.200640) -- (2.000000+\bw,0.200640) -- (2.000000+\bw,0.291871) -- (2.000000-\bw,0.291871) -- cycle;
		\draw[blue,opacity = 0.5] (2.000000-\bw,0.243320) -- (2.000000+\bw,0.243320);
		\node[color=blue] at (2.000000,0.240477) {.};
		\draw[blue,opacity=0.8] (0.000000,0.688855) -- (0.150000,0.560582) -- (0.300000,0.587084) -- (0.450000,0.584931) -- (0.750000,0.475774) -- (1.000000,0.411518) -- (1.500000,0.348193) -- (2.000000,0.240477);

		\draw[red,opacity = 0.5] (1.000000-\ww,0.438742) -- (1.000000+\ww,0.438742);\draw[red,opacity = 0.5] (1.000000-\ww,0.513283) -- (1.000000+\ww,0.513283);
\draw[red,opacity = 0.5] (1.000000,0.438742) -- (1.000000,0.455336);\draw[red,opacity = 0.5] (1.000000,0.499234) -- (1.000000,0.513283);		\draw[red,opacity = 0.5] (1.000000-\bw,0.455336) -- (1.000000+\bw,0.455336) -- (1.000000+\bw,0.499234) -- (1.000000-\bw,0.499234) -- cycle;
		\draw[red,opacity = 0.5] (1.000000-\bw,0.483674) -- (1.000000+\bw,0.483674);
		\node[color=red] at (1.000000,0.477120) {.};
		\draw[red,opacity=0.8] (1.000000,0.477120);

		\draw[black!50!red,opacity = 0.5] (1.000000-\ww,0.627090) -- (1.000000+\ww,0.627090);\draw[black!50!red,opacity = 0.5] (1.000000-\ww,0.671644) -- (1.000000+\ww,0.671644);
\draw[black!50!red,opacity = 0.5] (1.000000,0.627090) -- (1.000000,0.636220);\draw[black!50!red,opacity = 0.5] (1.000000,0.662005) -- (1.000000,0.671644);		\draw[black!50!red,opacity = 0.5] (1.000000-\bw,0.636220) -- (1.000000+\bw,0.636220) -- (1.000000+\bw,0.662005) -- (1.000000-\bw,0.662005) -- cycle;
		\draw[black!50!red,opacity = 0.5] (1.000000-\bw,0.647570) -- (1.000000+\bw,0.647570);
		\node[color=black!50!red] at (1.000000,0.648362) {.};
		\draw[black!50!red,opacity=0.8] (1.000000,0.648362);
		
		\def\xmax{2+2*\xm} 
		\def\xmin{0-2*\xm}
		\def\ymax{0.164238-9*\ym}
		\def\legh{0.4/5.5} 
		\def\indicatorw{0.3/3.67} 
		\input{\figs/legend}                                                          
\end{tikzpicture}

%% file: figs/vs_NC_NL3+4_new_withChristian.tex
\begin{tikzpicture}[xscale=0.0715,yscale=6.2]
	\def\ww{1.515152}\def\bw{1.515152}\def\xm{1.515152}\def\ym{\xm*0.0715/6.2}

	\draw (25.000000-2*\xm,0.346073-2*\ym) -- (125.000000+2*\xm,0.346073-2*\ym) -- (125.000000+2*\xm,0.927834+2*\ym) -- (25.000000-2*\xm,0.927834+2*\ym) -- cycle;
\node at (125.000000,0.346073-5*\ym) [anchor=north west]{\footnotesize $N_c$};

	\foreach \x/\n in {25/250,75/750,125/1250}{ \draw (\x,0.346073-\ym) -- (\x,0.346073-2*\ym) node[anchor=north,rotate=0]{\footnotesize $\n$}; \draw[thin,dotted,gray] (\x,0.346073-\ym) -- (\x,0.927834+2*\ym); }
	\foreach \y in {0.40,0.50,0.60,0.70,0.80,0.90}{ \draw (25.000000-\xm,\y) -- (25.000000-2*\xm,\y) node[anchor=east]{\footnotesize $\y$};  \draw[thin,dashed,gray] (25.000000-\xm,\y) -- (125.000000+2*\xm,\y); }
\node at (75.000000,0.927834+4*\ym) {\small F1-Scores ($\lambda: 0.50$, Noise Level: $ 3$, $\lambda_d: 0.00$)};

		\draw[black,opacity = 0.5] (25.000000-\ww,0.886597) -- (25.000000+\ww,0.886597);\draw[black,opacity = 0.5] (25.000000-\ww,0.920829) -- (25.000000+\ww,0.920829);
\draw[black,opacity = 0.5] (25.000000,0.886597) -- (25.000000,0.895075);\draw[black,opacity = 0.5] (25.000000,0.912863) -- (25.000000,0.920829);		\draw[black,opacity = 0.5] (25.000000-\bw,0.895075) -- (25.000000+\bw,0.895075) -- (25.000000+\bw,0.912863) -- (25.000000-\bw,0.912863) -- cycle;
		\draw[black,opacity = 0.5] (25.000000-\bw,0.903766) -- (25.000000+\bw,0.903766);
		\node[color=black] at (25.000000,0.903380) {.};
		\draw[black,opacity = 0.5] (75.000000-\ww,0.798175) -- (75.000000+\ww,0.798175);\draw[black,opacity = 0.5] (75.000000-\ww,0.826384) -- (75.000000+\ww,0.826384);
\draw[black,opacity = 0.5] (75.000000,0.798175) -- (75.000000,0.803371);\draw[black,opacity = 0.5] (75.000000,0.817159) -- (75.000000,0.826384);		\draw[black,opacity = 0.5] (75.000000-\bw,0.803371) -- (75.000000+\bw,0.803371) -- (75.000000+\bw,0.817159) -- (75.000000-\bw,0.817159) -- cycle;
		\draw[black,opacity = 0.5] (75.000000-\bw,0.809255) -- (75.000000+\bw,0.809255);
		\node[color=black] at (75.000000,0.810948) {.};
		\draw[black,opacity = 0.5] (125.000000-\ww,0.720190) -- (125.000000+\ww,0.720190);\draw[black,opacity = 0.5] (125.000000-\ww,0.758380) -- (125.000000+\ww,0.758380);
\draw[black,opacity = 0.5] (125.000000,0.720190) -- (125.000000,0.732648);\draw[black,opacity = 0.5] (125.000000,0.752475) -- (125.000000,0.758380);		\draw[black,opacity = 0.5] (125.000000-\bw,0.732648) -- (125.000000+\bw,0.732648) -- (125.000000+\bw,0.752475) -- (125.000000-\bw,0.752475) -- cycle;
		\draw[black,opacity = 0.5] (125.000000-\bw,0.738859) -- (125.000000+\bw,0.738859);
		\node[color=black] at (125.000000,0.739516) {.};
		\draw[black,opacity=0.8] (25.000000,0.903380) -- (75.000000,0.810948) -- (125.000000,0.739516);

		\draw[black!50!green,opacity = 0.5] (25.000000-\ww,0.894705) -- (25.000000+\ww,0.894705);\draw[black!50!green,opacity = 0.5] (25.000000-\ww,0.927834) -- (25.000000+\ww,0.927834);
\draw[black!50!green,opacity = 0.5] (25.000000,0.894705) -- (25.000000,0.906054);\draw[black!50!green,opacity = 0.5] (25.000000,0.923077) -- (25.000000,0.927834);		\draw[black!50!green,opacity = 0.5] (25.000000-\bw,0.906054) -- (25.000000+\bw,0.906054) -- (25.000000+\bw,0.923077) -- (25.000000-\bw,0.923077) -- cycle;
		\draw[black!50!green,opacity = 0.5] (25.000000-\bw,0.912821) -- (25.000000+\bw,0.912821);
		\node[color=black!50!green] at (25.000000,0.913165) {.};
		\draw[black!50!green,opacity = 0.5] (75.000000-\ww,0.811042) -- (75.000000+\ww,0.811042);\draw[black!50!green,opacity = 0.5] (75.000000-\ww,0.844806) -- (75.000000+\ww,0.844806);
\draw[black!50!green,opacity = 0.5] (75.000000,0.811042) -- (75.000000,0.817280);\draw[black!50!green,opacity = 0.5] (75.000000,0.837631) -- (75.000000,0.844806);		\draw[black!50!green,opacity = 0.5] (75.000000-\bw,0.817280) -- (75.000000+\bw,0.817280) -- (75.000000+\bw,0.837631) -- (75.000000-\bw,0.837631) -- cycle;
		\draw[black!50!green,opacity = 0.5] (75.000000-\bw,0.828628) -- (75.000000+\bw,0.828628);
		\node[color=black!50!green] at (75.000000,0.827332) {.};
		\draw[black!50!green,opacity = 0.5] (125.000000-\ww,0.746938) -- (125.000000+\ww,0.746938);\draw[black!50!green,opacity = 0.5] (125.000000-\ww,0.778296) -- (125.000000+\ww,0.778296);
\draw[black!50!green,opacity = 0.5] (125.000000,0.746938) -- (125.000000,0.752467);\draw[black!50!green,opacity = 0.5] (125.000000,0.771990) -- (125.000000,0.778296);		\draw[black!50!green,opacity = 0.5] (125.000000-\bw,0.752467) -- (125.000000+\bw,0.752467) -- (125.000000+\bw,0.771990) -- (125.000000-\bw,0.771990) -- cycle;
		\draw[black!50!green,opacity = 0.5] (125.000000-\bw,0.761166) -- (125.000000+\bw,0.761166);
		\node[color=black!50!green] at (125.000000,0.761891) {.};
		\draw[black!50!green,opacity=0.8] (25.000000,0.913165) -- (75.000000,0.827332) -- (125.000000,0.761891);

		\draw[blue,opacity = 0.5] (25.000000-\ww,0.750868) -- (25.000000+\ww,0.750868);\draw[blue,opacity = 0.5] (25.000000-\ww,0.819761) -- (25.000000+\ww,0.819761);
\draw[blue,opacity = 0.5] (25.000000,0.750868) -- (25.000000,0.766440);\draw[blue,opacity = 0.5] (25.000000,0.802752) -- (25.000000,0.819761);		\draw[blue,opacity = 0.5] (25.000000-\bw,0.766440) -- (25.000000+\bw,0.766440) -- (25.000000+\bw,0.802752) -- (25.000000-\bw,0.802752) -- cycle;
		\draw[blue,opacity = 0.5] (25.000000-\bw,0.782393) -- (25.000000+\bw,0.782393);
		\node[color=blue] at (25.000000,0.783990) {.};
		\draw[blue,opacity = 0.5] (75.000000-\ww,0.654692) -- (75.000000+\ww,0.654692);\draw[blue,opacity = 0.5] (75.000000-\ww,0.717711) -- (75.000000+\ww,0.717711);
\draw[blue,opacity = 0.5] (75.000000,0.654692) -- (75.000000,0.672811);\draw[blue,opacity = 0.5] (75.000000,0.702362) -- (75.000000,0.717711);		\draw[blue,opacity = 0.5] (75.000000-\bw,0.672811) -- (75.000000+\bw,0.672811) -- (75.000000+\bw,0.702362) -- (75.000000-\bw,0.702362) -- cycle;
		\draw[blue,opacity = 0.5] (75.000000-\bw,0.691941) -- (75.000000+\bw,0.691941);
		\node[color=blue] at (75.000000,0.688855) {.};
		\draw[blue,opacity = 0.5] (125.000000-\ww,0.602345) -- (125.000000+\ww,0.602345);\draw[blue,opacity = 0.5] (125.000000-\ww,0.664335) -- (125.000000+\ww,0.664335);
\draw[blue,opacity = 0.5] (125.000000,0.602345) -- (125.000000,0.615093);\draw[blue,opacity = 0.5] (125.000000,0.642402) -- (125.000000,0.664335);		\draw[blue,opacity = 0.5] (125.000000-\bw,0.615093) -- (125.000000+\bw,0.615093) -- (125.000000+\bw,0.642402) -- (125.000000-\bw,0.642402) -- cycle;
		\draw[blue,opacity = 0.5] (125.000000-\bw,0.625698) -- (125.000000+\bw,0.625698);
		\node[color=blue] at (125.000000,0.629007) {.};
		\draw[blue,opacity=0.8] (25.000000,0.783990) -- (75.000000,0.688855) -- (125.000000,0.629007);

		\draw[red,opacity = 0.5] (25.000000-\ww,0.606265) -- (25.000000+\ww,0.606265);\draw[red,opacity = 0.5] (25.000000-\ww,0.703765) -- (25.000000+\ww,0.703765);
\draw[red,opacity = 0.5] (25.000000,0.606265) -- (25.000000,0.631579);\draw[red,opacity = 0.5] (25.000000,0.684685) -- (25.000000,0.703765);		\draw[red,opacity = 0.5] (25.000000-\bw,0.631579) -- (25.000000+\bw,0.631579) -- (25.000000+\bw,0.684685) -- (25.000000-\bw,0.684685) -- cycle;
		\draw[red,opacity = 0.5] (25.000000-\bw,0.657626) -- (25.000000+\bw,0.657626);
		\node[color=red] at (25.000000,0.656423) {.};
		\draw[red,opacity = 0.5] (75.000000-\ww,0.438742) -- (75.000000+\ww,0.438742);\draw[red,opacity = 0.5] (75.000000-\ww,0.513283) -- (75.000000+\ww,0.513283);
\draw[red,opacity = 0.5] (75.000000,0.438742) -- (75.000000,0.455336);\draw[red,opacity = 0.5] (75.000000,0.499234) -- (75.000000,0.513283);		\draw[red,opacity = 0.5] (75.000000-\bw,0.455336) -- (75.000000+\bw,0.455336) -- (75.000000+\bw,0.499234) -- (75.000000-\bw,0.499234) -- cycle;
		\draw[red,opacity = 0.5] (75.000000-\bw,0.483674) -- (75.000000+\bw,0.483674);
		\node[color=red] at (75.000000,0.477120) {.};
		\draw[red,opacity = 0.5] (125.000000-\ww,0.346073) -- (125.000000+\ww,0.346073);\draw[red,opacity = 0.5] (125.000000-\ww,0.404413) -- (125.000000+\ww,0.404413);
\draw[red,opacity = 0.5] (125.000000,0.346073) -- (125.000000,0.363010);\draw[red,opacity = 0.5] (125.000000,0.391885) -- (125.000000,0.404413);		\draw[red,opacity = 0.5] (125.000000-\bw,0.363010) -- (125.000000+\bw,0.363010) -- (125.000000+\bw,0.391885) -- (125.000000-\bw,0.391885) -- cycle;
		\draw[red,opacity = 0.5] (125.000000-\bw,0.374256) -- (125.000000+\bw,0.374256);
		\node[color=red] at (125.000000,0.374687) {.};
		\draw[red,opacity=0.8] (25.000000,0.656423) -- (75.000000,0.477120) -- (125.000000,0.374687);

		\draw[black!50!red,opacity = 0.5] (25.000000-\ww,0.817597) -- (25.000000+\ww,0.817597);\draw[black!50!red,opacity = 0.5] (25.000000-\ww,0.872611) -- (25.000000+\ww,0.872611);
\draw[black!50!red,opacity = 0.5] (25.000000,0.817597) -- (25.000000,0.828633);\draw[black!50!red,opacity = 0.5] (25.000000,0.859611) -- (25.000000,0.872611);		\draw[black!50!red,opacity = 0.5] (25.000000-\bw,0.828633) -- (25.000000+\bw,0.828633) -- (25.000000+\bw,0.859611) -- (25.000000-\bw,0.859611) -- cycle;
		\draw[black!50!red,opacity = 0.5] (25.000000-\bw,0.847060) -- (25.000000+\bw,0.847060);
		\node[color=black!50!red] at (25.000000,0.846573) {.};
		\draw[black!50!red,opacity = 0.5] (75.000000-\ww,0.627090) -- (75.000000+\ww,0.627090);\draw[black!50!red,opacity = 0.5] (75.000000-\ww,0.671644) -- (75.000000+\ww,0.671644);
\draw[black!50!red,opacity = 0.5] (75.000000,0.627090) -- (75.000000,0.636220);\draw[black!50!red,opacity = 0.5] (75.000000,0.662005) -- (75.000000,0.671644);		\draw[black!50!red,opacity = 0.5] (75.000000-\bw,0.636220) -- (75.000000+\bw,0.636220) -- (75.000000+\bw,0.662005) -- (75.000000-\bw,0.662005) -- cycle;
		\draw[black!50!red,opacity = 0.5] (75.000000-\bw,0.647570) -- (75.000000+\bw,0.647570);
		\node[color=black!50!red] at (75.000000,0.648362) {.};
		\draw[black!50!red,opacity = 0.5] (125.000000-\ww,0.494249) -- (125.000000+\ww,0.494249);\draw[black!50!red,opacity = 0.5] (125.000000-\ww,0.537271) -- (125.000000+\ww,0.537271);
\draw[black!50!red,opacity = 0.5] (125.000000,0.494249) -- (125.000000,0.507028);\draw[black!50!red,opacity = 0.5] (125.000000,0.527264) -- (125.000000,0.537271);		\draw[black!50!red,opacity = 0.5] (125.000000-\bw,0.507028) -- (125.000000+\bw,0.507028) -- (125.000000+\bw,0.527264) -- (125.000000-\bw,0.527264) -- cycle;
		\draw[black!50!red,opacity = 0.5] (125.000000-\bw,0.521102) -- (125.000000+\bw,0.521102);
		\node[color=black!50!red] at (125.000000,0.517301) {.};
		\draw[black!50!red,opacity=0.8] (25.000000,0.846573) -- (75.000000,0.648362) -- (125.000000,0.517301);
		
		\draw[black,thick,dash pattern=on 2.3pt off 1.1pt on 0.8pt off 1.1pt] (25,0.896247) node {\footnotesize $\oplus$} -- (75,0.806990) node {\footnotesize $\oplus$} -- (125,0.725790) node {\footnotesize $\oplus$};

\end{tikzpicture}

\vspace{-10pt}

\begin{tikzpicture}[xscale=0.0715,yscale=4.75]
	\def\ww{1.515152}\def\bw{1.515152}\def\xm{1.515152}\def\ym{\xm*0.0715/4.75}

	\draw (25.000000-2*\xm,0.133491-2*\ym) -- (125.000000+2*\xm,0.133491-2*\ym) -- (125.000000+2*\xm,0.872611+2*\ym) -- (25.000000-2*\xm,0.872611+2*\ym) -- cycle;
\node at (125.000000,0.133491-5*\ym) [anchor=north west]{\footnotesize $N_c$};

	\foreach \x/\n in {25/250,75/750,125/1250}{ \draw (\x,0.133491-\ym) -- (\x,0.133491-2*\ym) node[anchor=north,rotate=0]{\footnotesize $\n$}; \draw[thin,dotted,gray] (\x,0.133491-\ym) -- (\x,0.872611+2*\ym); }
	\foreach \y in {0.15,0.30,0.45,0.60,0.72,0.85}{ \draw (25.000000-\xm,\y) -- (25.000000-2*\xm,\y) node[anchor=east]{\footnotesize $\y$};  \draw[thin,dashed,gray] (25.000000-\xm,\y) -- (125.000000+2*\xm,\y); }
\node at (75.000000,0.872611+4*\ym) {\small F1-Scores ($\lambda: 0.50$, Noise Level: $ 4$, $\lambda_d: 0.00$)};

		\draw[black,opacity = 0.5] (25.000000-\ww,0.799544) -- (25.000000+\ww,0.799544);\draw[black,opacity = 0.5] (25.000000-\ww,0.865036) -- (25.000000+\ww,0.865036);
\draw[black,opacity = 0.5] (25.000000,0.799544) -- (25.000000,0.811966);\draw[black,opacity = 0.5] (25.000000,0.850526) -- (25.000000,0.865036);		\draw[black,opacity = 0.5] (25.000000-\bw,0.811966) -- (25.000000+\bw,0.811966) -- (25.000000+\bw,0.850526) -- (25.000000-\bw,0.850526) -- cycle;
		\draw[black,opacity = 0.5] (25.000000-\bw,0.830195) -- (25.000000+\bw,0.830195);
		\node[color=black] at (25.000000,0.832586) {.};
		\draw[black,opacity = 0.5] (75.000000-\ww,0.684074) -- (75.000000+\ww,0.684074);\draw[black,opacity = 0.5] (75.000000-\ww,0.754560) -- (75.000000+\ww,0.754560);
\draw[black,opacity = 0.5] (75.000000,0.684074) -- (75.000000,0.703574);\draw[black,opacity = 0.5] (75.000000,0.736383) -- (75.000000,0.754560);		\draw[black,opacity = 0.5] (75.000000-\bw,0.703574) -- (75.000000+\bw,0.703574) -- (75.000000+\bw,0.736383) -- (75.000000-\bw,0.736383) -- cycle;
		\draw[black,opacity = 0.5] (75.000000-\bw,0.723418) -- (75.000000+\bw,0.723418);
		\node[color=black] at (75.000000,0.720593) {.};
		\draw[black,opacity = 0.5] (125.000000-\ww,0.613689) -- (125.000000+\ww,0.613689);\draw[black,opacity = 0.5] (125.000000-\ww,0.681618) -- (125.000000+\ww,0.681618);
\draw[black,opacity = 0.5] (125.000000,0.613689) -- (125.000000,0.638095);\draw[black,opacity = 0.5] (125.000000,0.670139) -- (125.000000,0.681618);		\draw[black,opacity = 0.5] (125.000000-\bw,0.638095) -- (125.000000+\bw,0.638095) -- (125.000000+\bw,0.670139) -- (125.000000-\bw,0.670139) -- cycle;
		\draw[black,opacity = 0.5] (125.000000-\bw,0.655338) -- (125.000000+\bw,0.655338);
		\node[color=black] at (125.000000,0.651121) {.};
		\draw[black,opacity=0.8] (25.000000,0.832586) -- (75.000000,0.720593) -- (125.000000,0.651121);

		\draw[black!50!green,opacity = 0.5] (25.000000-\ww,0.797889) -- (25.000000+\ww,0.797889);\draw[black!50!green,opacity = 0.5] (25.000000-\ww,0.865031) -- (25.000000+\ww,0.865031);
\draw[black!50!green,opacity = 0.5] (25.000000,0.797889) -- (25.000000,0.815618);\draw[black!50!green,opacity = 0.5] (25.000000,0.851695) -- (25.000000,0.865031);		\draw[black!50!green,opacity = 0.5] (25.000000-\bw,0.815618) -- (25.000000+\bw,0.815618) -- (25.000000+\bw,0.851695) -- (25.000000-\bw,0.851695) -- cycle;
		\draw[black!50!green,opacity = 0.5] (25.000000-\bw,0.834734) -- (25.000000+\bw,0.834734);
		\node[color=black!50!green] at (25.000000,0.833858) {.};
		\draw[black!50!green,opacity = 0.5] (75.000000-\ww,0.692979) -- (75.000000+\ww,0.692979);\draw[black!50!green,opacity = 0.5] (75.000000-\ww,0.757020) -- (75.000000+\ww,0.757020);
\draw[black!50!green,opacity = 0.5] (75.000000,0.692979) -- (75.000000,0.709302);\draw[black!50!green,opacity = 0.5] (75.000000,0.744758) -- (75.000000,0.757020);		\draw[black!50!green,opacity = 0.5] (75.000000-\bw,0.709302) -- (75.000000+\bw,0.709302) -- (75.000000+\bw,0.744758) -- (75.000000-\bw,0.744758) -- cycle;
		\draw[black!50!green,opacity = 0.5] (75.000000-\bw,0.731794) -- (75.000000+\bw,0.731794);
		\node[color=black!50!green] at (75.000000,0.726528) {.};
		\draw[black!50!green,opacity = 0.5] (125.000000-\ww,0.626498) -- (125.000000+\ww,0.626498);\draw[black!50!green,opacity = 0.5] (125.000000-\ww,0.684949) -- (125.000000+\ww,0.684949);
\draw[black!50!green,opacity = 0.5] (125.000000,0.626498) -- (125.000000,0.645648);\draw[black!50!green,opacity = 0.5] (125.000000,0.673904) -- (125.000000,0.684949);		\draw[black!50!green,opacity = 0.5] (125.000000-\bw,0.645648) -- (125.000000+\bw,0.645648) -- (125.000000+\bw,0.673904) -- (125.000000-\bw,0.673904) -- cycle;
		\draw[black!50!green,opacity = 0.5] (125.000000-\bw,0.660869) -- (125.000000+\bw,0.660869);
		\node[color=black!50!green] at (125.000000,0.658364) {.};
		\draw[black!50!green,opacity=0.8] (25.000000,0.833858) -- (75.000000,0.726528) -- (125.000000,0.658364);

		\draw[blue,opacity = 0.5] (25.000000-\ww,0.455083) -- (25.000000+\ww,0.455083);\draw[blue,opacity = 0.5] (25.000000-\ww,0.565459) -- (25.000000+\ww,0.565459);
\draw[blue,opacity = 0.5] (25.000000,0.455083) -- (25.000000,0.483721);\draw[blue,opacity = 0.5] (25.000000,0.539683) -- (25.000000,0.565459);		\draw[blue,opacity = 0.5] (25.000000-\bw,0.483721) -- (25.000000+\bw,0.483721) -- (25.000000+\bw,0.539683) -- (25.000000-\bw,0.539683) -- cycle;
		\draw[blue,opacity = 0.5] (25.000000-\bw,0.520732) -- (25.000000+\bw,0.520732);
		\node[color=blue] at (25.000000,0.514289) {.};
		\draw[blue,opacity = 0.5] (75.000000-\ww,0.304326) -- (75.000000+\ww,0.304326);\draw[blue,opacity = 0.5] (75.000000-\ww,0.409085) -- (75.000000+\ww,0.409085);
\draw[blue,opacity = 0.5] (75.000000,0.304326) -- (75.000000,0.339775);\draw[blue,opacity = 0.5] (75.000000,0.388802) -- (75.000000,0.409085);		\draw[blue,opacity = 0.5] (75.000000-\bw,0.339775) -- (75.000000+\bw,0.339775) -- (75.000000+\bw,0.388802) -- (75.000000-\bw,0.388802) -- cycle;
		\draw[blue,opacity = 0.5] (75.000000-\bw,0.361108) -- (75.000000+\bw,0.361108);
		\node[color=blue] at (75.000000,0.360360) {.};
		\draw[blue,opacity = 0.5] (125.000000-\ww,0.241732) -- (125.000000+\ww,0.241732);\draw[blue,opacity = 0.5] (125.000000-\ww,0.335533) -- (125.000000+\ww,0.335533);
\draw[blue,opacity = 0.5] (125.000000,0.241732) -- (125.000000,0.267048);\draw[blue,opacity = 0.5] (125.000000,0.316231) -- (125.000000,0.335533);		\draw[blue,opacity = 0.5] (125.000000-\bw,0.267048) -- (125.000000+\bw,0.267048) -- (125.000000+\bw,0.316231) -- (125.000000-\bw,0.316231) -- cycle;
		\draw[blue,opacity = 0.5] (125.000000-\bw,0.293066) -- (125.000000+\bw,0.293066);
		\node[color=blue] at (125.000000,0.291222) {.};
		\draw[blue,opacity=0.8] (25.000000,0.514289) -- (75.000000,0.360360) -- (125.000000,0.291222);

		\draw[red,opacity = 0.5] (25.000000-\ww,0.245352) -- (25.000000+\ww,0.245352);\draw[red,opacity = 0.5] (25.000000-\ww,0.332999) -- (25.000000+\ww,0.332999);
\draw[red,opacity = 0.5] (25.000000,0.245352) -- (25.000000,0.253829);\draw[red,opacity = 0.5] (25.000000,0.319101) -- (25.000000,0.332999);		\draw[red,opacity = 0.5] (25.000000-\bw,0.253829) -- (25.000000+\bw,0.253829) -- (25.000000+\bw,0.319101) -- (25.000000-\bw,0.319101) -- cycle;
		\draw[red,opacity = 0.5] (25.000000-\bw,0.287399) -- (25.000000+\bw,0.287399);
		\node[color=red] at (25.000000,0.289245) {.};
		\draw[red,opacity = 0.5] (75.000000-\ww,0.165125) -- (75.000000+\ww,0.165125);\draw[red,opacity = 0.5] (75.000000-\ww,0.216641) -- (75.000000+\ww,0.216641);
\draw[red,opacity = 0.5] (75.000000,0.165125) -- (75.000000,0.178203);\draw[red,opacity = 0.5] (75.000000,0.204735) -- (75.000000,0.216641);		\draw[red,opacity = 0.5] (75.000000-\bw,0.178203) -- (75.000000+\bw,0.178203) -- (75.000000+\bw,0.204735) -- (75.000000-\bw,0.204735) -- cycle;
		\draw[red,opacity = 0.5] (75.000000-\bw,0.193350) -- (75.000000+\bw,0.193350);
		\node[color=red] at (75.000000,0.191172) {.};
		\draw[red,opacity = 0.5] (125.000000-\ww,0.133491) -- (125.000000+\ww,0.133491);\draw[red,opacity = 0.5] (125.000000-\ww,0.169597) -- (125.000000+\ww,0.169597);
\draw[red,opacity = 0.5] (125.000000,0.133491) -- (125.000000,0.142222);\draw[red,opacity = 0.5] (125.000000,0.160610) -- (125.000000,0.169597);		\draw[red,opacity = 0.5] (125.000000-\bw,0.142222) -- (125.000000+\bw,0.142222) -- (125.000000+\bw,0.160610) -- (125.000000-\bw,0.160610) -- cycle;
		\draw[red,opacity = 0.5] (125.000000-\bw,0.150541) -- (125.000000+\bw,0.150541);
		\node[color=red] at (125.000000,0.149869) {.};
		\draw[red,opacity=0.8] (25.000000,0.289245) -- (75.000000,0.191172) -- (125.000000,0.149869);

		\draw[black!50!red,opacity = 0.5] (25.000000-\ww,0.817597) -- (25.000000+\ww,0.817597);\draw[black!50!red,opacity = 0.5] (25.000000-\ww,0.872611) -- (25.000000+\ww,0.872611);
\draw[black!50!red,opacity = 0.5] (25.000000,0.817597) -- (25.000000,0.828633);\draw[black!50!red,opacity = 0.5] (25.000000,0.859611) -- (25.000000,0.872611);		\draw[black!50!red,opacity = 0.5] (25.000000-\bw,0.828633) -- (25.000000+\bw,0.828633) -- (25.000000+\bw,0.859611) -- (25.000000-\bw,0.859611) -- cycle;
		\draw[black!50!red,opacity = 0.5] (25.000000-\bw,0.847060) -- (25.000000+\bw,0.847060);
		\node[color=black!50!red] at (25.000000,0.846573) {.};
		\draw[black!50!red,opacity = 0.5] (75.000000-\ww,0.627090) -- (75.000000+\ww,0.627090);\draw[black!50!red,opacity = 0.5] (75.000000-\ww,0.671644) -- (75.000000+\ww,0.671644);
\draw[black!50!red,opacity = 0.5] (75.000000,0.627090) -- (75.000000,0.636220);\draw[black!50!red,opacity = 0.5] (75.000000,0.662005) -- (75.000000,0.671644);		\draw[black!50!red,opacity = 0.5] (75.000000-\bw,0.636220) -- (75.000000+\bw,0.636220) -- (75.000000+\bw,0.662005) -- (75.000000-\bw,0.662005) -- cycle;
		\draw[black!50!red,opacity = 0.5] (75.000000-\bw,0.647570) -- (75.000000+\bw,0.647570);
		\node[color=black!50!red] at (75.000000,0.648362) {.};
		\draw[black!50!red,opacity = 0.5] (125.000000-\ww,0.494249) -- (125.000000+\ww,0.494249);\draw[black!50!red,opacity = 0.5] (125.000000-\ww,0.537271) -- (125.000000+\ww,0.537271);
\draw[black!50!red,opacity = 0.5] (125.000000,0.494249) -- (125.000000,0.507028);\draw[black!50!red,opacity = 0.5] (125.000000,0.527264) -- (125.000000,0.537271);		\draw[black!50!red,opacity = 0.5] (125.000000-\bw,0.507028) -- (125.000000+\bw,0.507028) -- (125.000000+\bw,0.527264) -- (125.000000-\bw,0.527264) -- cycle;
		\draw[black!50!red,opacity = 0.5] (125.000000-\bw,0.521102) -- (125.000000+\bw,0.521102);
		\node[color=black!50!red] at (125.000000,0.517301) {.};
		\draw[black!50!red,opacity=0.8] (25.000000,0.846573) -- (75.000000,0.648362) -- (125.000000,0.517301);

		\def\xmax{125+2*\xm} 
		\def\xmin{25-2*\xm}
		\def\ymax{0.133491-9*\ym}
		\def\legh{0.4/4.75} 
		\def\indicatorw{0.25/0.0715} 
		\input{\figs/legend_withChristian}
\end{tikzpicture}

%% file: figs/legend_withChristian.tex
\draw (\xmin,\ymax-\legh) rectangle (\xmax,\ymax);
  \foreach \num/\name/\c in {0.01/$g_k^{\mathrm{br3}}$/black!50!green,0.15/$g_k^{\mathrm{br1}}$/black,0.29/noise-free/black!50!red,0.5/noisy/red,0.64/deconvolution/blue}{
      \draw[\c, thick] (\num*\xmax-\num*\xmin+\xmin,\ymax-\legh/2) -- (\num*\xmax-\num*\xmin+\xmin+\indicatorw,\ymax-\legh/2) node [anchor=west,color=black] {\scriptsize \name};
  }
  \def\num{0.9} \def\name{human}
  \draw[black,dash pattern=on 2.3pt off 1.1pt on 0.8pt off 1.1pt,thick] (\num*\xmax-\num*\xmin+\xmin,\ymax-\legh/2) -- (\num*\xmax-\num*\xmin+\xmin+\indicatorw,\ymax-\legh/2) node [anchor=west,color=black] {\scriptsize \name};

%% file: figs/vs_NL_NC250+1250_new.tex
\begin{tikzpicture}[xscale=2.3485,yscale=5.0477]
	\def\ww{0.045455}\def\bw{0.045455}\def\xm{0.045455}\def\ym{\xm*2.3464/5.0477}

	\draw (1.000000-2*\xm,0.245352-2*\ym) -- (4.000000+2*\xm,0.245352-2*\ym) -- (4.000000+2*\xm,0.938734+2*\ym) -- (1.000000-2*\xm,0.938734+2*\ym) -- cycle;
\node at (4.000000,0.245352-5*\ym) [anchor=north west]{\footnotesize $\mathrm{NL}$};

	\foreach \x in {1,2,3,4}{ \draw (\x,0.245352-\ym) -- (\x,0.245352-2*\ym) node[anchor=north,rotate=0]{\footnotesize $\x$}; \draw[thin,dotted,gray] (\x,0.245352-\ym) -- (\x,0.938734+2*\ym); }
	\foreach \y in {0.25,0.37,0.50,0.65,0.80,0.95}{ \draw (1.000000-\xm,\y) -- (1.000000-2*\xm,\y) node[anchor=east]{\footnotesize $\y$};  \draw[thin,dashed,gray] (1.000000-\xm,\y) -- (4.000000+2*\xm,\y); }
\node at (2.500000,0.938734+4*\ym) {\small F1-Scores ($N_c: 250$, $\lambda: 0.50$, $\lambda_d: 0.00$)};

		\draw[black,opacity = 0.5] (1.000000-\ww,0.889566) -- (1.000000+\ww,0.889566);\draw[black,opacity = 0.5] (1.000000-\ww,0.926468) -- (1.000000+\ww,0.926468);
\draw[black,opacity = 0.5] (1.000000,0.889566) -- (1.000000,0.901879);\draw[black,opacity = 0.5] (1.000000,0.919255) -- (1.000000,0.926468);		\draw[black,opacity = 0.5] (1.000000-\bw,0.901879) -- (1.000000+\bw,0.901879) -- (1.000000+\bw,0.919255) -- (1.000000-\bw,0.919255) -- cycle;
		\draw[black,opacity = 0.5] (1.000000-\bw,0.910230) -- (1.000000+\bw,0.910230);
		\node[color=black] at (1.000000,0.909841) {.};
		\draw[black,opacity = 0.5] (2.000000-\ww,0.889800) -- (2.000000+\ww,0.889800);\draw[black,opacity = 0.5] (2.000000-\ww,0.928049) -- (2.000000+\ww,0.928049);
\draw[black,opacity = 0.5] (2.000000,0.889800) -- (2.000000,0.897541);\draw[black,opacity = 0.5] (2.000000,0.919255) -- (2.000000,0.928049);		\draw[black,opacity = 0.5] (2.000000-\bw,0.897541) -- (2.000000+\bw,0.897541) -- (2.000000+\bw,0.919255) -- (2.000000-\bw,0.919255) -- cycle;
		\draw[black,opacity = 0.5] (2.000000-\bw,0.911949) -- (2.000000+\bw,0.911949);
		\node[color=black] at (2.000000,0.909886) {.};
		\draw[black,opacity = 0.5] (3.000000-\ww,0.886597) -- (3.000000+\ww,0.886597);\draw[black,opacity = 0.5] (3.000000-\ww,0.920829) -- (3.000000+\ww,0.920829);
\draw[black,opacity = 0.5] (3.000000,0.886597) -- (3.000000,0.895075);\draw[black,opacity = 0.5] (3.000000,0.912863) -- (3.000000,0.920829);		\draw[black,opacity = 0.5] (3.000000-\bw,0.895075) -- (3.000000+\bw,0.895075) -- (3.000000+\bw,0.912863) -- (3.000000-\bw,0.912863) -- cycle;
		\draw[black,opacity = 0.5] (3.000000-\bw,0.903766) -- (3.000000+\bw,0.903766);
		\node[color=black] at (3.000000,0.903380) {.};
		\draw[black,opacity = 0.5] (4.000000-\ww,0.799544) -- (4.000000+\ww,0.799544);\draw[black,opacity = 0.5] (4.000000-\ww,0.865036) -- (4.000000+\ww,0.865036);
\draw[black,opacity = 0.5] (4.000000,0.799544) -- (4.000000,0.811966);\draw[black,opacity = 0.5] (4.000000,0.850526) -- (4.000000,0.865036);		\draw[black,opacity = 0.5] (4.000000-\bw,0.811966) -- (4.000000+\bw,0.811966) -- (4.000000+\bw,0.850526) -- (4.000000-\bw,0.850526) -- cycle;
		\draw[black,opacity = 0.5] (4.000000-\bw,0.830195) -- (4.000000+\bw,0.830195);
		\node[color=black] at (4.000000,0.832586) {.};
		\draw[black,opacity=0.8] (1.000000,0.909841) -- (2.000000,0.909886) -- (3.000000,0.903380) -- (4.000000,0.832586);

		\draw[black!50!green,opacity = 0.5] (1.000000-\ww,0.900218) -- (1.000000+\ww,0.900218);\draw[black!50!green,opacity = 0.5] (1.000000-\ww,0.937759) -- (1.000000+\ww,0.937759);
\draw[black!50!green,opacity = 0.5] (1.000000,0.900218) -- (1.000000,0.909091);\draw[black!50!green,opacity = 0.5] (1.000000,0.929752) -- (1.000000,0.937759);		\draw[black!50!green,opacity = 0.5] (1.000000-\bw,0.909091) -- (1.000000+\bw,0.909091) -- (1.000000+\bw,0.929752) -- (1.000000-\bw,0.929752) -- cycle;
		\draw[black!50!green,opacity = 0.5] (1.000000-\bw,0.917523) -- (1.000000+\bw,0.917523);
		\node[color=black!50!green] at (1.000000,0.918957) {.};
		\draw[black!50!green,opacity = 0.5] (2.000000-\ww,0.898462) -- (2.000000+\ww,0.898462);\draw[black!50!green,opacity = 0.5] (2.000000-\ww,0.938734) -- (2.000000+\ww,0.938734);
\draw[black!50!green,opacity = 0.5] (2.000000,0.898462) -- (2.000000,0.906504);\draw[black!50!green,opacity = 0.5] (2.000000,0.929752) -- (2.000000,0.938734);		\draw[black!50!green,opacity = 0.5] (2.000000-\bw,0.906504) -- (2.000000+\bw,0.906504) -- (2.000000+\bw,0.929752) -- (2.000000-\bw,0.929752) -- cycle;
		\draw[black!50!green,opacity = 0.5] (2.000000-\bw,0.914226) -- (2.000000+\bw,0.914226);
		\node[color=black!50!green] at (2.000000,0.917447) {.};
		\draw[black!50!green,opacity = 0.5] (3.000000-\ww,0.894705) -- (3.000000+\ww,0.894705);\draw[black!50!green,opacity = 0.5] (3.000000-\ww,0.927834) -- (3.000000+\ww,0.927834);
\draw[black!50!green,opacity = 0.5] (3.000000,0.894705) -- (3.000000,0.906054);\draw[black!50!green,opacity = 0.5] (3.000000,0.923077) -- (3.000000,0.927834);		\draw[black!50!green,opacity = 0.5] (3.000000-\bw,0.906054) -- (3.000000+\bw,0.906054) -- (3.000000+\bw,0.923077) -- (3.000000-\bw,0.923077) -- cycle;
		\draw[black!50!green,opacity = 0.5] (3.000000-\bw,0.912821) -- (3.000000+\bw,0.912821);
		\node[color=black!50!green] at (3.000000,0.913165) {.};
		\draw[black!50!green,opacity = 0.5] (4.000000-\ww,0.797889) -- (4.000000+\ww,0.797889);\draw[black!50!green,opacity = 0.5] (4.000000-\ww,0.865031) -- (4.000000+\ww,0.865031);
\draw[black!50!green,opacity = 0.5] (4.000000,0.797889) -- (4.000000,0.815618);\draw[black!50!green,opacity = 0.5] (4.000000,0.851695) -- (4.000000,0.865031);		\draw[black!50!green,opacity = 0.5] (4.000000-\bw,0.815618) -- (4.000000+\bw,0.815618) -- (4.000000+\bw,0.851695) -- (4.000000-\bw,0.851695) -- cycle;
		\draw[black!50!green,opacity = 0.5] (4.000000-\bw,0.834734) -- (4.000000+\bw,0.834734);
		\node[color=black!50!green] at (4.000000,0.833858) {.};
		\draw[black!50!green,opacity=0.8] (1.000000,0.918957) -- (2.000000,0.917447) -- (3.000000,0.913165) -- (4.000000,0.833858);

		\draw[blue,opacity = 0.5] (1.000000-\ww,0.785154) -- (1.000000+\ww,0.785154);\draw[blue,opacity = 0.5] (1.000000-\ww,0.843144) -- (1.000000+\ww,0.843144);
\draw[blue,opacity = 0.5] (1.000000,0.785154) -- (1.000000,0.794643);\draw[blue,opacity = 0.5] (1.000000,0.827740) -- (1.000000,0.843144);		\draw[blue,opacity = 0.5] (1.000000-\bw,0.794643) -- (1.000000+\bw,0.794643) -- (1.000000+\bw,0.827740) -- (1.000000-\bw,0.827740) -- cycle;
		\draw[blue,opacity = 0.5] (1.000000-\bw,0.811378) -- (1.000000+\bw,0.811378);
		\node[color=blue] at (1.000000,0.813642) {.};
		\draw[blue,opacity = 0.5] (2.000000-\ww,0.790854) -- (2.000000+\ww,0.790854);\draw[blue,opacity = 0.5] (2.000000-\ww,0.838001) -- (2.000000+\ww,0.838001);
\draw[blue,opacity = 0.5] (2.000000,0.790854) -- (2.000000,0.799076);\draw[blue,opacity = 0.5] (2.000000,0.827586) -- (2.000000,0.838001);		\draw[blue,opacity = 0.5] (2.000000-\bw,0.799076) -- (2.000000+\bw,0.799076) -- (2.000000+\bw,0.827586) -- (2.000000-\bw,0.827586) -- cycle;
		\draw[blue,opacity = 0.5] (2.000000-\bw,0.809951) -- (2.000000+\bw,0.809951);
		\node[color=blue] at (2.000000,0.812174) {.};
		\draw[blue,opacity = 0.5] (3.000000-\ww,0.750868) -- (3.000000+\ww,0.750868);\draw[blue,opacity = 0.5] (3.000000-\ww,0.819761) -- (3.000000+\ww,0.819761);
\draw[blue,opacity = 0.5] (3.000000,0.750868) -- (3.000000,0.766440);\draw[blue,opacity = 0.5] (3.000000,0.802752) -- (3.000000,0.819761);		\draw[blue,opacity = 0.5] (3.000000-\bw,0.766440) -- (3.000000+\bw,0.766440) -- (3.000000+\bw,0.802752) -- (3.000000-\bw,0.802752) -- cycle;
		\draw[blue,opacity = 0.5] (3.000000-\bw,0.782393) -- (3.000000+\bw,0.782393);
		\node[color=blue] at (3.000000,0.783990) {.};
		\draw[blue,opacity = 0.5] (4.000000-\ww,0.455083) -- (4.000000+\ww,0.455083);\draw[blue,opacity = 0.5] (4.000000-\ww,0.565459) -- (4.000000+\ww,0.565459);
\draw[blue,opacity = 0.5] (4.000000,0.455083) -- (4.000000,0.483721);\draw[blue,opacity = 0.5] (4.000000,0.539683) -- (4.000000,0.565459);		\draw[blue,opacity = 0.5] (4.000000-\bw,0.483721) -- (4.000000+\bw,0.483721) -- (4.000000+\bw,0.539683) -- (4.000000-\bw,0.539683) -- cycle;
		\draw[blue,opacity = 0.5] (4.000000-\bw,0.520732) -- (4.000000+\bw,0.520732);
		\node[color=blue] at (4.000000,0.514289) {.};
		\draw[blue,opacity=0.8] (1.000000,0.813642) -- (2.000000,0.812174) -- (3.000000,0.783990) -- (4.000000,0.514289);

		\draw[red,opacity = 0.5] (1.000000-\ww,0.805294) -- (1.000000+\ww,0.805294);\draw[red,opacity = 0.5] (1.000000-\ww,0.858064) -- (1.000000+\ww,0.858064);
\draw[red,opacity = 0.5] (1.000000,0.805294) -- (1.000000,0.818966);\draw[red,opacity = 0.5] (1.000000,0.847826) -- (1.000000,0.858064);		\draw[red,opacity = 0.5] (1.000000-\bw,0.818966) -- (1.000000+\bw,0.818966) -- (1.000000+\bw,0.847826) -- (1.000000-\bw,0.847826) -- cycle;
		\draw[red,opacity = 0.5] (1.000000-\bw,0.837810) -- (1.000000+\bw,0.837810);
		\node[color=red] at (1.000000,0.835437) {.};
		\draw[red,opacity = 0.5] (2.000000-\ww,0.777224) -- (2.000000+\ww,0.777224);\draw[red,opacity = 0.5] (2.000000-\ww,0.825293) -- (2.000000+\ww,0.825293);
\draw[red,opacity = 0.5] (2.000000,0.777224) -- (2.000000,0.787611);\draw[red,opacity = 0.5] (2.000000,0.809628) -- (2.000000,0.825293);		\draw[red,opacity = 0.5] (2.000000-\bw,0.787611) -- (2.000000+\bw,0.787611) -- (2.000000+\bw,0.809628) -- (2.000000-\bw,0.809628) -- cycle;
		\draw[red,opacity = 0.5] (2.000000-\bw,0.795556) -- (2.000000+\bw,0.795556);
		\node[color=red] at (2.000000,0.798493) {.};
		\draw[red,opacity = 0.5] (3.000000-\ww,0.606265) -- (3.000000+\ww,0.606265);\draw[red,opacity = 0.5] (3.000000-\ww,0.703765) -- (3.000000+\ww,0.703765);
\draw[red,opacity = 0.5] (3.000000,0.606265) -- (3.000000,0.631579);\draw[red,opacity = 0.5] (3.000000,0.684685) -- (3.000000,0.703765);		\draw[red,opacity = 0.5] (3.000000-\bw,0.631579) -- (3.000000+\bw,0.631579) -- (3.000000+\bw,0.684685) -- (3.000000-\bw,0.684685) -- cycle;
		\draw[red,opacity = 0.5] (3.000000-\bw,0.657626) -- (3.000000+\bw,0.657626);
		\node[color=red] at (3.000000,0.656423) {.};
		\draw[red,opacity = 0.5] (4.000000-\ww,0.245352) -- (4.000000+\ww,0.245352);\draw[red,opacity = 0.5] (4.000000-\ww,0.332999) -- (4.000000+\ww,0.332999);
\draw[red,opacity = 0.5] (4.000000,0.245352) -- (4.000000,0.253829);\draw[red,opacity = 0.5] (4.000000,0.319101) -- (4.000000,0.332999);		\draw[red,opacity = 0.5] (4.000000-\bw,0.253829) -- (4.000000+\bw,0.253829) -- (4.000000+\bw,0.319101) -- (4.000000-\bw,0.319101) -- cycle;
		\draw[red,opacity = 0.5] (4.000000-\bw,0.287399) -- (4.000000+\bw,0.287399);
		\node[color=red] at (4.000000,0.289245) {.};
		\draw[red,opacity=0.8] (1.000000,0.835437) -- (2.000000,0.798493) -- (3.000000,0.656423) -- (4.000000,0.289245);

		\draw[black!50!red,opacity = 0.5] (2.500000-\ww,0.817597) -- (2.500000+\ww,0.817597);\draw[black!50!red,opacity = 0.5] (2.500000-\ww,0.872611) -- (2.500000+\ww,0.872611);
\draw[black!50!red,opacity = 0.5] (2.500000,0.817597) -- (2.500000,0.828633);\draw[black!50!red,opacity = 0.5] (2.500000,0.859611) -- (2.500000,0.872611);		\draw[black!50!red,opacity = 0.5] (2.500000-\bw,0.828633) -- (2.500000+\bw,0.828633) -- (2.500000+\bw,0.859611) -- (2.500000-\bw,0.859611) -- cycle;
		\draw[black!50!red,opacity = 0.5] (2.500000-\bw,0.847060) -- (2.500000+\bw,0.847060);
		\node[color=black!50!red] at (2.500000,0.846573) {.};
		\draw[black!50!red,opacity=0.8] (2.500000,0.846573);

\end{tikzpicture}

\vspace{-10pt}

\begin{tikzpicture}[xscale=2.3485,yscale=5.3525]
	\def\ww{0.045455}\def\bw{0.045455}\def\xm{0.045455}\def\ym{\xm*2.3485/5.3525}

	\draw (1.000000-2*\xm,0.133491-2*\ym) -- (4.000000+2*\xm,0.133491-2*\ym) -- (4.000000+2*\xm,0.787394+2*\ym) -- (1.000000-2*\xm,0.787394+2*\ym) -- cycle;
\node at (4.000000,0.133491-4*\ym) [anchor=north west]{\footnotesize $\mathrm{NL}$};

	\foreach \x in {1,2,3,4}{ \draw (\x,0.133491-\ym) -- (\x,0.133491-2*\ym) node[anchor=north,rotate=0]{\footnotesize $\x$}; \draw[thin,dotted,gray] (\x,0.133491-\ym) -- (\x,0.787394+2*\ym); }
	\foreach \y in {0.20,0.30,0.40,0.50,0.60,0.70,0.80}{ \draw (1.000000-\xm,\y) -- (1.000000-2*\xm,\y) node[anchor=east]{\footnotesize $\y$};  \draw[thin,dashed,gray] (1.000000-\xm,\y) -- (4.000000+2*\xm,\y); }
\node at (2.500000,0.787394+4*\ym) {\small F1-Scores ($N_c: 1250$, $\lambda: 0.50$, $\lambda_d: 0.00$)};

		\draw[black,opacity = 0.5] (1.000000-\ww,0.724126) -- (1.000000+\ww,0.724126);\draw[black,opacity = 0.5] (1.000000-\ww,0.761821) -- (1.000000+\ww,0.761821);
\draw[black,opacity = 0.5] (1.000000,0.724126) -- (1.000000,0.738514);\draw[black,opacity = 0.5] (1.000000,0.754814) -- (1.000000,0.761821);		\draw[black,opacity = 0.5] (1.000000-\bw,0.738514) -- (1.000000+\bw,0.738514) -- (1.000000+\bw,0.754814) -- (1.000000-\bw,0.754814) -- cycle;
		\draw[black,opacity = 0.5] (1.000000-\bw,0.744678) -- (1.000000+\bw,0.744678);
		\node[color=black] at (1.000000,0.744974) {.};
		\draw[black,opacity = 0.5] (2.000000-\ww,0.724628) -- (2.000000+\ww,0.724628);\draw[black,opacity = 0.5] (2.000000-\ww,0.762117) -- (2.000000+\ww,0.762117);
\draw[black,opacity = 0.5] (2.000000,0.724628) -- (2.000000,0.734835);\draw[black,opacity = 0.5] (2.000000,0.754256) -- (2.000000,0.762117);		\draw[black,opacity = 0.5] (2.000000-\bw,0.734835) -- (2.000000+\bw,0.734835) -- (2.000000+\bw,0.754256) -- (2.000000-\bw,0.754256) -- cycle;
		\draw[black,opacity = 0.5] (2.000000-\bw,0.744336) -- (2.000000+\bw,0.744336);
		\node[color=black] at (2.000000,0.744283) {.};
		\draw[black,opacity = 0.5] (3.000000-\ww,0.720190) -- (3.000000+\ww,0.720190);\draw[black,opacity = 0.5] (3.000000-\ww,0.758380) -- (3.000000+\ww,0.758380);
\draw[black,opacity = 0.5] (3.000000,0.720190) -- (3.000000,0.732648);\draw[black,opacity = 0.5] (3.000000,0.752475) -- (3.000000,0.758380);		\draw[black,opacity = 0.5] (3.000000-\bw,0.732648) -- (3.000000+\bw,0.732648) -- (3.000000+\bw,0.752475) -- (3.000000-\bw,0.752475) -- cycle;
		\draw[black,opacity = 0.5] (3.000000-\bw,0.738859) -- (3.000000+\bw,0.738859);
		\node[color=black] at (3.000000,0.739516) {.};
		\draw[black,opacity = 0.5] (4.000000-\ww,0.613689) -- (4.000000+\ww,0.613689);\draw[black,opacity = 0.5] (4.000000-\ww,0.681618) -- (4.000000+\ww,0.681618);
\draw[black,opacity = 0.5] (4.000000,0.613689) -- (4.000000,0.638095);\draw[black,opacity = 0.5] (4.000000,0.670139) -- (4.000000,0.681618);		\draw[black,opacity = 0.5] (4.000000-\bw,0.638095) -- (4.000000+\bw,0.638095) -- (4.000000+\bw,0.670139) -- (4.000000-\bw,0.670139) -- cycle;
		\draw[black,opacity = 0.5] (4.000000-\bw,0.655338) -- (4.000000+\bw,0.655338);
		\node[color=black] at (4.000000,0.651121) {.};
		\draw[black,opacity=0.8] (1.000000,0.744974) -- (2.000000,0.744283) -- (3.000000,0.739516) -- (4.000000,0.651121);

		\draw[black!50!green,opacity = 0.5] (1.000000-\ww,0.751173) -- (1.000000+\ww,0.751173);\draw[black!50!green,opacity = 0.5] (1.000000-\ww,0.782121) -- (1.000000+\ww,0.782121);
\draw[black!50!green,opacity = 0.5] (1.000000,0.751173) -- (1.000000,0.760897);\draw[black!50!green,opacity = 0.5] (1.000000,0.778016) -- (1.000000,0.782121);		\draw[black!50!green,opacity = 0.5] (1.000000-\bw,0.760897) -- (1.000000+\bw,0.760897) -- (1.000000+\bw,0.778016) -- (1.000000-\bw,0.778016) -- cycle;
		\draw[black!50!green,opacity = 0.5] (1.000000-\bw,0.768862) -- (1.000000+\bw,0.768862);
		\node[color=black!50!green] at (1.000000,0.768550) {.};
		\draw[black!50!green,opacity = 0.5] (2.000000-\ww,0.747801) -- (2.000000+\ww,0.747801);\draw[black!50!green,opacity = 0.5] (2.000000-\ww,0.787394) -- (2.000000+\ww,0.787394);
\draw[black!50!green,opacity = 0.5] (2.000000,0.747801) -- (2.000000,0.759526);\draw[black!50!green,opacity = 0.5] (2.000000,0.783180) -- (2.000000,0.787394);		\draw[black!50!green,opacity = 0.5] (2.000000-\bw,0.759526) -- (2.000000+\bw,0.759526) -- (2.000000+\bw,0.783180) -- (2.000000-\bw,0.783180) -- cycle;
		\draw[black!50!green,opacity = 0.5] (2.000000-\bw,0.771687) -- (2.000000+\bw,0.771687);
		\node[color=black!50!green] at (2.000000,0.769317) {.};
		\draw[black!50!green,opacity = 0.5] (3.000000-\ww,0.746938) -- (3.000000+\ww,0.746938);\draw[black!50!green,opacity = 0.5] (3.000000-\ww,0.778296) -- (3.000000+\ww,0.778296);
\draw[black!50!green,opacity = 0.5] (3.000000,0.746938) -- (3.000000,0.752467);\draw[black!50!green,opacity = 0.5] (3.000000,0.771990) -- (3.000000,0.778296);		\draw[black!50!green,opacity = 0.5] (3.000000-\bw,0.752467) -- (3.000000+\bw,0.752467) -- (3.000000+\bw,0.771990) -- (3.000000-\bw,0.771990) -- cycle;
		\draw[black!50!green,opacity = 0.5] (3.000000-\bw,0.761166) -- (3.000000+\bw,0.761166);
		\node[color=black!50!green] at (3.000000,0.761891) {.};
		\draw[black!50!green,opacity = 0.5] (4.000000-\ww,0.626498) -- (4.000000+\ww,0.626498);\draw[black!50!green,opacity = 0.5] (4.000000-\ww,0.684949) -- (4.000000+\ww,0.684949);
\draw[black!50!green,opacity = 0.5] (4.000000,0.626498) -- (4.000000,0.645648);\draw[black!50!green,opacity = 0.5] (4.000000,0.673904) -- (4.000000,0.684949);		\draw[black!50!green,opacity = 0.5] (4.000000-\bw,0.645648) -- (4.000000+\bw,0.645648) -- (4.000000+\bw,0.673904) -- (4.000000-\bw,0.673904) -- cycle;
		\draw[black!50!green,opacity = 0.5] (4.000000-\bw,0.660869) -- (4.000000+\bw,0.660869);
		\node[color=black!50!green] at (4.000000,0.658364) {.};
		\draw[black!50!green,opacity=0.8] (1.000000,0.768550) -- (2.000000,0.769317) -- (3.000000,0.761891) -- (4.000000,0.658364);

		\draw[blue,opacity = 0.5] (1.000000-\ww,0.679182) -- (1.000000+\ww,0.679182);\draw[blue,opacity = 0.5] (1.000000-\ww,0.713736) -- (1.000000+\ww,0.713736);
\draw[blue,opacity = 0.5] (1.000000,0.679182) -- (1.000000,0.682974);\draw[blue,opacity = 0.5] (1.000000,0.702925) -- (1.000000,0.713736);		\draw[blue,opacity = 0.5] (1.000000-\bw,0.682974) -- (1.000000+\bw,0.682974) -- (1.000000+\bw,0.702925) -- (1.000000-\bw,0.702925) -- cycle;
		\draw[blue,opacity = 0.5] (1.000000-\bw,0.690907) -- (1.000000+\bw,0.690907);
		\node[color=blue] at (1.000000,0.693413) {.};
		\draw[blue,opacity = 0.5] (2.000000-\ww,0.673466) -- (2.000000+\ww,0.673466);\draw[blue,opacity = 0.5] (2.000000-\ww,0.708288) -- (2.000000+\ww,0.708288);
\draw[blue,opacity = 0.5] (2.000000,0.673466) -- (2.000000,0.681299);\draw[blue,opacity = 0.5] (2.000000,0.699761) -- (2.000000,0.708288);		\draw[blue,opacity = 0.5] (2.000000-\bw,0.681299) -- (2.000000+\bw,0.681299) -- (2.000000+\bw,0.699761) -- (2.000000-\bw,0.699761) -- cycle;
		\draw[blue,opacity = 0.5] (2.000000-\bw,0.687007) -- (2.000000+\bw,0.687007);
		\node[color=blue] at (2.000000,0.689348) {.};
		\draw[blue,opacity = 0.5] (3.000000-\ww,0.602345) -- (3.000000+\ww,0.602345);\draw[blue,opacity = 0.5] (3.000000-\ww,0.664335) -- (3.000000+\ww,0.664335);
\draw[blue,opacity = 0.5] (3.000000,0.602345) -- (3.000000,0.615093);\draw[blue,opacity = 0.5] (3.000000,0.642402) -- (3.000000,0.664335);		\draw[blue,opacity = 0.5] (3.000000-\bw,0.615093) -- (3.000000+\bw,0.615093) -- (3.000000+\bw,0.642402) -- (3.000000-\bw,0.642402) -- cycle;
		\draw[blue,opacity = 0.5] (3.000000-\bw,0.625698) -- (3.000000+\bw,0.625698);
		\node[color=blue] at (3.000000,0.629007) {.};
		\draw[blue,opacity = 0.5] (4.000000-\ww,0.241732) -- (4.000000+\ww,0.241732);\draw[blue,opacity = 0.5] (4.000000-\ww,0.335533) -- (4.000000+\ww,0.335533);
\draw[blue,opacity = 0.5] (4.000000,0.241732) -- (4.000000,0.267048);\draw[blue,opacity = 0.5] (4.000000,0.316231) -- (4.000000,0.335533);		\draw[blue,opacity = 0.5] (4.000000-\bw,0.267048) -- (4.000000+\bw,0.267048) -- (4.000000+\bw,0.316231) -- (4.000000-\bw,0.316231) -- cycle;
		\draw[blue,opacity = 0.5] (4.000000-\bw,0.293066) -- (4.000000+\bw,0.293066);
		\node[color=blue] at (4.000000,0.291222) {.};
		\draw[blue,opacity=0.8] (1.000000,0.693413) -- (2.000000,0.689348) -- (3.000000,0.629007) -- (4.000000,0.291222);

		\draw[red,opacity = 0.5] (1.000000-\ww,0.492370) -- (1.000000+\ww,0.492370);\draw[red,opacity = 0.5] (1.000000-\ww,0.535074) -- (1.000000+\ww,0.535074);
\draw[red,opacity = 0.5] (1.000000,0.492370) -- (1.000000,0.502033);\draw[red,opacity = 0.5] (1.000000,0.525518) -- (1.000000,0.535074);		\draw[red,opacity = 0.5] (1.000000-\bw,0.502033) -- (1.000000+\bw,0.502033) -- (1.000000+\bw,0.525518) -- (1.000000-\bw,0.525518) -- cycle;
		\draw[red,opacity = 0.5] (1.000000-\bw,0.517093) -- (1.000000+\bw,0.517093);
		\node[color=red] at (1.000000,0.514860) {.};
		\draw[red,opacity = 0.5] (2.000000-\ww,0.467703) -- (2.000000+\ww,0.467703);\draw[red,opacity = 0.5] (2.000000-\ww,0.512793) -- (2.000000+\ww,0.512793);
\draw[red,opacity = 0.5] (2.000000,0.467703) -- (2.000000,0.481387);\draw[red,opacity = 0.5] (2.000000,0.504244) -- (2.000000,0.512793);		\draw[red,opacity = 0.5] (2.000000-\bw,0.481387) -- (2.000000+\bw,0.481387) -- (2.000000+\bw,0.504244) -- (2.000000-\bw,0.504244) -- cycle;
		\draw[red,opacity = 0.5] (2.000000-\bw,0.494399) -- (2.000000+\bw,0.494399);
		\node[color=red] at (2.000000,0.491801) {.};
		\draw[red,opacity = 0.5] (3.000000-\ww,0.346073) -- (3.000000+\ww,0.346073);\draw[red,opacity = 0.5] (3.000000-\ww,0.404413) -- (3.000000+\ww,0.404413);
\draw[red,opacity = 0.5] (3.000000,0.346073) -- (3.000000,0.363010);\draw[red,opacity = 0.5] (3.000000,0.391885) -- (3.000000,0.404413);		\draw[red,opacity = 0.5] (3.000000-\bw,0.363010) -- (3.000000+\bw,0.363010) -- (3.000000+\bw,0.391885) -- (3.000000-\bw,0.391885) -- cycle;
		\draw[red,opacity = 0.5] (3.000000-\bw,0.374256) -- (3.000000+\bw,0.374256);
		\node[color=red] at (3.000000,0.374687) {.};
		\draw[red,opacity = 0.5] (4.000000-\ww,0.133491) -- (4.000000+\ww,0.133491);\draw[red,opacity = 0.5] (4.000000-\ww,0.169597) -- (4.000000+\ww,0.169597);
\draw[red,opacity = 0.5] (4.000000,0.133491) -- (4.000000,0.142222);\draw[red,opacity = 0.5] (4.000000,0.160610) -- (4.000000,0.169597);		\draw[red,opacity = 0.5] (4.000000-\bw,0.142222) -- (4.000000+\bw,0.142222) -- (4.000000+\bw,0.160610) -- (4.000000-\bw,0.160610) -- cycle;
		\draw[red,opacity = 0.5] (4.000000-\bw,0.150541) -- (4.000000+\bw,0.150541);
		\node[color=red] at (4.000000,0.149869) {.};
		\draw[red,opacity=0.8] (1.000000,0.514860) -- (2.000000,0.491801) -- (3.000000,0.374687) -- (4.000000,0.149869);

		\draw[black!50!red,opacity = 0.5] (2.500000-\ww,0.494249) -- (2.500000+\ww,0.494249);\draw[black!50!red,opacity = 0.5] (2.500000-\ww,0.537271) -- (2.500000+\ww,0.537271);
\draw[black!50!red,opacity = 0.5] (2.500000,0.494249) -- (2.500000,0.507028);\draw[black!50!red,opacity = 0.5] (2.500000,0.527264) -- (2.500000,0.537271);		\draw[black!50!red,opacity = 0.5] (2.500000-\bw,0.507028) -- (2.500000+\bw,0.507028) -- (2.500000+\bw,0.527264) -- (2.500000-\bw,0.527264) -- cycle;
		\draw[black!50!red,opacity = 0.5] (2.500000-\bw,0.521102) -- (2.500000+\bw,0.521102);
		\node[color=black!50!red] at (2.500000,0.517301) {.};
		\draw[black!50!red,opacity=0.8] (2.500000,0.517301);

		\def\xmax{4+2*\xm} 
		\def\xmin{1-2*\xm}
		\def\ymax{0.133491-8*\ym}
		\def\legh{0.4/5.3525} 
		\def\indicatorw{0.3/2.3485} 
		\input{\figs/legend}                   
\end{tikzpicture}

%% file: figs/EMDsvsNoise_integral.tex
\begin{tikzpicture}[xscale=2.435,yscale=1.35]
	\def\ww{0.045455}\def\bw{0.045455}\def\ym{0.083752}\def\xm{\ym*1.35/2.435}

	\draw (1.000000-2*\xm,2.349857-2*\ym) -- (4.000000+2*\xm,2.349857-2*\ym) -- (4.000000+2*\xm,5.281168+2*\ym) -- (1.000000-2*\xm,5.281168+2*\ym) node [anchor=south] {\footnotesize$\mathrm{pix}$} -- cycle;
\node at (4.000000,2.349857-5*\ym) [anchor=north west]{\footnotesize $\mathrm{NL}$};

	\foreach \x in {1,2,3,4}{ \draw (\x,2.349857-\ym) -- (\x,2.349857-2*\ym) node[anchor=north,rotate=0]{\footnotesize $\x$}; \draw[thin,dotted,gray] (\x,2.349857-\ym) -- (\x,5.281168+2*\ym); }
	\foreach \y in {2.5,3.0,3.5,4.0,4.5,5.0}{ \draw (1.000000-\xm,\y) -- (1.000000-2*\xm,\y) node[anchor=east]{\footnotesize $\y$};  \draw[thin,dashed,gray] (1.000000-\xm,\y) -- (4.000000+2*\xm,\y); }
\node at (2.500000,5.281168+4*\ym) {\small EMDs $\left(N_c: 250\mbox{, }\lambda: 0.5\mbox{, Kernel: }g_k^{\mathrm{b1r}}\right)$};

		\draw[black,opacity = 0.5] (1.000000-\ww,2.351953) -- (1.000000+\ww,2.351953);\draw[black,opacity = 0.5] (1.000000-\ww,3.133992) -- (1.000000+\ww,3.133992);
\draw[black,opacity = 0.5] (1.000000,2.351953) -- (1.000000,2.505788);\draw[black,opacity = 0.5] (1.000000,2.813945) -- (1.000000,3.133992);		\draw[black,opacity = 0.5] (1.000000-\bw,2.505788) -- (1.000000+\bw,2.505788) -- (1.000000+\bw,2.813945) -- (1.000000-\bw,2.813945) -- cycle;
		\draw[black,opacity = 0.5] (1.000000-\bw,2.715475) -- (1.000000+\bw,2.715475);
		\node[color=black] at (1.000000,2.733397) {.};
		\draw[black,opacity = 0.5] (2.000000-\ww,2.349857) -- (2.000000+\ww,2.349857);\draw[black,opacity = 0.5] (2.000000-\ww,3.126848) -- (2.000000+\ww,3.126848);
\draw[black,opacity = 0.5] (2.000000,2.349857) -- (2.000000,2.504474);\draw[black,opacity = 0.5] (2.000000,2.821208) -- (2.000000,3.126848);		\draw[black,opacity = 0.5] (2.000000-\bw,2.504474) -- (2.000000+\bw,2.504474) -- (2.000000+\bw,2.821208) -- (2.000000-\bw,2.821208) -- cycle;
		\draw[black,opacity = 0.5] (2.000000-\bw,2.712368) -- (2.000000+\bw,2.712368);
		\node[color=black] at (2.000000,2.736082) {.};
		\draw[black,opacity = 0.5] (3.000000-\ww,2.391548) -- (3.000000+\ww,2.391548);\draw[black,opacity = 0.5] (3.000000-\ww,3.094637) -- (3.000000+\ww,3.094637);
\draw[black,opacity = 0.5] (3.000000,2.391548) -- (3.000000,2.570150);\draw[black,opacity = 0.5] (3.000000,2.866055) -- (3.000000,3.094637);		\draw[black,opacity = 0.5] (3.000000-\bw,2.570150) -- (3.000000+\bw,2.570150) -- (3.000000+\bw,2.866055) -- (3.000000-\bw,2.866055) -- cycle;
		\draw[black,opacity = 0.5] (3.000000-\bw,2.722334) -- (3.000000+\bw,2.722334);
		\node[color=black] at (3.000000,2.764127) {.};
		\draw[black,opacity = 0.5] (4.000000-\ww,3.736934) -- (4.000000+\ww,3.736934);\draw[black,opacity = 0.5] (4.000000-\ww,5.281168) -- (4.000000+\ww,5.281168);
\draw[black,opacity = 0.5] (4.000000,3.736934) -- (4.000000,4.049963);\draw[black,opacity = 0.5] (4.000000,4.658952) -- (4.000000,5.281168);		\draw[black,opacity = 0.5] (4.000000-\bw,4.049963) -- (4.000000+\bw,4.049963) -- (4.000000+\bw,4.658952) -- (4.000000-\bw,4.658952) -- cycle;
		\draw[black,opacity = 0.5] (4.000000-\bw,4.257475) -- (4.000000+\bw,4.257475);
		\node[color=black] at (4.000000,4.414251) {.};
		\draw[black,opacity=0.8] (1.000000,2.733397) -- (2.000000,2.736082) -- (3.000000,2.764127) -- (4.000000,4.414251);

\end{tikzpicture}

%% file: figs/recovery_asigma_1.tex
\begin{tikzpicture}[xscale=\w,yscale=\h] 
\def\xm{1.902941} \def\ym{0.566074}

	\draw[->] (0.000000-2*\xm,0.000000-\ym) -- (64.700000+2*\xm,0.000000-\ym) node[anchor=west]{\footnotesize $\dsigma$};
	\draw[->] (0.000000-\xm,0.000000-2*\ym) -- (0.000000-\xm,11.321477+2*\ym);
	\node at (30,12) [anchor=south] { \small $a_{\pos_1}(\dsigma)$~~vs~~$\hat{a}_{\mathrm{opt},\pos_1}(\dsigma-\dsigma_\mathrm{b})$ };
	\foreach \x in {0,15,30,45,60}{ \draw (\x,0.000000-\ym) -- (\x,0.000000-2*\ym) node[anchor=north]{\footnotesize $\x$};  \draw[very thin,dotted,gray] (\x,0.000000-\ym) -- (\x,11.321477+2*\ym); }
	\foreach \y in {0,2,4,6,8,10,12}{ \draw (0.000000-\xm,\y) -- (0.000000-2*\xm,\y) node[anchor=east]{\footnotesize $\y$};  \draw[very thin,dotted,gray] (0.000000-\xm,\y) -- (64.700000+2*\xm,\y); }

		\draw[blue,fill=blue,opacity=0.6] (0.000000,0) -- (2.038037,0) -- (2.038037,3.744852) -- (0.000000,3.744852) -- cycle;
		\draw[blue,fill=blue,opacity=0.6] (2.038037,0) -- (4.076075,0) -- (4.076075,4.340902) -- (2.038037,4.340902) -- cycle;
		\draw[blue,fill=blue,opacity=0.6] (4.076075,0) -- (6.444840,0) -- (6.444840,4.533378) -- (4.076075,4.533378) -- cycle;
		\draw[blue,fill=blue,opacity=0.6] (6.444840,0) -- (8.646660,0) -- (8.646660,4.388751) -- (6.444840,4.388751) -- cycle;
		\draw[blue,fill=blue,opacity=0.6] (8.646660,0) -- (10.784280,0) -- (10.784280,4.065985) -- (8.646660,4.065985) -- cycle;
		\draw[blue,fill=blue,opacity=0.6] (10.784280,0) -- (12.889680,0) -- (12.889680,3.679127) -- (10.784280,3.679127) -- cycle;
		\draw[blue,fill=blue,opacity=0.6] (12.889680,0) -- (14.976455,0) -- (14.976455,3.284190) -- (12.889680,3.284190) -- cycle;
		\draw[blue,fill=blue,opacity=0.6] (14.976455,0) -- (17.172809,0) -- (17.172809,2.898160) -- (14.976455,2.898160) -- cycle;
		\draw[blue,fill=blue,opacity=0.6] (17.172809,0) -- (19.334521,0) -- (19.334521,2.537851) -- (17.172809,2.537851) -- cycle;
		\draw[blue,fill=blue,opacity=0.6] (19.334521,0) -- (21.472057,0) -- (21.472057,2.219692) -- (19.334521,2.219692) -- cycle;
		\draw[blue,fill=blue,opacity=0.6] (21.472057,0) -- (23.591989,0) -- (23.591989,1.941079) -- (21.472057,1.941079) -- cycle;
		\draw[blue,fill=blue,opacity=0.6] (23.591989,0) -- (25.779361,0) -- (25.779361,1.693740) -- (23.591989,1.693740) -- cycle;
		\draw[blue,fill=blue,opacity=0.6] (25.779361,0) -- (27.944161,0) -- (27.944161,1.475332) -- (25.779361,1.475332) -- cycle;
		\draw[blue,fill=blue,opacity=0.6] (27.944161,0) -- (30.091262,0) -- (30.091262,1.286541) -- (27.944161,1.286541) -- cycle;
		\draw[blue,fill=blue,opacity=0.6] (30.091262,0) -- (32.224201,0) -- (32.224201,1.122765) -- (30.091262,1.122765) -- cycle;
		\draw[blue,fill=blue,opacity=0.6] (32.224201,0) -- (34.345617,0) -- (34.345617,0.980137) -- (32.224201,0.980137) -- cycle;
		\draw[blue,fill=blue,opacity=0.6] (34.345617,0) -- (36.514442,0) -- (36.514442,0.853891) -- (34.345617,0.853891) -- cycle;
		\draw[blue,fill=blue,opacity=0.6] (36.514442,0) -- (38.669041,0) -- (38.669041,0.741973) -- (36.514442,0.741973) -- cycle;
		\draw[blue,fill=blue,opacity=0.6] (38.669041,0) -- (40.811668,0) -- (40.811668,0.643875) -- (38.669041,0.643875) -- cycle;
		\draw[blue,fill=blue,opacity=0.6] (40.811668,0) -- (42.944113,0) -- (42.944113,0.557489) -- (40.811668,0.557489) -- cycle;
		\draw[blue,fill=blue,opacity=0.6] (42.944113,0) -- (45.113881,0) -- (45.113881,0.480317) -- (42.944113,0.480317) -- cycle;
		\draw[blue,fill=blue,opacity=0.6] (45.113881,0) -- (47.271925,0) -- (47.271925,0.411212) -- (45.113881,0.411212) -- cycle;
		\draw[blue,fill=blue,opacity=0.6] (47.271925,0) -- (49.419781,0) -- (49.419781,0.349840) -- (47.271925,0.349840) -- cycle;
		\draw[blue,fill=blue,opacity=0.6] (49.419781,0) -- (51.558722,0) -- (51.558722,0.295109) -- (49.419781,0.295109) -- cycle;
		\draw[blue,fill=blue,opacity=0.6] (51.558722,0) -- (53.689813,0) -- (53.689813,0.245633) -- (51.558722,0.245633) -- cycle;
		\draw[blue,fill=blue,opacity=0.6] (53.689813,0) -- (55.851150,0) -- (55.851150,0.171289) -- (53.689813,0.171289) -- cycle;
		\draw[blue,fill=blue,opacity=0.6] (55.851150,0) -- (58.003562,0) -- (58.003562,0.092948) -- (55.851150,0.092948) -- cycle;
		\draw[blue,fill=blue,opacity=0.6] (58.003562,0) -- (60.148006,0) -- (60.148006,0.030569) -- (58.003562,0.030569) -- cycle;
		\draw[blue,fill=blue,opacity=0.6] (60.148006,0) -- (62.285305,0) -- (62.285305,0.000184) -- (60.148006,0.000184) -- cycle;
		\draw[blue,fill=blue,opacity=0.6] (62.285305,0) -- (64.448402,0) -- (64.448402,0.000000) -- (62.285305,0.000000) -- cycle;
		\draw[red,fill=red,opacity=0.6] (0.000000,0) -- (2.700000,0) -- (2.700000,11.321477) -- (0.000000,11.321477) -- cycle;
		\draw[red,fill=red,opacity=0.6] (2.700000,0) -- (6.700000,0) -- (6.700000,1.577836) -- (2.700000,1.577836) -- cycle;
		\draw[red,fill=red,opacity=0.6] (6.700000,0) -- (10.700000,0) -- (10.700000,3.728029) -- (6.700000,3.728029) -- cycle;
		\draw[red,fill=red,opacity=0.6] (10.700000,0) -- (20.700000,0) -- (20.700000,2.446013) -- (10.700000,2.446013) -- cycle;
		\draw[red,fill=red,opacity=0.6] (20.700000,0) -- (30.700000,0) -- (30.700000,1.230137) -- (20.700000,1.230137) -- cycle;
		\draw[red,fill=red,opacity=0.6] (30.700000,0) -- (40.700000,0) -- (40.700000,0.535791) -- (30.700000,0.535791) -- cycle;
		\draw[red,fill=red,opacity=0.6] (40.700000,0) -- (50.700000,0) -- (50.700000,0.261668) -- (40.700000,0.261668) -- cycle;
		\draw[red,fill=red,opacity=0.6] (50.700000,0) -- (64.700000,0) -- (64.700000,0.248033) -- (50.700000,0.248033) -- cycle;
\end{tikzpicture}

%% file: figs/recovery_asigma_2.tex
\begin{tikzpicture}[xscale=\w,yscale=\h]  
\def\xm{1.902941} \def\ym{0.752349} 

	\draw[->] (0.000000-2*\xm,0.000000-\ym) -- (64.700000+2*\xm,0.000000-\ym) node[anchor=west]{\footnotesize $\dsigma$};
	\draw[->] (0.000000-\xm,0.000000-2*\ym) -- (0.000000-\xm,15.046971+2*\ym);
	\node at (30,16) [anchor=south] { \small $a_{\pos_2}(\dsigma)$~~vs~~$a_{\mathrm{opt},\pos_2}(\dsigma-\dsigma_\mathrm{b})$ };
	\foreach \x in {0,15,30,45,60}{ \draw (\x,0.000000-\ym) -- (\x,0.000000-2*\ym) node[anchor=north]{\footnotesize $\x$};  \draw[very thin,dotted,gray] (\x,0.000000-\ym) -- (\x,15.046971+2*\ym); }
	\foreach \y in {0,2,4,6,8,10,12,14,16}{ \draw (0.000000-\xm,\y) -- (0.000000-2*\xm,\y) node[anchor=east]{\footnotesize $\y$};  \draw[very thin,dotted,gray] (0.000000-\xm,\y) -- (64.700000+2*\xm,\y); }

		\draw[blue,fill=blue,opacity=0.6] (0.000000,0) -- (2.038037,0) -- (2.038037,5.733948) -- (0.000000,5.733948) -- cycle;
		\draw[blue,fill=blue,opacity=0.6] (2.038037,0) -- (4.076075,0) -- (4.076075,5.654080) -- (2.038037,5.654080) -- cycle;
		\draw[blue,fill=blue,opacity=0.6] (4.076075,0) -- (6.444840,0) -- (6.444840,5.276507) -- (4.076075,5.276507) -- cycle;
		\draw[blue,fill=blue,opacity=0.6] (6.444840,0) -- (8.646660,0) -- (8.646660,4.710208) -- (6.444840,4.710208) -- cycle;
		\draw[blue,fill=blue,opacity=0.6] (8.646660,0) -- (10.784280,0) -- (10.784280,4.125264) -- (8.646660,4.125264) -- cycle;
		\draw[blue,fill=blue,opacity=0.6] (10.784280,0) -- (12.889680,0) -- (12.889680,3.577870) -- (10.784280,3.577870) -- cycle;
		\draw[blue,fill=blue,opacity=0.6] (12.889680,0) -- (14.976455,0) -- (14.976455,3.088580) -- (12.889680,3.088580) -- cycle;
		\draw[blue,fill=blue,opacity=0.6] (14.976455,0) -- (17.172809,0) -- (17.172809,2.650028) -- (14.976455,2.650028) -- cycle;
		\draw[blue,fill=blue,opacity=0.6] (17.172809,0) -- (19.334521,0) -- (19.334521,2.265177) -- (17.172809,2.265177) -- cycle;
		\draw[blue,fill=blue,opacity=0.6] (19.334521,0) -- (21.472057,0) -- (21.472057,1.940716) -- (19.334521,1.940716) -- cycle;
		\draw[blue,fill=blue,opacity=0.6] (21.472057,0) -- (23.591989,0) -- (23.591989,1.666416) -- (21.472057,1.666416) -- cycle;
		\draw[blue,fill=blue,opacity=0.6] (23.591989,0) -- (25.779361,0) -- (25.779361,1.429623) -- (23.591989,1.429623) -- cycle;
		\draw[blue,fill=blue,opacity=0.6] (25.779361,0) -- (27.944161,0) -- (27.944161,1.225218) -- (25.779361,1.225218) -- cycle;
		\draw[blue,fill=blue,opacity=0.6] (27.944161,0) -- (30.091262,0) -- (30.091262,1.051811) -- (27.944161,1.051811) -- cycle;
		\draw[blue,fill=blue,opacity=0.6] (30.091262,0) -- (32.224201,0) -- (32.224201,0.903686) -- (30.091262,0.903686) -- cycle;
		\draw[blue,fill=blue,opacity=0.6] (32.224201,0) -- (34.345617,0) -- (34.345617,0.776344) -- (32.224201,0.776344) -- cycle;
		\draw[blue,fill=blue,opacity=0.6] (34.345617,0) -- (36.514442,0) -- (36.514442,0.663713) -- (34.345617,0.663713) -- cycle;
		\draw[blue,fill=blue,opacity=0.6] (36.514442,0) -- (38.669041,0) -- (38.669041,0.548799) -- (36.514442,0.548799) -- cycle;
		\draw[blue,fill=blue,opacity=0.6] (38.669041,0) -- (40.811668,0) -- (40.811668,0.446116) -- (38.669041,0.446116) -- cycle;
		\draw[blue,fill=blue,opacity=0.6] (40.811668,0) -- (42.944113,0) -- (42.944113,0.359142) -- (40.811668,0.359142) -- cycle;
		\draw[blue,fill=blue,opacity=0.6] (42.944113,0) -- (45.113881,0) -- (45.113881,0.284688) -- (42.944113,0.284688) -- cycle;
		\draw[blue,fill=blue,opacity=0.6] (45.113881,0) -- (47.271925,0) -- (47.271925,0.221117) -- (45.113881,0.221117) -- cycle;
		\draw[blue,fill=blue,opacity=0.6] (47.271925,0) -- (49.419781,0) -- (49.419781,0.167630) -- (47.271925,0.167630) -- cycle;
		\draw[blue,fill=blue,opacity=0.6] (49.419781,0) -- (51.558722,0) -- (51.558722,0.122793) -- (49.419781,0.122793) -- cycle;
		\draw[blue,fill=blue,opacity=0.6] (51.558722,0) -- (53.689813,0) -- (53.689813,0.085433) -- (51.558722,0.085433) -- cycle;
		\draw[blue,fill=blue,opacity=0.6] (53.689813,0) -- (55.851150,0) -- (55.851150,0.054354) -- (53.689813,0.054354) -- cycle;
		\draw[blue,fill=blue,opacity=0.6] (55.851150,0) -- (58.003562,0) -- (58.003562,0.028893) -- (55.851150,0.028893) -- cycle;
		\draw[blue,fill=blue,opacity=0.6] (58.003562,0) -- (60.148006,0) -- (60.148006,0.008660) -- (58.003562,0.008660) -- cycle;
		\draw[blue,fill=blue,opacity=0.6] (60.148006,0) -- (62.285305,0) -- (62.285305,0.000000) -- (60.148006,0.000000) -- cycle;
		\draw[blue,fill=blue,opacity=0.6] (62.285305,0) -- (64.448402,0) -- (64.448402,0.000000) -- (62.285305,0.000000) -- cycle;
		\draw[red,fill=red,opacity=0.6] (0.000000,0) -- (2.700000,0) -- (2.700000,15.046971) -- (0.000000,15.046971) -- cycle;
		\draw[red,fill=red,opacity=0.6] (2.700000,0) -- (6.700000,0) -- (6.700000,2.006463) -- (2.700000,2.006463) -- cycle;
		\draw[red,fill=red,opacity=0.6] (6.700000,0) -- (10.700000,0) -- (10.700000,1.871244) -- (6.700000,1.871244) -- cycle;
		\draw[red,fill=red,opacity=0.6] (10.700000,0) -- (20.700000,0) -- (20.700000,0.498175) -- (10.700000,0.498175) -- cycle;
		\draw[red,fill=red,opacity=0.6] (20.700000,0) -- (30.700000,0) -- (30.700000,0.726216) -- (20.700000,0.726216) -- cycle;
		\draw[red,fill=red,opacity=0.6] (30.700000,0) -- (40.700000,0) -- (40.700000,0.987125) -- (30.700000,0.987125) -- cycle;
		\draw[red,fill=red,opacity=0.6] (40.700000,0) -- (50.700000,0) -- (50.700000,0.986972) -- (40.700000,0.986972) -- cycle;
		\draw[red,fill=red,opacity=0.6] (50.700000,0) -- (64.700000,0) -- (64.700000,0.848391) -- (50.700000,0.848391) -- cycle;
\end{tikzpicture}

%% file: secs/Discussion.tex
	Throughout this two-part paper, we have: 1) proposed an observation model in function spaces for image measurements
	of a 3D reaction-diffusion-adsorption-desorption system; 2) provided results that fully characterize our observation 
	model with respect to physical parameters and allow the generation of synthetic data; 3) proposed an optimization 
	problem in function spaces for inverse diffusion when the reactive term is spatially localized and temporally 
	continuous; 4) proposed sound methodology for solving the aforementioned optimization problem; 5) contributed a novel 
	proximal operator of the sum of two functions, i.e., that of the non-negative group-sparsity regularizer; 6) provided
	a discretization scheme that leads to practical, easy-to-approximate algorithms for synthesis and analysis of data based 
	on our methods and; 7) thoroughly examined the results of our optimization algorithm in terms of operational performance
	metrics and distributional recovery metrics.
	
		The proposed algorithm provides high-performing, unmatched SL detection results across a wide range of 
		realistic experimental conditions with remarkable and unique
		robustness to additive noise. Moreover, the source location estimates are very accurate, even when the 
		observed spot is the result of the combined emissions of several
		close sources. Additionally, although our algorithm requires an hyper-parameter $\lambda$, it is very robust to 
		its choice, which makes it a good candidate for practical use.
		
	Our study is not without limitations. In particular, because much of what concerns discretization of ill-posed 
	functional inverse problems is yet unknown, we provide no guideline
	for discretizations of our functional methods different than our own. In fact, even our own discretization is 
	argued intuitively, and theoretical results to strongly support it are left for further research. 
	Furthermore, because we focus on the analysis of the proposed optimization framework, we disregard the discussion 
	of convergence, relying only on the theoretical results on the convergence rate. In practice, however, theoretically
	sound approaches to speeding up the 
	convergence of proximal-gradient-based algorithms are available \cite{Xiao2013,Eghbali2017}, and many 
	heuristics can substantially reduce computations without a substantial loss in SL detection performance. 
	In particular,
	the masking function $\mu(\pos)$ in our optimization framework can be modified adaptively after some iterations 
	in order to
	discard regions that appear to have no cells, and thereby focus the algorithmic effort on more promising areas.
	Finally, because the computations involved in generating synthetic data are substantial, we limit the resolution 
	of the underlying source locations to $1~\mathrm{pix}$, and obtain the discretized kernels for generation from the hypothesis that 
	cells are pixel-centered.
	This could bias our analysis, since kernels within our algorithms are computed analogously. However, we consider 
	that this is unlikely, because the different discretization of the $\dsigma$-dimension in synthesis and analysis
	affects the kernels greatly, and because experimentation on real data yields similarly impressive results.
			
	In our work, we have also encountered paths for future research efforts. While discussing discretizations, we have 
	suggested representations with either thinner spatial grids or off-the-grid solutions to obtain super-resolution 
	location accuracy. Throughout the paper, we have suggested that further improvements in the estimation of the PSDR
	could enable the study and assessment of the per-cell secretion in a biochemical assay. These improvements could 
	plausibly be achieved through the weighting function $\xi(\sigma)$ that controls the group-sparsity regularizer, 
	which is supported in all of our theoretical results. Finally, through our discretized algorithm in this paper, 
	we have provided empirical evidence that tensor-based modeling of a matrix observation, through adequate group-sparsity
	coupling and non-negativity constraints, is a viable option for the reconstruction of highly-structured images.

%% file: secs/AppProxOps.tex
	In this appendix, we will provide and prove the results on proximal operators 
	upon which the proposed regularized algorithm relies.
	These will relate to the analysis of the functional
	$f$ in \eqref{eq:ffunction}, which represents both the 
	regularizer and the constraint imposed in \eqref{eq:InvDif:Regularised}. 
	Because this appendix includes the most technical functional-analytic derivations in the paper, we will 
	first introduce some extra notations, and urge the interested reader to 
	explore \cite{Luenberger1969,Bauschke2011} for details on optimization in function spaces and relevant references.
	
	\subsection{Notation} \label{ssec:AppNotation}
		
		Consider this section a continuation of Section \ref{ssec:Notation}.
		For any specific functional $f:\mathcal{Y}\rightarrow \reals$, 
		$f_-:\mathcal{Y}\rightarrow \reals$ is its negative part, i.e.,
		$
			f_-(y) = \min\lbrace f(y), 0 \rbrace\,,\forall y \in \mathcal{Y}\,,
		$
		and we have that $f=f_+ + f_-$.
		
		When discussing a Hilbert space $\X$, $\X^*$ is its dual space, and for any $x^*\in\X^*$,
		$r_{x^*}\in \X$ is its Riesz representation, i.e. $
			\linfunX{x^*}{x} = \prodX{r_{x^*}}{x}, \forall x \in \X$.
		Further, $\xp \in \X^*$ (and $\xn \in \X^*$) are the linear continuous functionals 
		represented by $[r_{x^*}]_+$ (and $[r_{x^*}]_-$, respectively), and $x^*=\xp + \xn$. We refer to $\xp$ and $\xn$ as dual-positive and dual-negative parts, respectively.
		
		When discussing a normed functional space, and given any strictly positive weight function, i.e., 
		$\xi \in \X_+$ such that $1/\xi = \xi^{-1} \in \X_+$, and $\gamma>0$, $
			\clellips{}{\xi}{\gamma} = \left\lbrace x\in\X : \normX{\xi^{-1} x}\leq \gamma\right\rbrace 
		$
		is the closed ellipsoid with constant $\xi^{-1}$-weighted norm under $\gamma$ and 
		$
			\clellips{*}{\xi}{\gamma} = \left\lbrace x^*\in\X^* : \normX{\xi^{-1} r_{x^*}}  \leq \gamma \right\rbrace 
		$
		is the closed dual ellipsoid with $\xi^{-1}$-weighted dual norm under $\gamma$. Additionally, 
		$\clellips{}{}{\gamma}$ is the closed ball in $\X$ with norm under $\gamma$.
		Finally, for any convex set $\mathcal{Z}\subset\X$, $\ProjOp{\mathcal{Z}}:\X\rightarrow \mathcal{Z}$
		is the projection operator onto it, i.e.,
		\begin{IEEEeqnarray*}{c}
			\Proj{\mathcal{Z}}{x} = \arg \underset{y\in \mathcal{Z}}{\min}\left[ \normX{y-x}^2 \right]\,.
		\end{IEEEeqnarray*}
	
	\subsection{Proximal Operator of the Positively-Constrained Weighted Norm in \texorpdfstring{$\LebTwo{\aleph}$}{L2(aleph)}} \label{ssec:ProxOp}

		Throughout this section, recall the weighting function $\xi\in\mathrm{L}_+^{\infty}\left[ 0, \sigmax\right]$ 
		introduced in Section~\ref{sec:Intro} and \cite[Section III]{AguilaPla2017}, and recall that we use $\aleph=\supp{\xi}$ and $\complement{\aleph}=[0,\sigmax]\setminus \aleph$. 
		Additionally, let $\X=\LebTwo{\aleph}$. 
		
		In Definition~\ref{def:PosContrNorm}, we introduce the functional that 
		characterizes the behavior of the constraint set and the regularizer in \eqref{eq:InvDif:Regularised} in the $\sigma$-dimension.
		
		\def\funcname{non-negative weighted norm in $\X$}
		\begin{definition}[Non-negative weighted norm in $\X$] \label{def:PosContrNorm}
			Define the the functional 
			\begin{IEEEeqnarray}{rl}
			\IEEEyesnumber \label{eq:relevantfunc} \IEEEyessubnumber* \label{eq:relevantfunc:l1}
					\vartheta:\X\, & \rightarrow \bar{\reals}_+ \\
					x\, & \mapsto \normX{\xi x} + \delta_{\X_+}(x), \forall x \in \X\,.
			\end{IEEEeqnarray}
		\end{definition}
		\def\sfuncname{scaled, \funcname}
		
		The main results of this appendix, which will be presented in Lemmas \ref{prop:ProxPosContrNorm}
		and \ref{prop:ProxScaPosContrNorm}, will provide the value of the
		proximal operator of $\gamma \vartheta$, i.e., $\prox_{\gamma \vartheta}(x)$, 
		$\forall x \in \X,\forall \gamma > 0$. This results are used in the proofs of 
		Theorems \ref{theorem:genvers} and \ref{theorem:ballvers}, which are presented at the end of the appendix. 
		In order to derive Lemmas \ref{prop:ProxPosContrNorm}
		and \ref{prop:ProxScaPosContrNorm}, we will follow a path similar to the classical proof of the proximal
		operator of a norm. We will first find the convex conjugate functional $\left(\gamma \vartheta\right)^*$
		in Lemma~\ref{prop:ConjFunctional}.
		Then, we will derive its proximal operator 
		$\prox_{(\gamma \vartheta)^*}(x^*)$, $\forall x^* \in \X^*,\forall \gamma > 0$ in
		Lemma~\ref{prop:ProxIndiDBall}. Finally, we will use
		this result and Moreau's identity to lead us into 
		Lemmas \ref{prop:ProxPosContrNorm} and \ref{prop:ProxScaPosContrNorm}. 
		
		\begin{lemma}[Fenchel conjugate of the \sfuncname] \label{prop:ConjFunctional}
			Consider the functional $\vartheta$ in Definition~\ref{def:PosContrNorm}.
			Then, $\forall \gamma >0$, we have that the convex conjugate functional of 
			$\gamma \vartheta$ is 
			\begin{IEEEeqnarray*}{rl}
				\left(\gamma \vartheta\right)^* : \X^* & \rightarrow \bar{\reals} \\
				x^* & \mapsto \delta_{\clellips{*}{\xi}{\gamma}}( \xp )\,, \forall x^* \in \X^*,
			\end{IEEEeqnarray*}
			with $\clellips{*}{\xi}{\gamma}$ as defined in the previous section.
		\end{lemma}
			\begin{IEEEproof}
				Here, we will instead show that the Fenchel conjugate of $\delta_{\clellips{*}{\xi}{\gamma}}( \xp )$ is
				the functional $\hat{\vartheta}:\X\rightarrow \bar{\reals}$ such that
				\begin{IEEEeqnarray*}{c}
					\hat{\vartheta} = \left[\delta_{\clellips{*}{\xi}{\gamma}}(\xp)\right]^* = \gamma \vartheta\,.
				\end{IEEEeqnarray*}
				The Fenchel-Moreau theorem then allows us to conclude that, because $\vartheta$ is convex, proper and lower 
				semi-continuous, $\delta_{\clellips{*}{\xi}{\gamma}}( \xp )$ is the Fenchel conjugate of $\gamma \vartheta$.
				
				Starting now from the definition of $\hat{\vartheta}$ we obtain 
				\begin{IEEEeqnarray*}{rl}
					\hat{\vartheta}(x) \,\,& = 
					\sup_{x^*\in\X^*}\left\lbrace \linfunX{x^*}{x} - \delta_{\clellips{*}{\xi}{\gamma}}(\xp) \right\rbrace\\
					&= \sup_{x^*\in\X^*}\left\lbrace \linfunX{\xn}{x} + \linfunX{\xp}{x} - \delta_{\clellips{*}{\xi}{\gamma}}(\xp) \right\rbrace\,.
				\end{IEEEeqnarray*}
				Here, we can readily determine that, if $x\not\in\X_+$, $\hat{\vartheta}(x)=+\infty$. Indeed, if $\exists S \subset \aleph$ such that 
				$x<0~\ae$ in $S$, we have that for $\hat{\vartheta}(x)\geq \sup_{x^*\in\X^*}\linfunX{\xn}{x} \geq \sup_{K<0} K \int_S x = +\infty$. 
				
				We continue by noting that, if $x\in\X_+$, then $\linfunX{\xn}{x}\leq0$, and thus, it will be enough to consider the case $x^* = \xp$, i.e., $\xn=0$,
				to determine $\hat{\vartheta}(x)$. Therefore, for any $x\in \X_+$ we can use the Cauchy-Schwartz inequality to show that
				\begin{IEEEeqnarray*}{rl}
					\hat{\vartheta}(x) & = \sup_{x^*\in\X^*}\left\lbrace \linfunX{\xp}{x} - \delta_{\clellips{*}{\xi}{\gamma}}(\xp) \right\rbrace \\
									   & = \sup_{x^*\in\X^*}\left\lbrace \linfunX{x^*}{x} - \delta_{\clellips{*}{\xi}{\gamma}}(x^*) \right\rbrace \\
									   & = \sup_{x^*\in \clellips{*}{\xi}{\gamma}}\left\lbrace \linfunX{x^*}{x} \right\rbrace \\
									   & = \sup_{r_{x^*}\in \clellips{}{\xi}{\gamma}}\left\lbrace \prodX{\xi^{-1} r_{x^*}}{\xi x} \right\rbrace
									    = \gamma \normX{\xi x}\,.
				\end{IEEEeqnarray*}
				
				In conclusion, then, 
				\begin{IEEEeqnarray*}{rl}
					\hat{\vartheta}(x) & = \left\lbrace 
										 \begin{array}{ll}
					                     	+\infty & \mbox{ if }x\not\in\X_+, \\
					                     	\gamma \normX{\xi x} & \mbox{ if }x\in\X.
					                     \end{array}
					                     \right\rbrace \\
					                     & = \gamma \normX{\xi x} + \delta_{\X_+}(x) = \gamma \vartheta\,,
				\end{IEEEeqnarray*}
				which finishes our proof.

			\end{IEEEproof}

		Similarly to what happens with the dual of a norm, the dual functional $\left( \gamma \vartheta \right)^*$
		is a simple indicator. This makes its proximal operator 
		in Lemma~\ref{prop:ProxIndiDBall} a combination of simple, standard operations, such as 
		dual-positive and dual-negative parts, and projections onto convex sets.
		
		\begin{lemma}[Projection of the positive part on the dual ellipsoid] \label{prop:ProxIndiDBall}
			Consider the functional 
			\begin{IEEEeqnarray*}{rl}
					\zeta:\X^* & \rightarrow\lbrace0,+\infty\rbrace \\
					x^* & \mapsto \delta_{\clellips{*}{\xi}{\gamma}}(\xp),\, \forall x^*\in\X^*\,,
			\end{IEEEeqnarray*} 
			i.e., $\zeta = (\gamma \vartheta)^*$.
			Then,
			\begin{IEEEeqnarray*}{c}
				\prox_{\zeta}(x^*) = \xn + \Proj{\clellips{*}{\xi}{\gamma}}{\xp}\,.
			\end{IEEEeqnarray*}
		\end{lemma}
			\begin{IEEEproof}
					Recall here that obtaining $\prox_\zeta(x^*)$ is obtaining the minimizer of
					\begin{IEEEeqnarray}{c} \label{eq:optProbProxIndi}
						\underset{y^*\in\X^*}{\min}
														\left[ \frac{1}{2}\normXs{y^*-x^*}^2 +  \delta_{\clellips{*}{\xi}{\gamma}}(y^*_\mathrm{p})   \right]\,.
					\end{IEEEeqnarray}
					\def\yopt{y^*_{\mathrm{ opt}}}
					In this proof, let us refer to this minimizer as $\yopt$.
					The solution to \eqref{eq:optProbProxIndi} is intuitively simple. On one hand, because the value $\zeta(y^*)$ does not vary 
					with changes in the negative part of $y^*$, we will have that the minimization of the
					term $\normXs{y^*-x^*}^2$ will dominate the negative part of the optimal solution
					and $y^*_{\mathrm{ opt},\mathrm{ n}} = \xn$. 
					On the other hand, the positive part of $y^*$ is only constrained to be in the ellipsoid 
					$\clellips{*}{\xi}{\gamma}$ and, thus, the positive part of the solution will be the point at minimum distance 
					from $\xp$ inside the ellipsoid, i.e., $
						y^*_{\mathrm{ opt},\mathrm{ p}} = \Proj{\clellips{*}{\xi}{\gamma}}{\xp}
					$.
					
						Let us now formalize this by considering any element $x^*\in \X^*$, and letting  \def\Nx{N_{x^*}}
						$
							\Nx = \lbrace \sigma \in \aleph: r_{x^*}(\sigma) < 0 \rbrace = \supp{r_{x_n^*}}
						$.
						For any $y^* \in \X^*$, let $p_{y^*},n_{y^*}\in\X$ be such that 
						\begin{IEEEeqnarray}{c} \label{eq:supports}
							\supp{p_{y^*}}\subset\complement{\Nx}, \,\supp{n_{y^*}}\subset\Nx,
						\end{IEEEeqnarray}
						with $\complement{\Nx}=\aleph \setminus \Nx$, and
						\begin{IEEEeqnarray}{c} \label{eq:sumtoy}
							r_{y^*} = p_{y^*} + n_{y^*}\,.
						\end{IEEEeqnarray}
						Then,
						\begin{IEEEeqnarray*}{rl}
							\normXs{y^*-x^*}^2 & = \int_{\complement{\Nx}} \left(r_{\xp} -p_{y^*} \right)^2 \nonumber \\
							& + \int_{\Nx} \left(r_{\xn} - n_{y^*} \right)^2 \nonumber \\
							& = \normX{r_{\xp} -p_{y^*}}^2 + \normX{r_{\xn} - n_{y^*}}^2
							\,. \label{eq:DNormInTwo}
						\end{IEEEeqnarray*}
						Therefore, \eqref{eq:optProbProxIndi}
						is equivalent to 
						\begin{IEEEeqnarray*}{c}
							\underset{\underset{\mbox{\tiny s.t. } \left[p_{y^*}\right]_+ +\left[n_{y^*}\right]_+ \in \clellips{}{\xi}{\gamma}}{p_{y^*},n_{y^*}\in\X}}{\min}\left[ 
							\frac{1}{2}\normX{r_{\xp} - p_{y^*}}^2 + \frac{1}{2}\normX{r_{\xn} - n_{y^*}}^2  \right] 
						\end{IEEEeqnarray*}
						as long as \eqref{eq:supports} and \eqref{eq:sumtoy} are fulfilled.
				
						We will now prove that $n_{y^*_{\mathrm{ opt}}}  \leq 0~\ae$ in $\aleph$, which will decouple 
						the minimization of the two summands in the problem above. 
						Assume that $y^*_{\mathrm{ opt}} \in \X^*$
						is an optimal point of  \eqref{eq:optProbProxIndi} that does not fulfill this condition, i.e., that if
						\begin{IEEEeqnarray*}{c}
							\Theta =\left\lbrace \sigma \in \Nx : n_{y^*_{\mathrm{ opt}}}(\sigma) > 0\right\rbrace\mbox{, then }\int_\Theta  n_{y^*_{\mathrm{ opt}}}>0\,.
						\end{IEEEeqnarray*}
						Let $y^*_1 \in \X^*$ such that
						$
							p_{y_1^*} = p_{y^*_{\mathrm{ opt}}}$, and 
						$	n_{y_1^*} = \left[n_{y^*_{\mathrm{ opt}}}\right]_-$.
						Then, because $y^*_{\mathrm{ opt}}$ was a feasible point, i.e.,
						$y^*_{\mathrm{ opt},\mathrm{ p}} \in \clellips{*}{\xi}{\gamma}$, we have that
						\begin{IEEEeqnarray*}{rl}
							\gamma^2 & \geq \int_{\aleph} \xi^{-2} 
							\left( \left[p_{y_{\mathrm{ opt}}^*}\right]_+ + \left[n_{y_{\mathrm{ opt}}^*}\right]_+ \right)^2 \\
							& \geq 
							\int_{\aleph} \xi^{-2} \left[p_{y_{\mathrm{ opt}}^*}\right]^2_+ = \|y^*_{1,\mathrm{ p}}\|^2_{\X^*}\,,
						\end{IEEEeqnarray*}
						i.e., $y^*_{1,\mathrm{ p}} \in \clellips{*}{\xi}{\gamma}$ and $y^*_{1}$ is a feasible point. 
						Moreover, $\forall \sigma \in \Theta $, we have that 
						$\left|r_{\xn}(\sigma) -  n_{y^*_{\mathrm{ opt}}}(\sigma)\right| > \left| r_{\xn}(\sigma) \right|$
						and thus
						\begin{IEEEeqnarray}{rl}
							\normX{r_{\xn} - n_{y^*_{\mathrm{ opt}}}}^2 &= \int_{\Nx}  
							\left(r_{\xn} -  n_{y^*_{\mathrm{ opt}}} \right)^2   \nonumber \\
							&> \int_{\Nx\setminus \Theta }  
							\left(r_{\xn} -  n_{y^*_{\mathrm{ opt}}} \right)^2 
							+ \int_{\Theta } r_{\xn}^2 \nonumber \\
							&= \normX{r_{\xn} - n_{y^*_{1}}}^2\,, \label{eq:refertotrick}
						\end{IEEEeqnarray}
						which implies that $
							\normXs{y_{\mathrm{ opt}}^*-x^*}^2 > \normXs{y_{1}^*-x^*}^2
						$.
						This contradicts the optimality of $y_{\mathrm{ opt}}^*$. 
						Thus, an optimal point $y^*_{\mathrm{ opt}}$ must fulfill
						$n_{y^*_{\mathrm{ opt}}} \leq 0~\ae$ in $\aleph$.
				
						Therefore, \hyperref[eq:constrainedPart]{$\left(\ref*{eq:bothparts}\right)$} is equivalent to  \eqref{eq:optProbProxIndi}, 
						as long as conditions \eqref{eq:supports} and \eqref{eq:sumtoy} are fulfilled.
						\begin{IEEEeqnarray}{c} \IEEEyesnumber \label{eq:bothparts} \IEEEyessubnumber* \label{eq:constrainedPart}
							\underset{\underset{\mbox{\tiny s.t. } \left[p_{y^*}\right]_+ \in \clellips{}{\xi}{\gamma}}{p_{y^*}\in\X}}{\min}\left[ 
							\frac{1}{2}\normX{r_{\xp} - p_{y^*}}^2   \right] \\
						 \label{eq:unconstrainedPart}
							\underset{n_{y^*}\in\X}{\min}\left[ \frac{1}{2}\normX{r_{\xn} - n_{y^*}}^2 \right]
						\end{IEEEeqnarray}
						 \eqref{eq:unconstrainedPart} is an unconstrained norm minimization, and has its minimum at ${n_{y_{\mathrm{ opt}}^*}=r_{\xn}}$,
						which fulfills \eqref{eq:supports}.
						Using an argument parallel to the one that lead to \eqref{eq:refertotrick}, we have that  
						$r_{\xp} \geq 0~\ae$ in $\aleph$ implies that $p_{y_{\mathrm{ opt}}^*}\geq 0~\ae$ 
						in $\aleph$ too. Thus,  \eqref{eq:constrainedPart} is equivalent to
						\begin{IEEEeqnarray*}{c}
							\underset{p_{y^*}\in \clellips{}{\xi}{\gamma}}{\min}\left[ 
							\frac{1}{2}\normX{r_{\xp} - p_{y^*}}^2   \right] \,,
						\end{IEEEeqnarray*}
						and thus, $
							p_{y_{\mathrm{ opt}}^*} = \Proj{\clellips{}{\xi}{\gamma}}{r_{\xp}}$.
						In Property~\ref{prop:ProjEli}, we obtain an expression for $\Proj{\clellips{}{\xi}{\gamma}}{x}$ for any $x\in\X$ that shows that
						$\supp{\Proj{\clellips{}{\xi}{\gamma}}{x}} \subset \supp{x}$
						and, thus, \eqref{eq:supports} is fulfilled. 
						Then, the solution to \eqref{eq:optProbProxIndi} is given by \eqref{eq:sumtoy} as the $y_{\mathrm{ opt}}^*\in\X^*$ 			
						represented by $
							r_{y_{\mathrm{ opt}}^*} = r_{\xn} + \Proj{\clellips{}{\xi}{\gamma}}{r_{\xp}}
						$, 
						i.e.,
						\begin{IEEEeqnarray*}{c}
							y_{\mathrm{ opt}}^* = \xn + \Proj{\clellips{*}{\xi}{\gamma}}{\xp}\,.
						\end{IEEEeqnarray*}			
			\end{IEEEproof}
			
		We now can use the relation between the proximal operator of a functional and that of its convex conjugate to finally achieve the desired result
		in Lemma~\ref{prop:ProxPosContrNorm}.  
					
		\begin{lemma}[Proximal operator of the \sfuncname]
		\label{prop:ProxPosContrNorm}
			Consider the functional $\vartheta$ in Definition~\ref{def:PosContrNorm}.
			Then, $\forall \gamma >0$, we have that the proximal operator of the functional $\gamma \vartheta$ is
			\begin{IEEEeqnarray*}{c}
				\prox_{\gamma \vartheta}(x) = x_+ - \Proj{\clellips{}{\xi}{\gamma}}{x_+}, \forall x\in\X\,.
			\end{IEEEeqnarray*}
		\end{lemma}
			\begin{IEEEproof}
					Lemmas \ref{prop:ConjFunctional} and \ref{prop:ProxIndiDBall} grant that 
					$
						\prox_{\left(\gamma \vartheta \right)^*}(x^*) = \xn + \Proj{\clellips{*}{\xi}{\gamma}}{\xp}
					$.
					A well-known generalization of Moreau's decomposition theorem for projection on convex cones in Hilbert 
					spaces is that
					\begin{IEEEeqnarray}{c} \label{eq:Moreau}
						\prox_{\gamma \vartheta}(x) + \prox_{\left(\gamma \vartheta \right)^*} (x) = x\,.
					\end{IEEEeqnarray}
					Note here that we abuse the notation by identifying $\X$ with its dual $\X^*$ and 
						$\prox_{\left(\gamma \vartheta \right)^*} (x)$
					with 
					$	r_{\prox_{\left(\gamma \vartheta \right)^*} (x^*)}\in\X$
					such that $x^*\in\X^*$ is represented by $r_{x^*} = x$.
					Directly from \eqref{eq:Moreau}, then, we obtain that
					\begin{IEEEeqnarray*}{rl}
						\prox_{\gamma \vartheta}(x) &= x - \prox_{\left(\gamma \vartheta \right)^*} (x) \\
						&= x_+ - \Proj{\clellips{}{\xi}{\gamma}}{x_+}\,.
					\end{IEEEeqnarray*}
			\end{IEEEproof}
		
		Although we now have our result compactly expressed in terms of simple, known operations,
		the inherent optimization problem in the term $\Proj{\clellips{}{\xi}{\gamma}}{x_+}$
		is known to have no closed-form solution. For completeness, we include this result in Property~\ref{prop:ProjEli}.
		
		\begin{property}[Projection on an ellipsoid] \label{prop:ProjEli}
			The projection of a functional $x\in \X$ onto the closed ellipsoid
			$\clellips{}{\xi}{\gamma}$ is
			\begin{IEEEeqnarray*}{c}
				\Proj{\clellips{}{\xi}{\gamma}}{x} =
				\begin{cases}
					x & \mbox{ if } x\in \clellips{}{\xi}{\gamma}, \\
					 \frac{\xi^2 }{\xi^2 + 2\lambda} x & \mbox{ if }
					x \in \X\setminus\clellips{}{\xi}{\gamma},					
				\end{cases}
			\end{IEEEeqnarray*}
			with $\lambda \geq 0$ such that
			\begin{IEEEeqnarray*}{c}
				\normX{ \frac{\xi}{\xi^2 + 2\lambda} x} = \gamma\,.
			\end{IEEEeqnarray*}
		\end{property}
			\begin{IEEEproof}
				Recall here that the projection operator is defined as
				\begin{IEEEeqnarray}{c} \label{eq:ProblemProjEli}
					\Proj{\clellips{}{\xi}{\gamma}}{x} =
					\arg \underset{y\in \clellips{}{\xi}{\gamma}}{\min}\left[ 
							\frac{1}{2}\normX{x - y}^2   \right]\,.
				\end{IEEEeqnarray}
				Because $\X$ is complete and $\clellips{}{\xi}{\gamma}$ is convex and closed, the projection
				operator is well defined and strong Lagrange duality is granted. Note that the convexity of 
				$\clellips{}{\xi}{\gamma}$ is granted by the convexity of the weighted norm 
					$\normX{\xi^{-1} \,\cdot\,}$,
				which follows directly from the convexity of the norm $\normX{\cdot}$. 
				
				The Lagrangian for this problem is
				\begin{IEEEeqnarray*}{rl}
					L(y,\lambda) &= \frac{1}{2}\normX{x - y}^2 +
					\lambda \left (\normX{\xi^{-1} y}^2 - \gamma^2\right) \\
					& = \prodX{y\left[ \frac{1}{2}+\lambda \xi^{-2} \right]}{y} + 
					  \frac{1}{2}\prodX{x}{x} - \prodX{x}{y} - \lambda \gamma^2
				\end{IEEEeqnarray*}
				with $\lambda \geq 0$. Because the Lagrangian $L(y,\lambda)$ is 
				convex and Fréchet differentiable with respect to $y\in\X$, and its Fréchet derivative is
				$
					\nabla_y L(y,\lambda) = 
					2 y \left[\frac{1}{2}+\lambda \xi^{-2} \right] -x
				$, 
				its minimizer $y_{\mathrm{ opt}}\in\X$ is 
				\begin{IEEEeqnarray*}{c}
					y_{\mathrm{ opt}}(\lambda) = \frac{1}{2} \frac{ x}{\frac{1}{2} + \lambda \xi^{-2}}
					= \frac{\xi^2}{\xi^2 + 2\lambda}x\,.
				\end{IEEEeqnarray*}
				
				The dual function for  \eqref{eq:ProblemProjEli} is
				\begin{IEEEeqnarray*}{rl}
					h(\lambda) &= L(y_{\mathrm{ opt}}, \lambda) \\
					&= -\lambda \gamma^2 +\frac{1}{2}\left[ \normX{x}^2 -\prodX{x}{\frac{1}{2} \frac{ x}{\frac{1}{2} + \lambda \xi^{-2}}} \right] \\
					&= \frac{\normX{x}^2}{2} -\lambda \gamma^2 
					- \int_{\aleph} \frac{x^2}{4} \frac{1}{\frac{1}{2} + \lambda \xi^{-2}}\,,
				\end{IEEEeqnarray*}
				which is concave in $\lambda \geq 0$ and, thus, has its maximum at 
				either $\lambda_{\mathrm{ opt},1} = 0$ or at that 
				$\lambda_{\mathrm{ opt},2}$ that yields
				\begin{IEEEeqnarray*}{c}
					\frac{\partial}{\partial \lambda}h  
					= - \gamma^2 + \int_\aleph \frac{x^2}{4} \frac{\xi^{-2}}{\left(\frac{1}{2}+\lambda_{\mathrm{ opt},2} \xi^{-2}\right)^2}
					 = 0\,,
				\end{IEEEeqnarray*}
				i.e.,
					$\normX{\xi^{-1} y_{\mathrm{ opt}}(\lambda_{\mathrm{ opt},2})} = \gamma$.
				Generally, the value of $\lambda_{\mathrm{ opt},2}$ cannot be obtained in closed form.
				
				If $x\in \clellips{}{\xi}{\gamma}$, we know that the optimal value
				for  \eqref{eq:ProblemProjEli} is $0$ and is achieved at $y_{\mathrm{ opt}}=x$,
				which implies that the optimal Lagrange multiplier is $\lambda=
				\lambda_{\mathrm{ opt},1} = 0$. If 
				$x\in \X\setminus\clellips{}{\xi}{\gamma}$, we know that the optimal
				value for  \eqref{eq:ProblemProjEli} must be larger than zero,
				which by strong duality implies that $\lambda \neq 0$ and, thus,
				that the optimal Lagrange multiplier is 
				$\lambda = \lambda_{\mathrm{ opt},2}$ and the optimal 
				primal point is $y_{\mathrm{ opt}}(\lambda_{\mathrm{ opt},2})$, 
				which is primal-feasible by definition.
			\end{IEEEproof}
		
		This result determines the shape of $\Proj{\clellips{}{\xi}{\gamma}}{x_+}$, 
		but it does not give a closed-form expression for it. To find this projection, the value of 
		$\lambda$ in Property~\ref{prop:ProjEli} has to be found. Several numerical methods have been developed to find this value
		or otherwise compute the projection on an ellipsoid \cite{Jia2017}. Using these in the context of our problem, however, is outside the scope of our paper. 
		We opt instead for particularizing in Property~\ref{prop:ball} to cases in which the weighting function is constant $\ae$ in $\aleph$.
		This makes the projection to be computed, without loss of generality, $\Proj{\clellips{}{}{\gamma}}{x_+}$, the projection onto a closed ball in $\X$.
		
		\begin{property}[Projection on a ball] \label{prop:ball}
			The projection of a functional $x\in \X$ onto the closed ball of norm under $\gamma$, i.e.,
			$\clellips{}{}{\gamma}$, is
			\begin{IEEEeqnarray*}{c}
				\Proj{\clellips{}{}{\gamma}}{x} =
				\begin{cases}
					x & \mbox{ if } x\in \clellips{}{}{\gamma}, \\
					 \frac{\gamma}{\normX{x}} x & \mbox{ if }
					x \in \complement{\clellips{}{}{\gamma}}.
				\end{cases}
			\end{IEEEeqnarray*}
		\end{property}
			\begin{IEEEproof}
				Note that this is nothing but a particular case of Property~\ref{prop:ProjEli} in which
				$\xi = 1~\ae$ in $\aleph$. Then, the case $x\in \clellips{}{}{\gamma}$ is trivial. For 
				$x \in \complement{\clellips{}{}{\gamma}}$, the equation for $\lambda \geq 0$ in Property 
				\ref{prop:ProjEli} can be solved in closed form, yielding
				\begin{IEEEeqnarray*}{c}
					\normX{ \frac{ x}{1 + 2\lambda}} =\left|\frac{1}{1+2\lambda}\right| \normX{x} = \frac{1}{1 + 2 \lambda} \normX{x}
					= \gamma\,,
				\end{IEEEeqnarray*}
				i.e.,
					$\lambda = \frac{1}{2}\left(\frac{\normX{x}}{\gamma}-1\right)$,
				and
				\begin{IEEEeqnarray*}{c}
					y_{\mathrm{ opt}}(\lambda) = \frac{\xi^2 x}{\xi^2 + 2\lambda} =
					\frac{ x}{1 + 2 \frac{\frac{\normX{x}}{\gamma}-1}{2}} = \frac{\gamma}{\normX{x}} x\,.
				\end{IEEEeqnarray*}
			\end{IEEEproof}
		
		This allows us to obtain a closed-form version of Lemma~\ref{prop:ProxPosContrNorm} for this specific choice of $\xi$, i.e., Lemma~\ref{prop:ProxScaPosContrNorm}.
		
		\begin{lemma}[Proximal operator of the scaled, non-negative norm in $\X$]
		\label{prop:ProxScaPosContrNorm}
			Consider the functional $\vartheta$ in Definition~\ref{def:PosContrNorm} when $\xi=1~\ae$ in $\aleph$.
			Then, $\forall \gamma >0$, we have that the proximal operator of the functional $\gamma \vartheta$ is
			\begin{IEEEeqnarray*}{c}
				\prox_{\gamma \vartheta}(x) = x_+ \left( 1 - \frac{\gamma}{\normX{x_+}}  \right)_+\,.
			\end{IEEEeqnarray*}
		\end{lemma}
			\begin{IEEEproof}
				Using Lemma~\ref{prop:ProxPosContrNorm} and Property~\ref{prop:ball} we obtain that
				\begin{IEEEeqnarray*}{rl}
					\prox_{\gamma \vartheta}(x) &= x_+ - \Proj{\clellips{}{}{\gamma}}{x_+} \\
					&= \begin{cases}
							0 & \mbox{ if } \normX{x_+} / \gamma \leq 1, \\
							x_+ \left( 1 - \frac{\gamma}{\normX{x_+}} \right) & \mbox{ if } \normX{x_+} / \gamma > 1,
					   \end{cases} \\
					&= x_+ \left( 1 - \frac{\gamma}{\normX{x_+}} \right)_+\,.
				\end{IEEEeqnarray*}
			\end{IEEEproof}
		
		We now use the results above to prove Theorems \ref{theorem:genvers} and \ref{theorem:ballvers}, which constitute the backbone of Alg.~\ref{algs:AccProxGradforRegInvDif}.
		
		\begin{IEEEproof}[Proof - Theorem~\ref{theorem:genvers} (Proximal operator of the non-negative weighted group-sparsity regularizer)]
				Consider the functional $\vartheta$ in Definition~\ref{def:PosContrNorm}. Then, recalling the functional $f$ in \eqref{eq:PositiveGroupSparsity}, we have
				\begin{IEEEeqnarray}{rl}
					\gamma f(a) & = \delta_{\ASpace_+}(a) + \gamma \lambda \left\| \left\|\xi a_\pos\right\|_{\X} \right\|_{\mathrm{L}^1\left(\reals^2\right)}
					\IEEEyesnumber \label{eq:pthgen:normonlyaleph} \IEEEyessubnumber* \label{eq:pthgen:normonlyaleph:l1}\\
					&= \delta_{\ASpace_+}(a) + \gamma \lambda \int_{\reals^2} \left\|\xi a_{\aleph,\pos}\right\|_{\X} \dint\pos \\
					& = \int_{\reals^2} \left( \delta_{\mathrm{L}_+^2[0,\sigmax]}(a_\pos) +\gamma  \lambda \left\|\xi a_{\aleph,\pos}\right\|_{\X}\right) \dint\pos 
					\IEEEyesnumber \label{eq:pthgen:indicatorsplay} \IEEEyessubnumber* \label{eq:pthgen:indicatorsplay:l1}\\
					& = \int_{\reals^2} \left( \delta_{\LebTwoP{\complement{\aleph}}}(a_{\complement{\aleph},\pos}) +
						\gamma \lambda \vartheta\left(a_{\aleph,\pos}\right) \right) \dint\pos\,.
				\end{IEEEeqnarray}
				Here, \hyperref[eq:pthgen:normonlyaleph:l1]{(\ref*{eq:pthgen:normonlyaleph})} uses that $\xi a_\pos=0~\ae$ in $\complement{\aleph}$,
				and \hyperref[eq:pthgen:indicatorsplay:l1]{(\ref*{eq:pthgen:indicatorsplay})} uses that $a\in\ASpace_+$ is equivalent to $a_\pos \in \mathrm{L}_+^2[0,\sigmax]$ for almost
				any $\pos\in\reals^2$, and that is equivalent to $a_{\complement{\aleph},\pos} \in \LebTwoP{\complement{\aleph}}$ and $a_{\aleph,\pos} \in \LebTwoP{\aleph} = \X_+$
				for almost any $\pos\in\reals^2$.
				
				$\prox_{\gamma f}(a)$ is the minimizer to
				\begin{IEEEeqnarray*}{c}
					\min_{b\in \ASpace} \left[ 
						\frac{1}{2}\normA{b-a}^2 + \gamma f(b)
					\right]\,,
				\end{IEEEeqnarray*}
				or, equivalently,
				\begin{IEEEeqnarray*}{rl}
					\min_{b\in \ASpace} \Bigg[ 
						\int_{\reals^2} \bigg( 
						&\frac{1}{2}\normLebTwo{\complement{\aleph}}{b_{\pos,\complement{\aleph}} - a_{\complement{\aleph},\pos}}^2
						+ \delta_{\LebTwoP{\complement{\aleph}}}(b_{\complement{\aleph},\pos}) \\
						{+}\:&\frac{1}{2}\normX{b_{\pos,\aleph} - a_{\aleph,\pos}}^2 +
						\gamma \lambda \vartheta\left(b_{\aleph,\pos}\right) \bigg) \dint\pos
					\Bigg]\,.				 
				\end{IEEEeqnarray*}
				By linearity of the integral, then, the optimization of $b$ for $\sigma\in\aleph$ and $\sigma\in\complement{\aleph}$
				is completely decoupled. Furthermore, if obtaining the minimizer $b_{\pos,\mathrm{opt}}\in\mathrm{L}[0,\sigmax]$ 
				of the term inside the integral for each $\pos\in\reals^2$ and constructing $b_{\mathrm{opt}}$ such that
				$b_{\mathrm{opt}}(\sigma,\pos) = b_{\pos,\mathrm{opt}}(\sigma), \forall \sigma \in [0,\sigmax]$ yields $b_{\mathrm{opt}} \in \ASpace$,
				$b_{\mathrm{opt}}$ will be optimal with respect to the problem above. 
				In this light, we consider first the optimization for $\sigma\in\complement{\aleph}$ and a specific $\pos\in\reals^2$, i.e.,
				\begin{IEEEeqnarray*}{c}
					\arg \min_{b_{\pos,\complement{\aleph}} \in \LebTwoP{\complement{\aleph}}} 
					\left[ \frac{1}{2}\normLebTwo{\complement{\aleph}}{b_{\pos,\complement{\aleph}} - a_{\complement{\aleph},\pos}}^2 \right] = \left[ a_{\complement{\aleph},\pos} \right]_+\,,
				\end{IEEEeqnarray*}
				and see that it is resolved by a simple non-negative projection.
				Then, we observe that the optimization for $\sigma\in\aleph$ and a specific $\pos\in\reals^2$ is of the form considered in Lemma \ref{prop:ProxPosContrNorm},
				which resolved it to $\left[ a_{\aleph,\pos} \right]_+ - \Proj{\clellips{}{\xi}{\gamma\lambda}}{\left[ a_{\aleph,\pos} \right]_+}$. 
				We then have that
				\begin{IEEEeqnarray*}{rl}
					b_{\pos,\mathrm{opt}} &= \left[ a_{\complement{\aleph},\pos} \right]_+ + \left[ a_{\aleph,\pos} \right]_+ - \Proj{\clellips{}{\xi}{\gamma\lambda}}{\left[ a_{\aleph,\pos} \right]_+} \\
					& = \left[ a_{\pos} \right]_+ - \Proj{\clellips{}{\xi}{\gamma\lambda}}{\left[ a_{\aleph,\pos} \right]_+}\,.
				\end{IEEEeqnarray*}
				Note that 
				\begin{IEEEeqnarray}{c} \label{eq:fulfill}
					\left\|b_{\pos,\mathrm{opt}}\right\|^2_{\mathrm{L}^2[0,\sigmax]} \leq \left\| \left[a_\pos\right]_+\right\|^2_{\mathrm{L}^2[0,\sigmax]} \leq 
					\left\|a_\pos \right\|^2_{\mathrm{L}^2[0,\sigmax]},\,\,\,\,\,\,\,\,
				\end{IEEEeqnarray}
				and, thus, $a\in\ASpace$ implies that $b_{\mathrm{opt}}\in \ASpace$,
				which completes the proof of Theorem \ref{theorem:genvers}.
		\end{IEEEproof}
		
		\begin{IEEEproof}[Proof - Theorem~\ref{theorem:ballvers} (Proximal operator of the non-negative group-sparsity regularizer on $\aleph$)]
			Using the same proof structure as in Theorem~\ref{theorem:genvers}, but using Lemma \ref{prop:ProxScaPosContrNorm} for the optimization for $\sigma \in \aleph$ and a
			specific $\pos \in \reals^2$, we obtain that
			\begin{IEEEeqnarray*}{c}
				b_{\pos,\mathrm{opt}} 
				= \left[ a_{\complement{\aleph},\pos} \right]_+ + \left[ a_{\aleph,\pos} \right]_+  \left( 1 - \frac{\gamma \lambda}{\normLebTwo{\aleph}{\left[a_{\aleph,\pos}\right]_+}} \right)_+\,,
			\end{IEEEeqnarray*}
			and \eqref{eq:fulfill} implies that $b_{\mathrm{opt}}\in \ASpace$, which concludes the proof.
		\end{IEEEproof}

	\ifbool{tot}{}{
		\bibliographystyle{IEEEtran}
		\bibliography{IEEEabrv,\bib/multi_deconv}
	}